\documentclass{aa}  

\usepackage{graphicx}
\usepackage{txfonts}
\usepackage{xcolor}
\begin{document}

   \title{The GAPS Programme at TNG. XLI. The climate of KELT-9b revealed with a new approach to high spectral resolution phase curves.}
    \titlerunning{A HRS phase curve of KELT-9b}

   \subtitle{}

   \author{L. Pino
          \inst{1}
          \and
          M. Brogi\inst{2,3,4}
          \and
          J. M. D\'esert\inst{5}
          \and
          V. Nascimbeni\inst{6}
          \and
          A. S. Bonomo\inst{3}
          \and
          E. Rauscher\inst{7}
          \and
          M. Basilicata\inst{8}
          \and
          K. Biazzo\inst{9}
          \and
          A. Bignamini\inst{10}
          \and
          F. Borsa\inst{11}
          \and
          R. Claudi\inst{6}
          \and
          E. Covino\inst{12}
          \and
          M. P. Di Mauro\inst{13}
          \and
          G. Guilluy\inst{3}
          \and
          A. Maggio\inst{14}
          \and
          L. Malavolta\inst{15}
          \and
          G. Micela\inst{14}
          \and
          E. Molinari\inst{16}
          \and
          M. Molinaro\inst{10}
          \and
          M. Montalto\inst{9}
          \and
          D. Nardiello\inst{6, 17}
          \and
          M. Pedani\inst{18}
          \and
          G. Piotto\inst{15}
          \and
          E. Poretti\inst{18, 11}
          \and
          M. Rainer\inst{11}
          \and
          G. Scandariato\inst{9}
          \and
          D. Sicilia\inst{9}
          \and
          A. Sozzetti\inst{3}
          }

   \institute{INAF -- Osservatorio Astrofisico di Arcetri,
              Largo Enrico Fermi 5, 50125 Firenze\\
              \email{lorenzo.pino@inaf.it} 
        \and
            Department of Physics, University of Warwick, Coventry, CV4 7AL, UK 
        \and
            INAF -- Osservatorio Astrofisico di Torino, Via Osservatorio 20, I-10025, Pino Torinese, Italy 
        \and
            Centre for Exoplanets and Habitability, University of Warwick, Coventry, CV4 7AL, UK 
        \and
            Anton Pannekoek Institute for Astronomy, University of Amsterdam, 1098 XH Amsterdam, The Netherlands 
        \and
            INAF -- Osservatorio Astronomico di Padova, Padova 35122, Italy 
        \and
            Department of Astronomy, University of Michigan, Ann Arbor, MI 48109, USA 
        \and
            Department of Physics, University of Rome ``Tor Vergata'', Via della Ricerca Scientifica 1, I-00133 Rome, Italy 
        \and
            INAF -- Osservatorio Astrofisico di Catania, Via Santa Sofia 78, I-95123 Catania, Italy; 
        \and
            INAF -- Osservatorio Astronomico di Trieste, via Tiepolo 11, 34143 Trieste, Italy 
        \and
            INAF -- Osservatorio Astronomico di Brera, Via E. Bianchi 46, 23807 Merate, Italy 
        \and
            INAF -- Osservatorio Astronomico di Capodimonte, Salita Moiariello 16, 80131, Napoli, Italy 
        \and
            INAF -- IAPS Istituto di Astrofisica e Planetologia Spaziali, Via del Fosso del Cavaliere 100, 00133, Roma, Italy 
        \and
            INAF -- Osservatorio Astronomico di Palermo, Piazza del Parlamento, 1, 90134, Palermo, Italy 
        \and
            Dip. di Fisica e Astronomia Galileo Galilei -- Universit\`a di Padova, Vicolo dell’Osservatorio 2, 35122 Padova, Italy 
        \and
            INAF -- Osservatorio di Cagliari, via della Scienza 5, 09047 Selargius, Italy 
        \and
            Aix Marseille Univ, CNRS, CNES, LAM, Marseille, France 
        \and
            Fundaci\'on Galileo Galilei -- INAF, Rambla Jos\'e Ana Fernandez P\'erez 7, 38712 Bre\~na Baja, TF, Spain 
    } 
   \date{}

\abstract
{}
{We present a novel method to study the thermal emission of exoplanets as a function of orbital phase at very high spectral resolution, and apply it to investigate the climate of the ultra-hot Jupiter KELT-9b.}
{We combine 3 nights of HARPS-N and 2 nights of CARMENES optical spectra, covering orbital phases between quadratures ($0.25 < \varphi < 0.75$), when the planet shows its day-side hemisphere with different geometries. We co-add the signal of thousands of \ion{Fe}{i} lines through cross-correlation, which we map to a likelihood function. We investigate the phase-dependence of two separate observable quantities: (i) the line depths of \ion{Fe}{i}, and (ii) their Doppler shifts, by introducing a new method that exploits the very high spectral resolution of our observations. 
}
{We confirm a previous detection of \ion{Fe}{i} emission, and demonstrate a precision of $0.5~\mathrm{km}~\mathrm{s}^{-1}$ on the orbital properties of KELT-9b when combining all nights of observations. By studying the phase-resolved Doppler shift of \ion{Fe}{i} lines, we detect an anomaly in the planet's orbital radial velocity well-fitted with a slightly eccentric orbital solution ($e = 0.016\pm0.003$, $\omega = 150^{+13\,\circ}_{-11}$, 5$\sigma$ preference). However, we argue that such anomaly is caused by atmospheric circulation patterns, and can be explained if neutral iron gas is advected by day-to-night atmospheric wind flows of the order of a few ${\mathrm{km~s^{-1}}}$. We additionally show that the \ion{Fe}{i} emission line depths are symmetric around the substellar point within $10^{\circ}$ ($2\sigma$), possibly indicating the lack of a large hot-spot offset at the altitude probed by neutral iron emission lines. Finally, we do not obtain a significant preference for models with a strong phase-dependence of the \ion{Fe}{i} emission line strength. We show that these results are qualitatively compatible with predictions from general circulation models (GCMs) for ultra-hot Jupiter planets.}
{Very high-resolution spectroscopy phase curves have the sensitivity to reveal a phase dependence in both the line depths and their Doppler shifts throughout the orbit. They constitute an under-exploited treasure trove of information that is highly complementary to space-based phase curves obtained with HST and JWST, and open a new window into the still poorly understood climate and atmospheric structure  of the hottest planets known.}



   \maketitle
%




\section{Introduction}

Ultra-hot Jupiters (UHJs) are gas giants with orbital periods of hours to days and typical day-side temperatures of $2500~\mathrm{K}$ or more. They form a continuum with the broader population of hot Jupiters, of which they constitute the high temperature end \citep{Baxter2020, Mansfield2021}. As a result of their temperatures, which push towards stellar values, they have distinct atmospheric properties that make them ideal laboratories of physics and chemistry. Indeed, their day-sides are expected to be cloud-free and close to chemical equilibrium \citep{Kitzmann2018, Lothringer2018, Parmentier2018, Helling2019_w18}. Clouds may still form in their night-side, but they do not completely sequester heavy elements, which are present in their transmission (e.g., \citealt{Hoeijmakers2018_k9}) and emission spectrum (e.g., \citealt{Pino2020}). As a result, UHJs currently constitute the only class of planets for which it is possible to measure elemental abundances with different volatility, including rock forming elements. This makes them very promising benchmarks for planet formation theories \citep{Lothringer2021}.

Observational campaigns with a variety of instruments from both ground and space have so far mainly focused on the day-side (through emission spectroscopy) and terminator regions (through transmission spectroscopy) of UHJs, because they yield the best signal-to-noise ratio (S/N). These methods have been successful in probing the local properties of their atmospheres. However, because of their short orbital periods, UHJs are expected to be tidally locked. Therefore, the global properties of UHJ atmospheres may deviate from the local ones, and it is necessary to account for three-dimensional (3D) effects (e.g., \citealt{Feng2016}). Observationally, this has so far been best achieved through phase curves.

Phase curves record the flux emitted from the planet throughout the orbit. In photometry, a relative calibration to the stellar flux registered during the secondary eclipse permits the measurement of the phase-resolved effective temperature of the planet, including its day-to-night contrast, and the longitude of its atmospheric hot spot. When observed through a low-resolution spectrograph (e.g., HST WFC3), the resulting spectrophotometric phase curve can resolve these properties in altitude, and additionally reveal the phase-dependent chemistry \citep{Stevenson2014, MikalEvans2022}. 

Tens of hot and ultra-hot Jupiters have been observed through multi-wavelength photometric phase curves and a handful of them with spectrophotometric phase curves. These observations have revealed that the climate of UHJs may differ from that of their cooler counter-parts. Relatively small hot-spot offsets and small day-to-night contrasts (albeit observed with some scatter) indicate that heat transport likely occurs in the lack of strong equatorial jets, but is still efficient \citep{ParmentierCrossfield2018}. These observations contrast with trends observed for gas giants colder than $2500~\mathrm{K}$ \citep{Zhang2018, Wong2020}, which have weaker hot spot offsets but, unlike their hotter counter-parts, display larger day-to-night contrasts with increasing temperature, indicating a reduced heat transport efficiency. Theoretical studies have identified two key additional ingredients that emerge in UHJs, and are necessary to explain their different climate: thermal recombination of atomic hydrogen to $\mathrm{H}_2$ in their night-side, which leads to an increase in the efficiency of heat transport \citep{Bell2018}, and atmospheric drag, which affects their atmospheric circulation patterns. Indeed, the interaction of the significant planetary ionospheres, produced by the high temperatures, and of magnetic fields, has the capability to dampen waves that would otherwise lead to the formation of an equatorial jet \citep{Perna2010, Beltz2022a}. This would lead to smaller hot spot offsets, and potentially to a transition to day-to-night atmospheric flow \citep{Tan2019}. While the theory was moderately successful in explaining the overall properties of climate of UHJs, results obtained on individual UHJs still blurry this picture (e.g., \citealt{Addison2021}). Reconciling optical (e.g., TESS) and infrared (e.g., Spitzer) phase curves proves particularly challenging in some cases \citep{Wong2020, VonEssen2020}. 

Very high-resolution spectroscopy (HRS; $R\gtrsim100000$) offers a unique observational angle to tackle the open questions in this field. Indeed, by individually resolving spectral lines, it is able to directly measure the Doppler-shift of spectral lines, and thus probe the strength of winds moving masses of gas around the atmosphere of target planets. \cite{Ehrenreich2020} showcased the power of this technique by measuring phase-resolved transmission spectroscopy of UHJ WASP-76b. They reveal a complex Doppler shift pattern, sensitive to temperature structure, rotation, dynamics, and night-side condensation \citep{Wardenier2021, Savel2022}. However, HRS transmission spectra generally probe above the planet photosphere (e.g., \citealt{Hoeijmakers2019}), and can only access a limited longitudinal region of the planetary atmosphere \citep{Wardenier2022}. 

A complementary approach is to observe parts of a planetary phase curve with a high-resolution spectrograph. When observed through a high-resolution spectrograph, features in the planet continuum are lost, but the phase-resolved Doppler shift and Doppler broadening of emission/reflection lines uniquely constrains the orbit of the planet, and its rotation and dynamics \citep{CollierCameron1999, Kawahara2012, Brogi2013, Snellen2014}. This is supported by dedicated theoretical studies that, applying radiative transfer to General Circulation Models (GCMs), show that the combination of inhomogeneous specific intensity, rotation, and winds imprints kilometre-per-second level Doppler shifts and significant distortions in the line shapes and strengths \citep{Zhang2017, Beltz2022}. However, two challenges have so far hindered the use of this technique. First, the signal observed in this configuration is smaller compared to transmission spectroscopy, which has led most authors to stack spectra across planet phases to increase the significance of detections. Unfortunately, this operation washes out the phase-dependency of the line profiles, which is crucial to probe planetary climate. Second, HRS observations require complex data reduction methods, which makes the retrieval of atmospheric properties a non-trivial operation. 

Both challenges can now be surpassed. The application of likelihood-based frameworks of recent development \citep{Brogi2019, Gibson2020, Pino2020} allows us to reliably determine atmospheric properties and their error bars starting from HRS emission spectra. Exploiting this key technical novelty, \cite{Beltz2021} presented the first indication that the HRS phase curve of HD209458b observed with CRIRES has the sensitivity to probe 3D effects in the planetary atmosphere. In addition, thanks to their relatively large planet-to-star flux ratio, UHJs offer the opportunity to observe HRS phase curves at unprecedented S/N. Indeed, \cite{Herman2022} and \cite{vanSluijs2022} present the first parameterized studies of phase-dependent \ion{Fe}{i} and CO line strengths targeting a UHJ, WASP-33b, for which \cite{Cont2021} had already suggested that three-dimensional effects are required to explain the phase-stacked optical HRS emission spectrum of TiO and \ion{Fe}{i}. In addition, \cite{Borsa2022} found evidence for a different chemistry and thermal profile in pre- and post-eclipse phases of UHJ KELT-20b. 

In this paper, we further extend these previous studies to include, for the first time, a simultaneous parameterized analysis of phase-dependent line intensity and Doppler shift of the hottest UHJ: KELT-9b ($T_\mathrm{eq} = 3900~\mathrm{K}$; \citealt{Borsa2019}). In section \ref{sec:methods}, we present our data-reduction pipeline and custom model suite, designed to be able to capture in a parameterized way the phase-dependence of both \ion{Fe}{i} line intensity and Doppler shift. In section \ref{sec:results}, we present results from our retrievals, with a focus on phase-resolved, 3D properties of the atmosphere of KELT-9b. In section \ref{sec:discussion}, we discuss our results in the context of photometric phase-curve observations of KELT-9b and in the context of other observations and theory of UHJ climate with low and high-resolution spectroscopy; we additionally discuss the prospects of HRS phase curve studies with current and upcoming instrumentation, including unprecedented possibilities to study the orbital and atmospheric dynamics of UHJs. We present our conclusions in section \ref{sec:conclusions}.

\section{Methods}
\label{sec:methods}

The key methodological novelty introduced by this work, is that we extract information on the phase dependence of planetary atmospheric properties only by comparing planetary spectra at difference phases, without relying on external input such as a known systemic velocity, or the overall stellar flux level. In this section, we present all the crucial aspects to successfully apply this new approach to HRS phase curves.

\subsection{Ephemeris of KELT-9b}
\label{sec: Ephemeris of KELT-9b}
Knowing a reliable orbital ephemeris for KELT-9b is essential for our analysis, given the crucial impact it has for associating an accurate orbital phase to each spectrum. Unfortunately, even a quick look at recent transit photometry shows that the prediction by the only ephemeris available until recently (from \citealt{Gaudi2017}) disagrees with the observations by more than ten minutes at present epoch. In order to compute an updated orbital period $P$ and reference transit time $T_0$, we extract the TESS short-cadence PDCSAP light curves of KELT-9 from Sectors 14, 15 and 41 (2019 to 2021) and perform a JKTEBOP \citep{Southworth2008} transit fit to them, assuming a linear ephemeris and computing the error bars through a residual-permutation technique. The final result that we adopt for the subsequent analysis is: 
\begin{equation}
\left \{\begin{array}{l}
     P = 1.4811188 \pm 0.0000003 \textrm{ d}  \\
     T_0 =  2\,459\,006.3289 \pm  0.0001 \textrm{ BJD}_\textrm{TDB} 
\end{array}\right .\textrm{ ,}
\end{equation}
where the chosen reference frame and time standard is the Barycentric Julian Day in Barycentric Dynamical Time, following the prescription by \citet{Eastman2010}. It is worth noting that our new estimate of $P$ is perfectly consistent with those recently published by \citet{PaiAsnodkar2022} and \citet{Ivshina2022}.

\subsection{Observations and data reduction}
\label{sec:methods:observations_and_data_reduction}
We focus our analysis on neutral iron lines, which offer the highest S/N in KELT-9b thanks to their large cross-section, and have well known line positions and strengths. This is crucial to identify astrophysical variations in line strengths and positions with accuracy. In the following, we describe a custom pipeline that we developed to extract the planet spectrum, model iron emission lines including simplified pseudo-3D effects, and how we compare models to data in a statistical framework.

\subsubsection{Night selection}
\begin{figure}
  \resizebox{\hsize}{!}
    {
    \includegraphics[trim={0cm 1cm 2cm 0}]{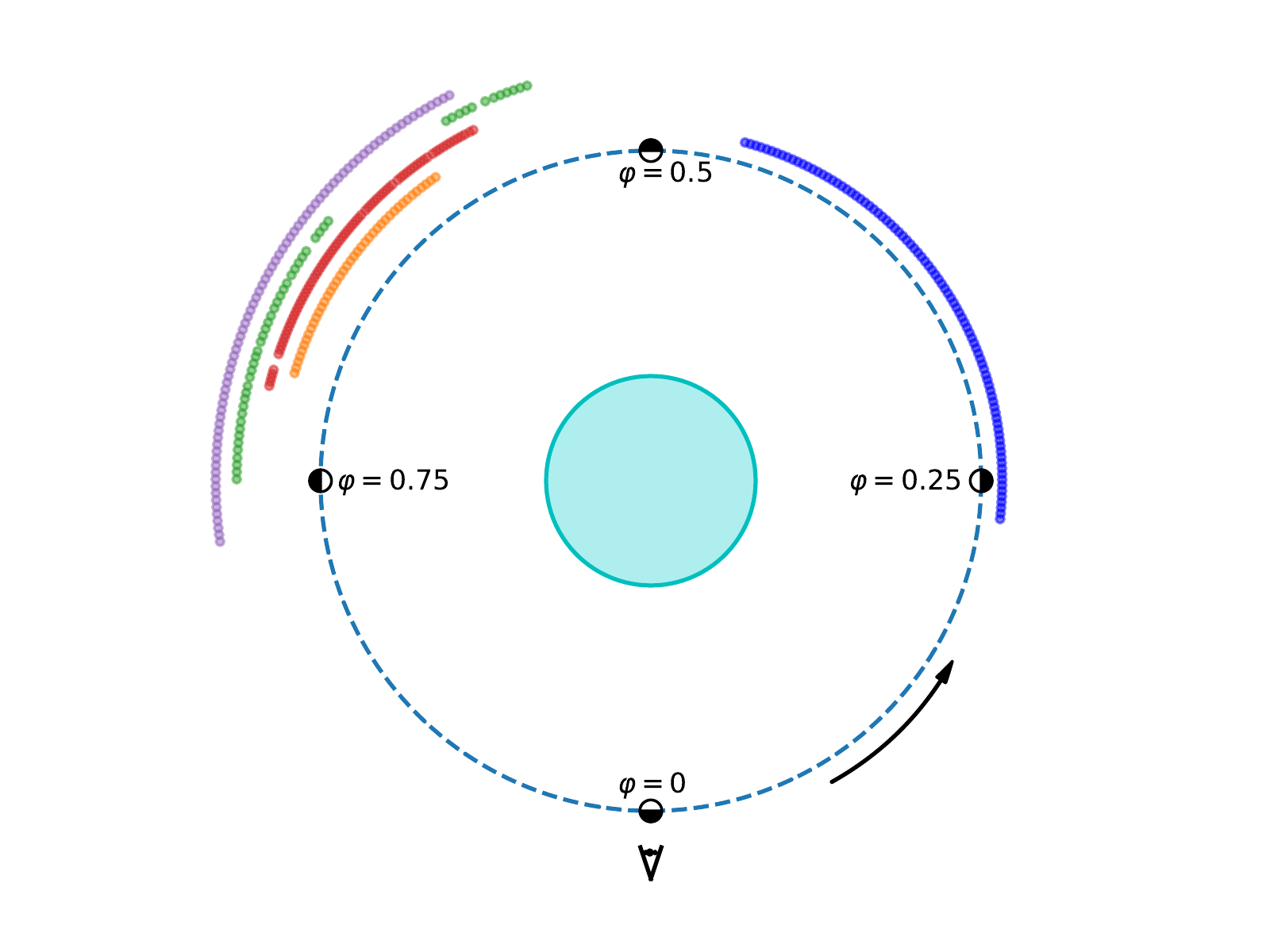}}
      \caption{Schematic representation of the orbit of KELT-9~b to scale, with the orbital phases covered in this study colour-coded by observing night (Blue: HARPS-N N1; Orange: HARPS-N N2; Purple: HARPS-N N3; Red: CARMENES N1; Green: CARMENES N2). The exoplanet orbits counterclockwise (curved arrow), and the observer is positioned at the bottom of the figure.}
      \label{fig:orbital_coverage}
\end{figure}

Our goal for this study was to obtain the most complete phase coverage between quadratures (orbital phases $0.25 < \varphi < 0.75$), while at the same time minimizing the size of the dataset to keep it computationally feasible to run Markov-Chain Monte Carlos on the data. We note that the secondary eclipse (SE) of KELT-9b occurs between orbital phases $0.44<\varphi<0.56$, and the spectra falling within this interval were excluded from the analysis.
Overall, we combined three nights of proprietary HARPS-N (\citealt{Cosentino2012}; $R=115000$, spectral coverage $390$--$690~\mathrm{nm}$) data\footnote{Observed in GIARPS (GIANO-B + HARPS-N) mode \citep{Claudi2017}; but we only use the HARPS-N data in this work.} 
with two nights of public CARMENES (\citealt{Quirrenbach2016, Quirrenbach2018}; $R\sim94600$, spectral coverage $520$--$960~\mathrm{nm}$ in the optical arm) data\footnote{http://caha.sdc.cab.inta-csic.es/calto/jsp/searchform.jsp}. 
We define our naming convention for the observing nights and list relevant parameters in  Table~\ref{table:observing_nights}. We visualize the orbital phases covered
in Figure~\ref{fig:orbital_coverage}. We only used the optical arm of CARMENES, where we expect most of the \ion{Fe}{i} signal to arise.

\begin{table*}
\caption{Details of the five observing nights included in this study.}
\label{table:observing_nights}
\centering
\begin{tabular}{cccccccc}
\hline \hline
Name & Date & $N_\mathrm{spectra}$ & Orbital phase & Airmass & $t_\mathrm{exp}$ & S/N & Instrument \\ 
\hline
HARPS-N N1 & 2018-07-22 & 84 & 0.233-0.457 & 1.02--1.63 & 300~s$^\star$ & 97-112-126 & HARPS-N \\
HARPS-N N2 & 2019-11-16 & 42 & 0.598-0.703 & 1.05--1.63 & 300~s & 55-63-89 & HARPS-N \\
CARMENES N1 & 2019-05-28 & 72 & 0.587-0.715 & 1.00--1.80 & 172~s & 65-96-109 & CARMENES \\
CARMENES N2 & 2018-09-02 & 44 & 0.557-0.741 & 1.01--1.80 & 300~s & 78-113-141 & CARMENES \\
HARPS-N N3 & 2021-09-05 & 78 & 0.577-0.772 & 1.02--2.09 & 300~s & 71-103-114 & HARPS-N \\
\hline
\end{tabular}
\tablefoot{Columns list the naming convention followed in this paper, the observing date, the number of observed spectra, the orbital phase range covered, the airmass range, the exposure time, typical S/N of the observations and the instrument. The quoted range in S/N is the 10-50-90 percentile of the median S/N calculated at the central pixel of each order.\\
$^\star$ In \cite{Pino2020} we reported an incorrect exposure time of 180 s. The correct exposure time is 300 s.}
\end{table*}

The HARPS-N nights were obtained within a Large Program (PI: Micela) awarded to the GAPS Collaboration \citep{Poretti2016}. This program already yielded results on atmospheric characterization (e.g., \citealt{Borsa2019, Pino2020, Guilluy2020, Rainer2021, Scandariato2021}) and is described in \cite{Guilluy2022}. We initially included the HARPS-N night published in \citet{Pino2020}, covering phases pre-SE, and a second HARPS-N night post-SE observed at lower S/N. To compensate for the latter, we added the two best archival nights of CARMENES post-SE. While performing the analysis, we observed a third HARPS-N night, also post-SE, which we integrated into the analysis to test for consistency between different instrument and observing dates.

The two CARMENES nights included spectra at very low S/N and very high airmass. Four spectra at visibly low S/N between phase 0.55 and 0.60 were removed from the night of 2018 September 2. To do so, a threshold of 35\% of the maximum median S/N calculated across all wavelengths was applied. Furthermore, we excluded from the analysis spectra taken at airmass greater than 1.8. With this choice, we minimized the size of the confidence intervals on the retrieved planetary orbital velocity. However, the exact choice of the airmass threshold does not significantly affect the combined analysis. Finally, we remove the 10 reddest orders of CARMENES from the analysis due to very low S/N.

At our unusually high level of precision obtained in the planet's radial velocity, it is important to account for the fact that the timestamps stored in the headers of both CARMENES and HARPS-N spectra contain BJD dates computed in the UTC frame rather than in the TDB frame. Indeed, between 2017 and 2021 included, the difference BJD UTC --  BJD TDB is fixed and amounts to 69.2 s, which is non-negligible for us. All the dates in the file headers are correctly weighted by the photometric barycentre of the exposure, i.e. they account for the fact that varying flux during the exposure can effectively offset the point where half of the flux is collected from the middle of the exposure.

All in all, our full set of data consists of 320 spectra covering orbital phases $\varphi$ = 0.233-0.772 excluding the secondary eclipse as mentioned above. While the reliance on just one HARPS-N night pre-SE might seem unbalanced, the exceptionally high quality of the pre-SE HARPS-N night results in similar confidence intervals from the pre-SE and the post-SE analysed separately.

\subsubsection{Data reduction}
\label{Sec: data reduction}
We design a pipeline that we can uniformly apply to HARPS-N and CARMENES. Both HARPS-N (we used the e2ds spectra) and CARMENES spectra are provided with a wavelength solution in the rest frame of the observer, in air for HARPS-N and in vacuum for CARMENES. We shift the HARPS-N wavelength solution to vacuum to be able to compare them with models. After deblazing and color-correcting the spectra, we thus proceed to shift them into the barycentric reference frame of the Solar System by applying the barycentric correction included in the respective file headers. We followed \citep{Pino2020} for this part of the analysis.

We then apply a telluric and stellar lines correction adapted from \citet{Giacobbe2021} and originally designed to clean near-infrared data from the more severe effects of telluric and instrumental systematic effects. It consists in the application of a Principal Component Analysis (PCA) algorithm to blindly identify ``state vectors'' in common between spectral channels and use them to describe the time-evolution of telluric lines, and remove them. We work in the barycentric rest-frame, where stellar lines are stationary and telluric lines are quasi-stationary. In this way, we slightly privilege correction of stellar lines over telluric lines, although practically the algorithm removes both components. This choice is driven by our focus on neutral iron lines that are typically present both in the stellar and planetary spectrum, but not in the Earth's atmosphere. Since in this work there are some differences with the algorithm in \citet{Giacobbe2021}, we fully describe our implementation of the PCA. We note that in exoplanet literature sometimes PCA is performed by only using the Singular Value Decomposition (SVD) algorithm, while in our case SVD is just one of the three main steps of the analysis. While conceptually very similar, the full PCA algorithm described here and the SVD-only algorithm differ in the way the wavelength channels in the data are weighted. 



We describe the data as a cube with three axes: order number, time (or phase), and wavelength. One cube is stored for each night of observations. Due to the fixed layout of the echellogram for both CARMENES and HARPS-N, the dimensions along the order axis is always 61 for CARMENES and 69 for HARPS-N. The corresponding number of pixels along the wavelength dimension is 4096 for both spectrographs.

The PCA algorithm in this analysis is entirely written in Python and based on {\tt Numpy} methods. The code is run on each night and each order separately, recognizing not only that the atmosphere behaves differently on each night, but also that different orders are affected by different time-correlated sources, both astrophysical and instrumental. The input matrix for the algorithm is always a two-dimensional array, with time on the vertical axis (along a column) and wavelength on the horizontal axis (along a row).

The PCA algorithm, applied independently to each spectral order and each observing night, is outlined as follows:
\begin{enumerate}
    \item Initial masking: non-finite flux values ({\tt NaN}), as well as values below 2\% of the median flux level (low-S/N) of the spectrum are flagged and a mask is created to keep track of such pixels. Spectral channels (data columns) with more than 1 invalid pixel are entirely masked.
    \item Standardization: each column (each spectral channel) has its mean subtracted and is divided by its standard deviation. At the end of the process, each column has thus mean zero and unit standard deviation. This step makes sure that the PCA weights all wavelength bins equally while looking for common modes in the next step. 
    \item Singular Value Decomposition (SVD): this is computed via {\tt numpy.linalg.svd()} on the standardized matrix (step 2), with the option {\tt full\_matrices = False}. SVD decomposes the array in a set of orthogonal eigenvectors and corresponding eigenvalues. In our case, there are as many eigenvectors(values) as the number of spectral channels, i.e. the dimension of the wavelength axis. In PCA jargon, this choice corresponds to running the PCA in the ``time domain'', i.e. with time (or orbital phase) as an independent variable. Note that we only feed the SVD algorithm the columns of data that were not masked at step 1.
    \item Component selection: SVD ranks eigenvectors according to their contribution to the data variance. Therefore, only the first few eigenvectors are needed to describe most of the flux variations in telluric lines. In our case, by visual inspection, we determined that 2 eigenvectors (also called ``components'') are sufficient to suppress telluric lines below the level of the noise for all the spectral orders, and we use those in the next step. 
    \item Multilinear regression: the eigenvalues obtained via SVD do not correctly describe the observed data, because they have been computed on the standardized flux array (step 2) rather than on the measured flux array (step 1). Therefore, we run a multilinear regression (MLR) between the set of two eigenvectors $\vec{v}_1(t), \vec{v}_2(t)$ determined at step 4 and the matrix stored at step 1. For each spectral channel $i$ in the matrix, the MLR calculates two eigenvalues $c_{1i}, c_{2i}$ plus an offset $c_{0i}$.
    \item For each spectral channel $i$, we calculate the linear combination 
    $l_i(t) = c_{0i}+c_{1i}\vec{v}_1(t) + c_{2i}\vec{v}_2(t)$ and divide the corresponding column of the matrix at point 1 through it, obtaining a residual matrix.
    \item High-pass filtering: columns in the residual matrix (step 6) that deviate by more than 3 times the standard deviation of the whole matrix are masked. Such mask is then merged via boolean {\tt OR} with the mask at step 1, and stored for future use during model reprocessing (Section~\ref{sec:modelling}). Low-order variations in each spectrum, i.e. along the wavelength axis, are fitted with a second order polynomial (excluding the masked pixels) and divided out.
    \item The mean of each spectrum obtained in step 7 is subtracted out, which is necessary to compute the log-likelihood function. We note that the mean is calculated by excluding the masked pixels; these are then manually reset to zero so that they do not contribute to the log-likelihood function.
\end{enumerate}
The 5 resulting data cubes (one per observing night) are then compared to models. 

\subsection{Models}\label{sec:modelling}
We produce a set of nested models to test the presence of 3D effects in the atmosphere of KELT-9b, generalizing the best-fit 1D model by \cite{Pino2020}. This is calculated employing a custom line-by-line radiative transfer code which solves the radiative transfer equation in its integral form using a ``linear in optical depth approximation'' for the source function \citep{Toon1989}. It assumes the temperature-pressure profile calculated by \cite{Lothringer2018}. Volume mixing ratios are calculated under the assumption of equilibrium chemistry using \texttt{FastChem} version 2 \citep{Stock2018}, with stellar (solar in KELT-9's case) composition. Following \cite{Pino2020} we only include lines from neutral iron, whose opacities we compute starting from the VALD3 database \citep{Piskunov1995, Ryabchikova1997, Kupka1999, Kupka2000, Ryabchikova2015}\footnote{Additional references specific to the used \ion{Fe}{i} line list: \cite{K14, BKK, BK, BPM, BWL, FMW}.}, and the bound-free ${\mathrm{H}^-}$ opacity from \cite{John1988}. We convolve the model with two kernels: the former is a Gaussian kernel with FWHM matching the velocity resolution of the HARPS-N/CARMENES spectrographs; the latter is a rotational kernel calculated assuming a rigidly rotating planet with equatorial velocity of 6.64~$\mathrm{km~s^{-1}}$. This value is obtained by assuming synchronous rotation and the planet radius from \cite{Gaudi2017}. We refer to \cite{Pino2020} for other details about the radiative transfer code.

Our one-dimensional, fiducial model 1C is built using the best-fit spectrum by \cite{Pino2020} as a template, and is described by three parameters: $K_\mathrm{p}$, $v_\mathrm{sys}$ and $\mathrm{S}$. The parameter $\mathrm{S}$ is a multiplicative scaling factor that accounts for a mismatch between the intensity of model lines and lines in observations \citep{Brogi2019}. In the lack of any additional contribution to Doppler shift (e.g., atmospheric winds), the parameters $K_\mathrm{p}$ and $v_\mathrm{sys}$ are physically interpreted as the maximum Keplerian orbital velocity of the planet (assuming a circular orbit in the case of model 1C) and the systemic velocity of the system, i.e. the constant component of the radial velocity of the system’s centre of mass relative to the solar system barycentre, which may also include an instrument-dependent radial velocity offset \citep{Deeg2018}. We keep nomenclature consistent with the literature, but caution the reader that additional velocity contributions present in the data will be captured by this model, changing the physical interpretation of these parameters (e.g., see Sec. \ref{sec: discussion_climate}). The same caveat applies to the following models. 

Model 1C contains limited information about the phase-dependence of the planet spectrum. Indeed, it has a constant shape across phases, and has a varying Doppler shift with planetary phase that is forced to follow a Keplerian, circular orbit around the system centre of mass. This model has been used by \cite{Pino2020}, who show that it reproduces well the average \ion{Fe}{i} emission line of the planet in HARPS-N N1. We generalize this model by relaxing two separate assumptions: (1) the rest frame from which the signal arises is revolving around the star KELT-9 with a circular motion; (2) the flux emitted from the photosphere of the planet is uniform with longitude. Table \ref{table:models adopted} summarizes the models and their parameters, and in the following we describe them more in detail.

We design an eccentric orbit model 1E for KELT-9b using the parameters $h=\sqrt{e}\sin\omega$ and $k=\sqrt{e}\cos\omega$. $e$ is the eccentricity, while $\omega$ is the argument of periastron. This parameterization performs comparable or faster than alternatives in an MCMC, and, in addition, it is free of the burden of dealing with angular parameters. Finally, it naturally sets a uniform prior in $e$ and $\omega$ that is regarded as the most sensible choice for hot Jupiters \citep{Eastman2013}. This model is described by five parameters: $K_\mathrm{p}$, $v_\mathrm{sys}$ and $\mathrm{S}$ that are the same as for the circular orbit model 1C, and $h$ and $k$ are in addition.

Finally, we employ two simple modelling approaches to capture the possible variation in intensity of the iron lines as a function of longitude in the planet atmosphere. In both approaches, we neglect latitudinal variations of atmospheric properties, thus properties measured at a given longitude should be seen as a latitudinal average. In the first approach, the two four-scale factor models 4C (circular orbit) and 4E (eccentric orbit) divide the orbital range in 4 parts ($0.25 < \varphi < 0.35$, $0.35 < \varphi < 0.45$, $0.55 < \varphi < 0.65$, $0.65 < \varphi < 0.75$). We assign a different scale factor ($S_1$, $S_2$, $S_3$, $S_4$) to each of these phase ranges, while the orbital parameters ($K_\mathrm{p}$, $v_\mathrm{sys}$, and additionally $h$ and $k$ in the case of an eccentric orbit) are in common. With this choice, our models are able to capture an inhomogeneous intensity of \ion{Fe}{i} lines. For instance, if the \ion{Fe}{i} signal mostly comes from the planet day-side, $S_1$ and $S_4$ should be smaller than $S_2$ and $S_3$, since a larger portion of the night-side contributes to determine them. The shortcoming of this modelling approach is that it mixes information coming from different longitudinal slices of the planet atmosphere in a suboptimal way. Indeed, a given longitudinal slice of the planet atmosphere could be observable in multiple phase ranges. The second modelling approach that we consider in this work overcomes this limitation. We adopt three models based on reflection off a Lambert sphere \citep{CollierCameron2002}. While this is likely unrealistic, we argue that these models have the desirable feature that line intensity has a maximum (whose position is a parameter model Loff), and then decreases symmetrically when moving further away in longitude. This would, for instance, reproduce a shallower thermal gradient profile, or a decrease of neutral iron abundance while moving towards the night-side. In this model, we substitute the constant multiplicative scale factor $S$ with the phase function $g(\varphi)$:
\begin{align}
\label{Eq.: Lambert_scale_factor}
&g(\varphi) = \frac{S}{\pi} + \frac{S_\mathrm{Lambert} \cdot \left[ \sin\alpha + \left( \pi - \alpha\right)\cdot \cos\alpha \right]}{\pi}\ ,\\
&\cos\alpha = -\cos\left(2\pi\varphi - 2\pi\varphi_0\right)\ ,
\end{align}
where we have assumed a perfectly edge-on orbit and $\varphi$ varies between 0 (inferior conjunction) and 1. In model Lbase, the free parameters are $K_\mathrm{p}$, $v_\mathrm{sys}$, $S$ (constant scale factor contributing emission independent of planetary phase), $S_\mathrm{Lambert}$ (scale factor of additional phase-dependent \ion{Fe}{i} line emission), and $h$ and $k$, and we fix $\varphi_0 = 0$ (offset of maximum scale factor from sub-stellar point). We additionally test model L where we also suppress the phase-independent emission ($S=0$, $\varphi=0$).

All our models rely on several simplifying assumptions, and adopt a parameterized approach that is agnostic of the underlying physics. In future work, we will consider more physically motivated, parameterized models able to partially account for the interplay between stellar irradiation, atmospheric dynamics and structure, and thus predict variations in thermal profile and chemistry as a function of longitude (e.g., \citealt{DobbsDixon2022}).

\begin{table*}
\caption{Model description, parameters and priors. See text for detailed explanation of parameters.}
\label{table:models adopted}
\centering
\begin{tabular}{cclc}
\hline \hline
Model name & Parameters & Short description & Priors\\ 
\hline
\textbf{1C} & $K_\mathrm{p}$ & Circular orbit, & [100, 300] km s$^{-1}$ \\ 
(3 parameters) & $v_\mathrm{sys}$ & single scale factor & [$-70$, 20] km s$^{-1}$ \\
& $S$ & (as \citealt{Pino2020}) & [0.01, 100] \\ 
\hline
\textbf{4C} & $K_\mathrm{p}$ & Circular orbit, & [100, 300] km s$^{-1}$ \\ 
(6 parameters) & $v_\mathrm{sys}$ & four scale factors & [$-70$, 20] km s$^{-1}$ \\ 
& $S_1$, $S_2$, $S_3$, $S_4$ & & [0.01, 100] \\ 
\hline
\textbf{1E} & Same as 1C + & Eccentric orbit, & \\ 
(5 parameters) & $h=\sqrt{e}\sin\omega$, $k=\sqrt{e}\cos\omega$ & single scale factor & [$-1, 1$] \\ 
\hline
\textbf{4E} & Same as 4C + & Eccentric orbit, &  \\ 
(8 parameters) & $h=\sqrt{e}\sin\omega$, $k=\sqrt{e}\cos\omega$ & four scale factor & [$-1, 1$] \\ 
\hline 
\textbf{L} & $K_\mathrm{p}$ &  Lambert sphere, & [100, 300] km s$^{-1}$ \\ 
(5 parameters) & $v_\mathrm{sys}$ & eccentric orbit & [$-70$, 20] km s$^{-1}$\\ 
& $S_\mathrm{lambert} / \pi$ & & [0, 10] \\ 
& $h=\sqrt{e}\sin\omega$, $k=\sqrt{e}\cos\omega$ &  & [$-1, 1$] \\ 
\hline 
\textbf{Loff} & Same as L + &  Lambert sphere, & \\ 
(6 parameters) & $\varphi_0$ & eccentric orbit, & [$-0.15,0.15$]\\ 
&  & offset from substellar point &  \\ 
\hline 
\textbf{Lbase} & Same as L + &  Lambert sphere, & \\ 
(6 parameters) & $S/\pi$ & eccentric orbit, & [0, 10]\\ 
&  & with baseline constant emission &  \\ 
& &  &  \\ 
\end{tabular} 
\end{table*}

\subsection{Cross-correlation-likelihood mapping}
\cite{Pino2020} used two different methods to stack the signal from thousands of iron lines, yielding the first detection of iron in the emission spectrum of KELT-9b: (1) a weighted mask method, and (2) a cross-correlation-likelihood mapping scheme by \cite{Brogi2019}. In this paper, we employ the cross-correlation-likelihood mapping method, and provide our rationale for this choice in Appendix \ref{sec_appendix: mask_vs_mapping}. We stress that each method has strengths and weaknesses, and the choice should be driven by the science goals.

The mapping is achieved through the use of the log-likelihood function
\begin{equation}\label{eq:log-L}
    \log L = -\frac{N}{2}\log(s_f^2+s_g^2-2R)\,,
\end{equation}
where $N$ is the number of valid (i.e. unmasked) spectral channels on each spectrum and each order, $s_f^2$ and $s_g^2$ are the data and model variances, respectively, and $R$ is the cross-covariance between model and data. While cross-correlation does not appear explicitly in this formula, its numerator ($R$) and denominator $(s_fs_g)$ are present in the argument of the natural logarithm. We partition Eq. (\ref{eq:log-L}) by calculating it on each order and frame. This correctly weights for noise variations, directly estimated from the sample variance, across orders and frames. This methodology has been shown to produce unbiased and accurate results \citep{Brogi2019, Gibson2022}.

As pointed out since the early application of the log-likelihood framework \citep{Brogi2019,Pino2020,Gibson2020}, the model spectrum cannot be compared to the data directly, i.e. by just scaling and shifting it. In order to avoid biases in the retrieved parameters, it is important to reproduce on the model any possible alteration of the planet signal induced by the data reduction process. This is achieved by running a parallel data analysis chain (steps 1-6 in Section~\ref{Sec: data reduction}) on a data cube where the model is injected on top of the observed data at a reduced amplitude of 1\%\footnote{We tested that optimizing this additional scaling factor between 10\% and 0.001\% of the model flux has no impact on the result}, so that the injection is not altering the analysis, in particular the ranking of the SVD components at step 3. Each injected spectrum $F_\mathrm{inj}$ is obtained from the observed spectrum $F_\mathrm{obs}$ via\footnote{Except for a constant offset of 1, subtracting or dividing through the PCA-processed data at this stage differs by a factor of order $(F_\mathrm{pl}/F_\mathrm{\star})^2$, that is a difference of the order of $10^{-8}$ in absolute flux (or $10^{-4}$ in relative flux). Here we opt for subtracting to avoid possible issues with division by values close to zero.}
\begin{equation}
    F_\mathrm{inj} = F_\mathrm{obs} \left( 1 + 0.01\frac{F_\mathrm{pl}(\vec\theta)}{F_\star}\right),
\end{equation}
where $\vec\theta$ is the state vector containing all the model parameters at the current MCMC step, and the model is expressed in planet/star units by normalizing through a stellar black body at the effective temperature of KELT-9.

\begin{figure*}
  \resizebox{\hsize}{!}
    {
    \includegraphics[trim={0 0 3cm 0}]{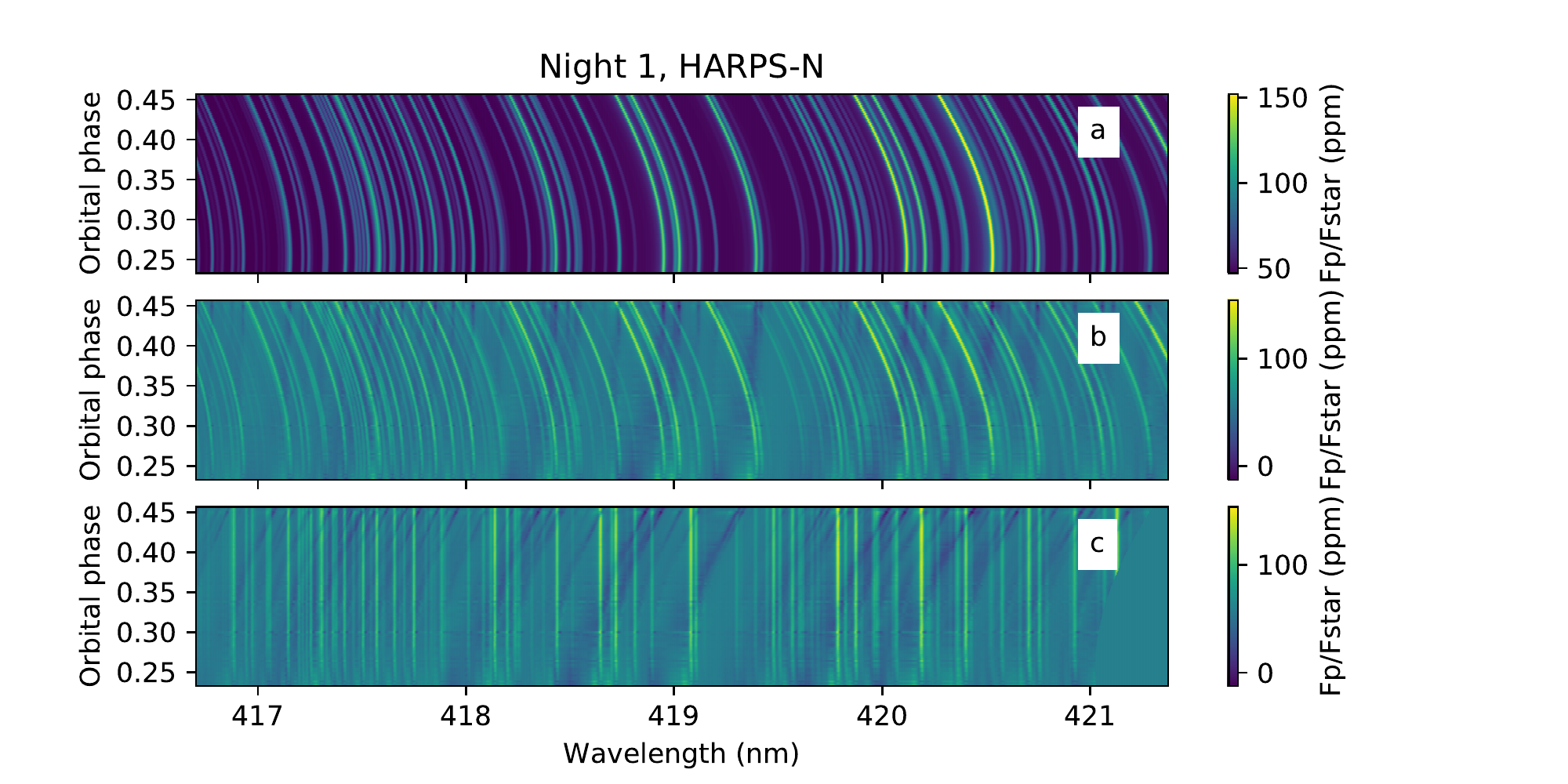}
    \includegraphics[width=0.42\textwidth]{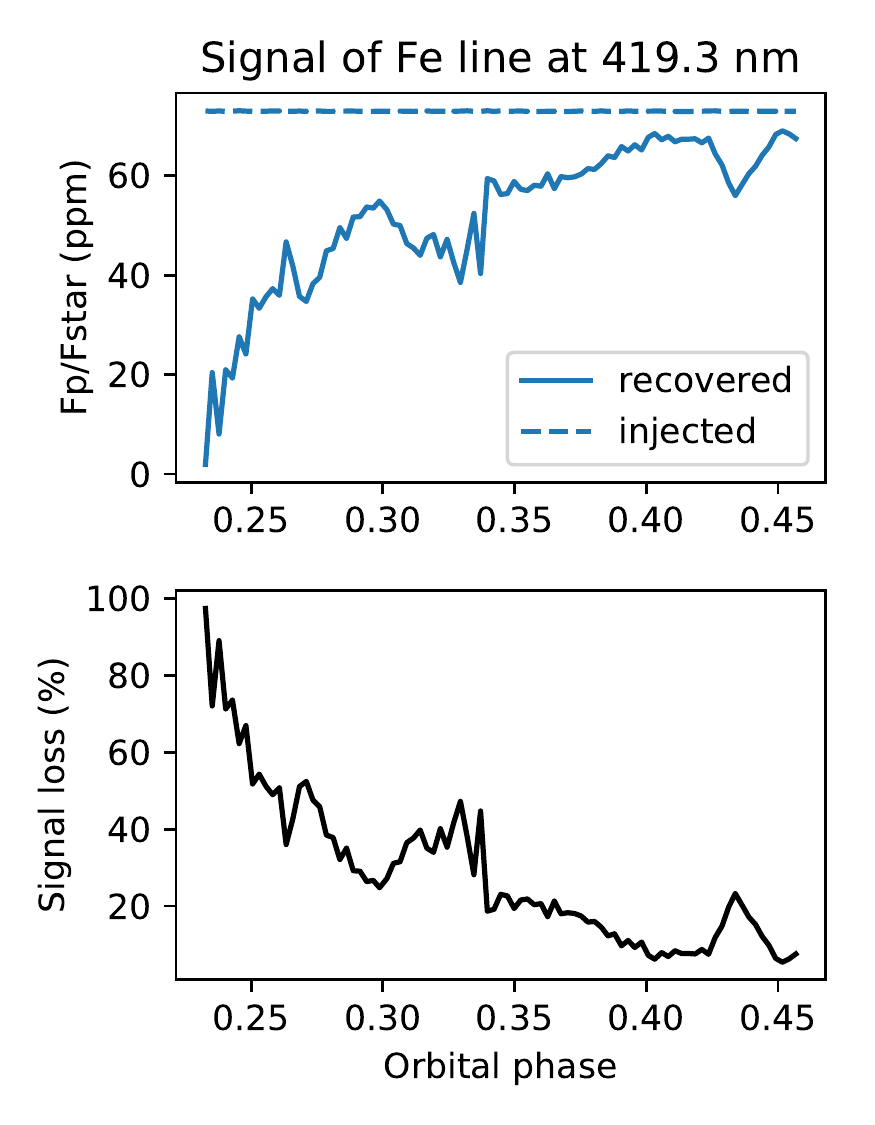}
    }
      \caption{Effects of telluric removal on the signal from KELT-9 b. {\sl Left:} the best-fit \ion{Fe}{i} emission model is Doppler-shifted based on the orbital parameters of KELT-9 b (panel a), injected on top of the 1st night of HARPS-N data, and passed through the analysis described in Section~\ref{Sec: data reduction}. The signal is recovered with evident alterations, as shown in the reference frame of the observer (panel b) and in the planet's rest frame (panel c). {\sl Top-right:} recovered amplitude of a strong injected Fe line as a function of orbital phase (solid) versus its amplitude at injection (dashed). {\sl Bottom-right:} percent loss of the Fe signal due to the telluric-removal analysis. Nearly 100\% of the signal is lost around quadrature.}
      \label{fig:processing}
\end{figure*}

After step 6 of the analysis, the observed dataset at the same stage is subtracted out, so that the resulting data cube represents the (noiseless) processed model. The latter is then multiplied by 100 to restore its original amplitude, mean-subtracted as in step 8, and used to calculate the log-likelihood values through Eq.~\ref{eq:log-L} after masking (i.e. setting to zero) the same pixels as for the observed data. We skip the application of a high-pass filter (step 7) on the model. While early tests indicate no impact on the final results, we believe that it is conceptually different to apply the filter on observed data (dominated by instrumental and photon noise) than on the noiseless model (dominated by signal variations). 
The difference between the model injected at step 1 and the model processed as above is shown in Fig.\ref{fig:processing}, and it shows how even by visual inspection there are evident alterations in the depth and shape of the planetary lines that need to be accounted for. As expected, the effect is phase-dependent and particularly severe close to quadrature, where the planet trail is nearly parallel to telluric and stellar lines. In Appendix \ref{sec_appendix: injection}, we show that our reprocessing methodology is capable of recovering a phase-dependent signal accurately at our level of precision.

Currently, model reprocessing is the bottleneck of our retrieval code. In order to optimise it, we follow the practice of \citet{Gibson2022} and reuse for model reprocessing the same eigenvectors calculated by the SVD algorithm on the observed spectra (step 3). We also reuse the same mask for both analysis and reprocessing. With these choices, we speed up the calculation by a factor of about 2.


\section{Results}

\label{sec:results}

\subsection{Confirmation of iron emission lines from the day-side of KELT-9b}
We run an MCMC using model 1C, the same that was adopted by \cite{Pino2020}, on each individual night. We detect neutral iron in each of the unpublished HARPS-N N2 and N3, and in each of the two CARMENES N1 and N2. 
This confirms the results by \cite{Pino2020} and \cite{Kasper2021}, based on pre-eclipse observations, and shows that neutral iron is present also at post-eclipse phases. Our best single-night detection remains HARPS-N N1, which featured better observing conditions and twice the exposures compared to HARPS-N N2. In HARPS-N N1, we measure a $K_\mathrm{p}$ which is incompatible at $2\sigma$ with \cite{Pino2020}, despite using the same data. This discrepancy is due to the different ephemeris adopted in this work (see Sec. \ref{sec: Ephemeris of KELT-9b}). We directly verified that our new PCA pipeline, when applied using the same ephemeris adopted by \cite{Pino2020}, reproduces their best-fit planetary orbital velocities. HARPS-N N3 features a similar total S/N compared to HARPS-N N1, but provides looser constraints on $v_\mathrm{sys}$ and $K_\mathrm{p}$, and a marginally lower scale factor. Furthermore, CARMENES provides looser confidence intervals on the parameters compared to HARPS-N. For comparison, the precision reached on all parameters in the best of the two CARMENES nights (CARMENES N1) is comparable to the precision reached in the worst HARPS-N night (HARPS-N N2). 

It is not immediately clear why HARPS-N outperforms CARMENES in our analysis, and multiple factors could be at play. First, the S/N estimate for each CARMENES and HARPS-N dataset is provided by the respective pipelines, and are thus not necessarily comparable. A proper assessment of the S/N would require us to homogeneously reduce the raw frames of both instruments, which is out of the scope of this paper. Second, the amount of signal intrinsically carried in different wavelength ranges could differ, because of the difference in strength and number of \ion{Fe}{i} lines, and in planet-to-star flux ratio. In Appendix \ref{sec_appendix: intrisic_information_content} we show that bluer wavelengths (probed by HARPS-N) seem to carry more information about \ion{Fe}{i} compared to redder wavelengths (probed by CARMENES) in KELT-9b. Finally, telluric correction is more severe at the redder wavelengths probed by CARMENES. We can not exclude that, if present, the cumulative effect of slight telluric residuals left behind by our analysis, despite being buried in the photon noise, would more likely negatively impact the CARMENES observations.

  \begin{figure*}
  \resizebox{\hsize}{!}
            {\includegraphics{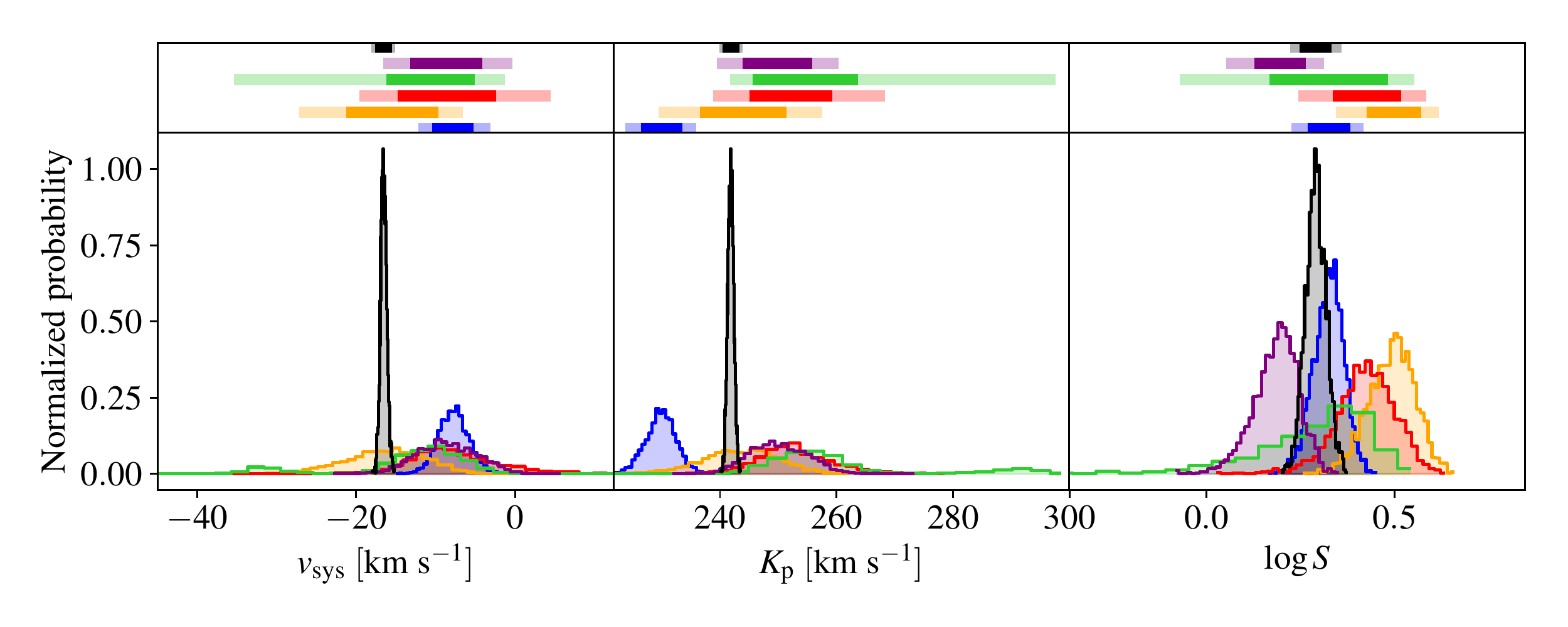}}
      \caption{Posterior distributions of parameters of model 1C fit with an MCMC to HARPS-N N1 (blue), HARPS-N N2 (orange), CARMENES N1 (red), CARMENES N2 (green), HARPS-N N3 (purple), and all 5 nights combined (black). The lower panels show the normalized posteriors (higher peak for better constrained parameters), and the marked improvement obtained when combining all data sets. The upper panels show the $1\sigma$ (dark color) and $2\sigma$ (light colour) confidence intervals obtained from the corresponding posteriors in the lower panels.
      }\label{fig:night_comparison}
  \end{figure*}

Most of the individual posteriors are compatible with each other at $1\sigma$ (see Fig. \ref{fig:night_comparison} and Appendix \ref{sec_appendix: individual nights}). The only exceptions are: the posteriors for the scaling factor in HARPS-N N2 (orange posterior) and HARPS-N N3 (purple posterior), marginally incompatible with each other; the posteriors for $K_\mathrm{p}$ in HARPS-N N1 (blue posterior), which is only compatible with HARPS-N N2 (orange posterior). Since the discrepancy in the posteriors for the scaling factor between HARPS-N N2 and HARPS-N N3 (orange and purple posterior) is little more than $2\sigma$, we do not consider it significant. We thus refrain from interpreting it as an astrophysical signal, also in light of the significant impact of the PCA on the strengths of \ion{Fe}{i} lines (see Fig. \ref{fig:processing}) that needs to be very accurately captured by our injection method for a correct interpretation. We do note, however, that HARPS-N N3 was observed at least two years after the other epochs, so that any variability in the planet atmosphere on year-long timescales would be visible. On the other hand, the discrepancy in $K_\mathrm{p}$ is more likely real. It concerns the only pre-eclipse night that we observed, and could be indicative of a pre-eclipse / post-eclipse asymmetry. We exclude that this is caused by the nodal precession of KELT-9b \citep{Stephan2022}, which would only change the inclination of the orbit by about $1^\circ$ in three years. Its effect would only have a relative impact on $K_\mathrm{p}$ at the $10^{-4}$ level. We will come back to our interpretation of the variation of $K_\mathrm{p}$ in the next section and in section \ref{sec: discussion_climate}. 

%
We also perform a combined analysis of the nights. 
The combined analysis leads to a marked improvement in the precision of each parameter (the relative corner plot is displayed in Fig. \ref{fig:MCMC_eccentric_versus_circular}). Our best fit parameters are $v_\mathrm{sys} = -16.5^{+0.4}_{-0.4}~\mathrm{km~s^{-1}}$, $K_\mathrm{p} = 241.8^{+0.5}_{-0.5}~\mathrm{km~s^{-1}}$, and $\log S =  0.29^{+0.03}_{-0.03}$ (see Table \ref{table:retrieval results}). As \cite{Pino2020}, we find that our model under-predicts the Fe I line intensity by a factor of about 2. This could be due to an underestimation of the temperature of the upper atmosphere (e.g., \citealt{Fossati2020, Fossati2021}), an underestimation of the \ion{Fe}{i} volume mixing ratio, an overestimation of the stellar flux, the presence of other atomic and molecular species in the spectrum that partially mask neutral iron spectral features \citep{Kasper2021}, or a combination of these. Understanding the source of this discrepancy is out of the scope of this paper, and our conclusions do not hinge on reproducing exactly the Fe I line depth with our one-dimensional model template.


\subsection{Neutral iron lines trace a displacement from a circular orbit for KELT-9b}
\label{sec: results, eccentric_orbit}
\begin{figure*}
  \resizebox{\hsize}{!}
  {\includegraphics[width=\hsize]{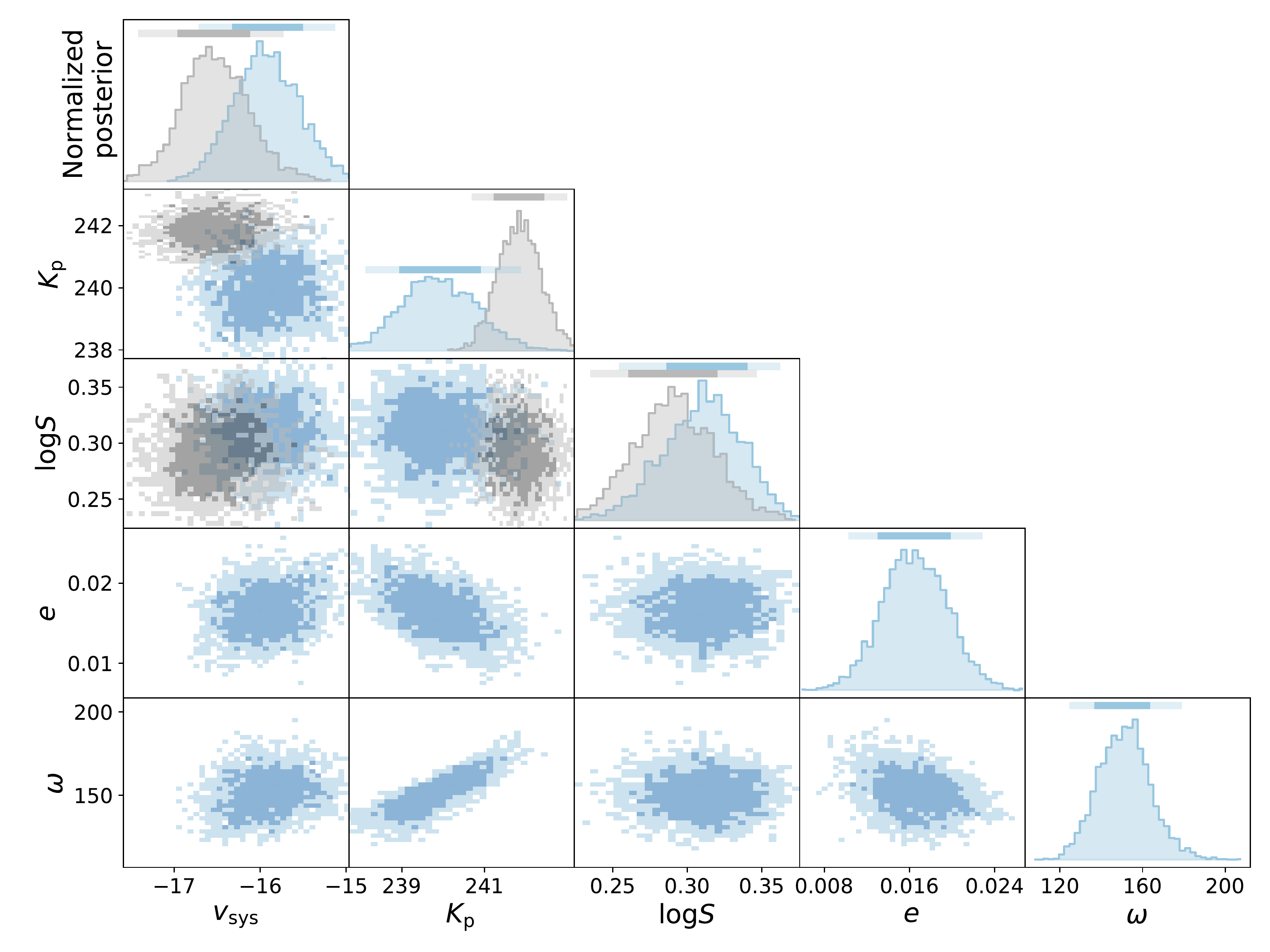}}
      \caption{Posterior for the MCMC fit of models 1C (circular orbit, black histograms) and 1E (eccentric orbit, blue histograms). The $1\sigma$ and $2\sigma$ confidence intervals are indicated with darker and lighter shades, respectively.}
         \label{fig:MCMC_eccentric_versus_circular}
  \end{figure*}

Encouraged by the sub-kilometre-per-second precision reached on 5 combined nights with model 1C, we perform a fit using an eccentric orbit (model 1E, see Table \ref{table:models adopted}). We obtain bound posteriors for all parameters of model 1E. We convert the posteriors in $h$ and $k$ to posteriors in $e$ and $\omega$ by converting every MCMC sample of $h$ and $k$ using $e = h^2 + k^2$ and $\omega = \arctan (h/k)$, and building the corresponding probability distribution. We show the posteriors in Fig. \ref{fig:MCMC_eccentric_versus_circular}, and our best fit parameters are: $v_\mathrm{sys} = -15.9^{+0.35}_{-0.4}~\mathrm{km~s^{-1}}$, $K_\mathrm{p} = 239.9^{+0.8}_{-1}~\mathrm{km~s^{-1}}$, $\log S =  0.31^{+0.02}_{-0.03}$, $e = 0.016^{+0.003}_{-0.003}$, and $\omega = 150^{+13\,\circ}_{-11}$ (see Table \ref{table:retrieval results}).

This result is surprising at face value, and in apparent contrast with \cite{Wong2020}. Indeed, they report a tight upper limit to the eccentricity of the planet of $e<0.007$ at $2\sigma$ by using TESS photometry, and additionally find that the addition of parameters $e$ and $\omega$ is disfavoured by BIC and AIC in their analysis, indicating a likely circular orbit, as is expected for KELT-9b. In sections \ref{sec: discussion_eccentricity} and \ref{sec: discussion_climate} we reconcile our result with \cite{Wong2020} by interpreting our measured eccentric as apparent, and actually due to atmospheric dynamics (winds). For now, we just remark that model 1E captures a significant deviation from a circular orbit traced by \ion{Fe}{i} lines, independently of the source of the anomaly.

Since model 1E has additional parameters compared to model 1C, we perform model comparison to determine whether the additional complexity is justified by the data, using several methods: likelihood ratio test, Akaike Information Criterion (AIC) and Bayesian Information Criterion (BIC)\footnote{These values should be computed in the maximum likelihood, which we have approximated with the median of the posterior. With a flat prior and perfectly symmetrical posteriors, this is a correct assumption.}. Since model 1C and 1E are nested, we apply the likelihood ratio test (as described in Appendix D in \citealt{Pino2020}) with 2 fixed parameters and obtain a $5\sigma$ preference for the eccentric model. We also obtain the apparently contrasting results that the BIC test strongly favours the circular orbit solution ($\mathrm{BIC_{1E}} - \mathrm{BIC_{1C}}= 7.9$), while AIC strongly favours the eccentric orbit solution ($\mathrm{AIC_{1E}} - \mathrm{AIC_{1C}}= -24.5$, which corresponds to a $4.5\sigma$ preference\footnote{Calculated from the relative likelihood of model 1C $\exp (\mathrm{AIC_{1E}} - \mathrm{AIC_{1C}})/2$ converted to $\sigma$ level using a two-tailed normal distribution test} for model 1E). In reality, this is not surprising: the BIC test penalizes higher complexity models much more compared to the AIC test, and this difference is more marked for larger data-sets. We fit simultaneously $n_\mathrm{tot}$ = 71978272 pixels. Thus, for a change of 2 in the number of parameters, the penalty term in the BIC test is $2\log(n_\mathrm{tot}) = 36.2$, 
compared to 4 for the AIC test. In other words, the BIC test is much more demanding in terms of quality of fit for higher complexity models.

We now compare the posterior distributions for models 1C and 1E (see Fig. \ref{fig:MCMC_eccentric_versus_circular}). Eccentricity and argument of periastron are partially degenerate with $K_\mathrm{p}$. As a result, the marginalized distribution for $K_\mathrm{p}$ is slightly broader in model 1E.
 All common parameters between models 1C and 1E are compatible at $2\sigma$. The eccentricity that we measure has a significance of more than $5\sigma$ level, which, in combination with the AIC test, suggests that the higher complexity model introduced by model 1E over 1C is justified.

\begin{figure*}
  \resizebox{\hsize}{!}
            {\includegraphics{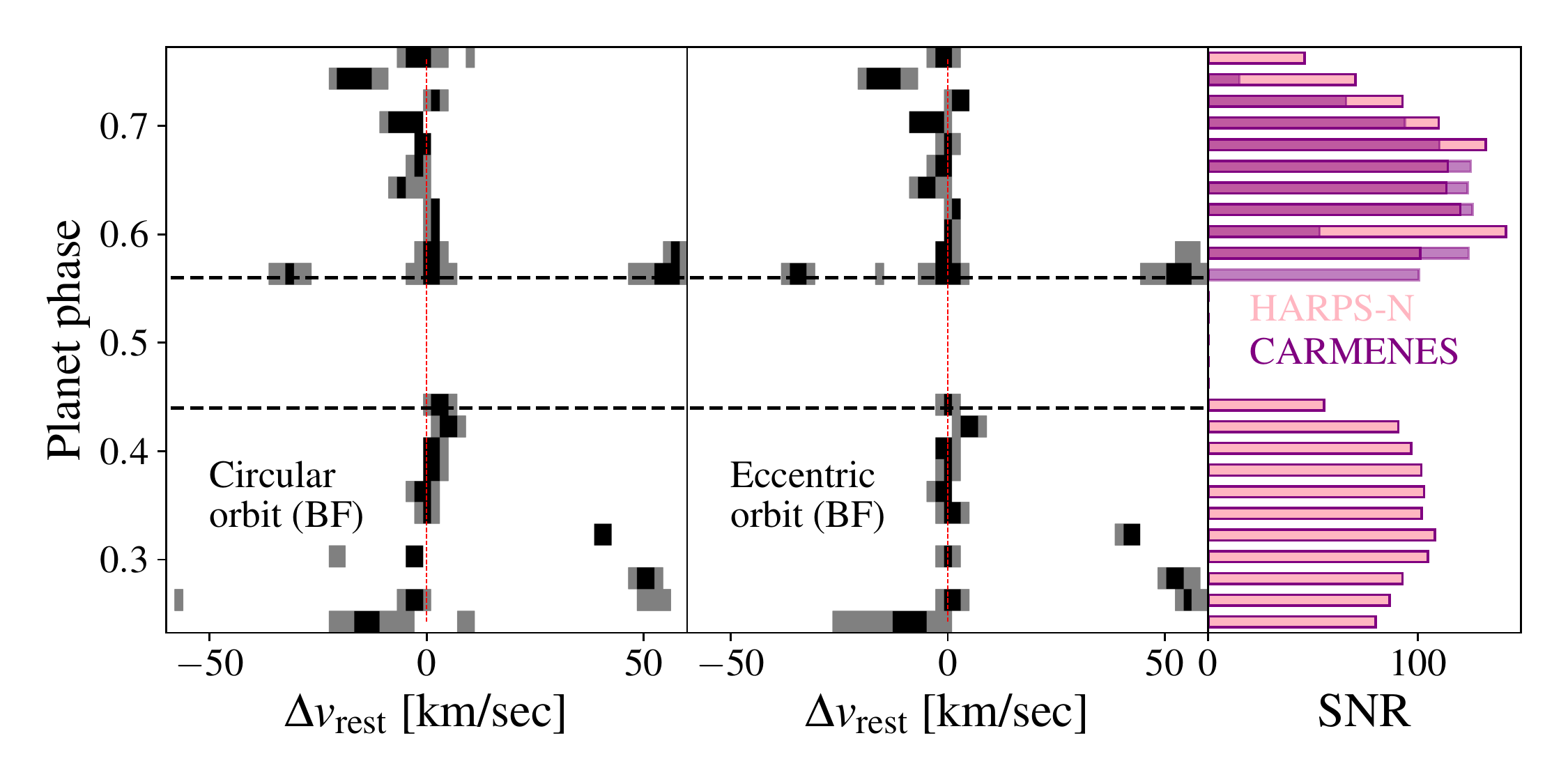}}
      \caption{Time-resolved confidence intervals as a function of the planet rest-frame velocity obtained by binning (i.e. co-adding) the individual likelihood functions by 0.02 in orbital phase. Black represents $1\sigma$ deviations from the best-fit rest-frame velocity in each phase bin and gray represents $2\sigma$ deviations. The left and centre panels show results for the 1C (circular) and 1E (eccentric) model, respectively, when fixing all the model parameters at their best-fit value, i.e. the median of the posterior distribution. Horizontal black dashed lines indicate phases corresponding to ingress and egress, and a vertical red dashed line indicates the best fit planet rest-frame. The right panel represents the total S/N achieved within each phase bin, with HARPS-N (lighter pink) and CARMENES (darker purple). \label{fig:trail}}
  \end{figure*}

We developed a new method to display the radial velocity displacement from the best-fit planet's rest frame ($\Delta v_\mathrm{rest}$) as a function of planet phase (planet trail) in the context of the likelihood framework by \cite{Brogi2019}. Conceptually, we can build the planet trail for a given solution as the conditional probability\footnote{As opposed to marginal probability: in this case we are interested in the value of $\Delta v_\mathrm{rest}$ for the given best fit solution, rather than trying to identify the most likely values of $\Delta v_\mathrm{rest}$ independently of the other parameters.} of $\Delta v_\mathrm{rest}$ at every planetary phase, given the best fit values for the model considered. In this representation, the confidence interval obtained at every phase is completely independent of all other phases, and confidence intervals of adjacent phases are not related. Practically, we divide the covered phase range in 0.02-wide phase bins, and for each phase bin we calculate the conditional likelihood as:
\begin{equation}
\label{Eq:planet_trail}
\mathcal{L}_\mathrm{trail}\left( \Delta v_\mathrm{rest}; \varphi \right) = \prod_\mathrm{exposure\,1, \varphi}^{\mathrm{exposure\,n}, \varphi} \mathcal{L}\left( K_\mathrm{p\,BF}, v_\mathrm{sys\, BF} + \Delta v_\mathrm{rest},  S_\mathrm{BF}; \varphi \right),
\end{equation}
where each phase bin is sampled by n exposures (n could differ in every phase bin), the subscript BF indicates the best fit value for a parameter, and $h$ and $k$ appear as additional parameters evaluated in the best fit position in the eccentric orbit case. The assumptions behind Eq. \ref{Eq:planet_trail} are that there is no atmospheric variability, and that each phase bin is small enough that we can neglect the planet motion within it. Finally, in every bin, we calculate the $1\sigma$ and $2\sigma$ deviations from the maximum likelihood in $\Delta v_\mathrm{rest}$ using Wilk's theorem. Figure \ref{fig:trail} displays the resulting trail in the best fit planetary rest frame using a box-plot \citep{Hyndman1996} for models 1C and 1E.  Multi-peaked solutions likely correspond to phase bins where the best-fit solution is not very well constrained, so that $1\sigma$ and $2\sigma$ deviations capture additional spurious peaks. In the eccentric orbit case, the trail appears as a vertical line (except a few noisy phase bins), which is the expectation in the case that we are in the rest frame where the lines are originated. On the opposite, in the circular orbit case, the trail appears slanted. This indicates that a circular orbit is not fully capable of capturing the morphology of the planet trail or, in other words, that the rest frame from which the neutral iron signal is generated deviates from a circular orbit.

Concluding, our results markedly favour model 1E over model 1C, indicating the presence of a significant radial velocity anomaly detected for \ion{Fe}{i} lines. As we already mentioned, and further address in sections \ref{sec: discussion_eccentricity} and \ref{sec: discussion_climate}, such radial velocity anomaly is unlikely due to the orbital motion of the planet, and is more likely the result of atmospheric dynamics in KELT-9b.

\subsection{Symmetric intensity of iron lines around secondary eclipse}
\label{sec: results_symmetry}
\begin{figure*}
  \resizebox{\hsize}{!}
            {\includegraphics{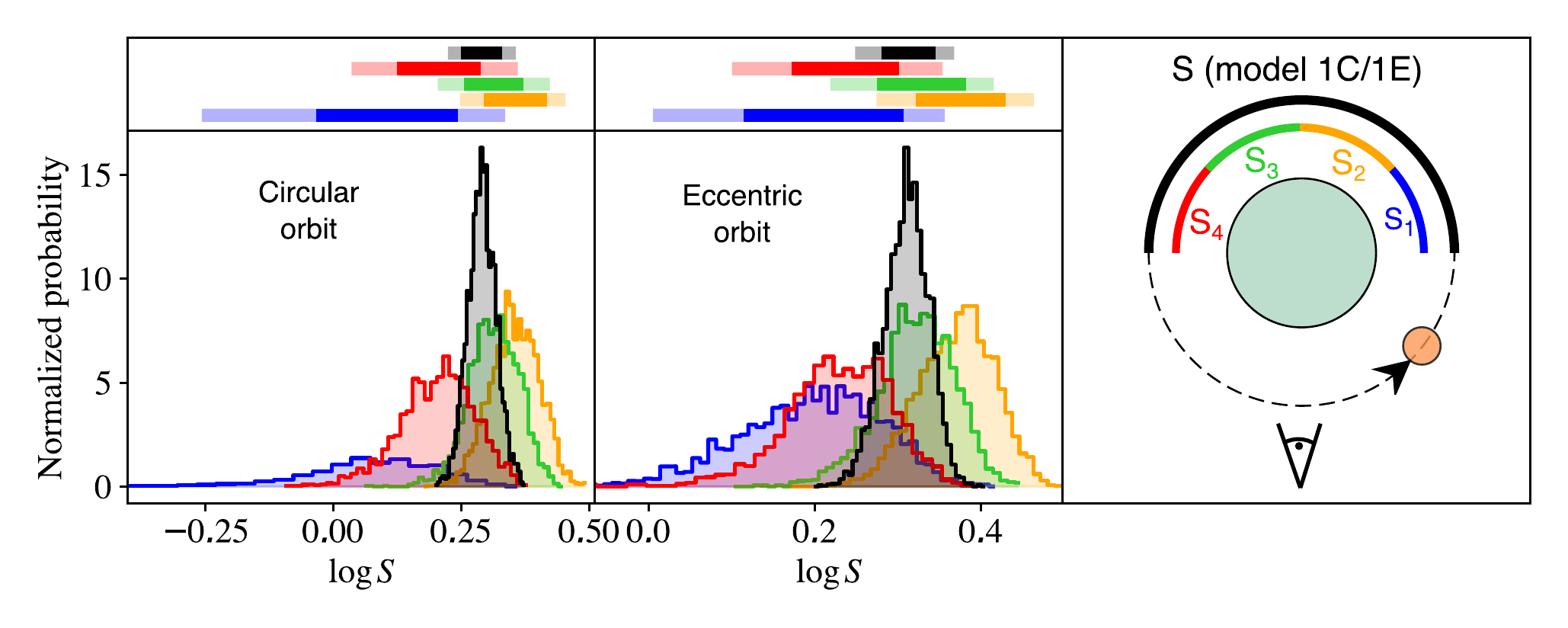}}
      \caption{Posterior distribution for the scale factors of models 1C and 4C (left panel), 1E and 4E (middle panel). The right panel shows the phase ranges corresponding to each of the 5 parameters for each panel, color-coded. \label{fig: single_versus_four_scale_factors}}
  \end{figure*}

We then search a variation in the intensity of the observed neutral iron lines while the planet orbits around its host. First, we assume a circular orbit and fit a separate scale factor for 4 mutually exclusive ranges in phase (model 4C). Figure \ref{fig: single_versus_four_scale_factors}, left panel, shows the resulting posteriors for the scale factors. We show the full corner plots in Appendix \ref{sec_appendix:phase_dependent_intensity_corner_plots} (Fig. \ref{fig:4C_vs_1C}), and report best fit parameters and their errors in Table \ref{table:retrieval results}. The MCMC finds well constrained posteriors for all parameters, indicating that our data contains sufficient information to perform this kind of study. The posteriors of $K_\mathrm{p}$ and $v_\mathrm{sys}$ are in good agreement with those for model 1C. We attribute this to the fact that the scale factor (hence, neutral iron line intensity) and the orbital parameters (hence, neutral iron line Doppler shift) are not correlated at our level of precision. 

We additionally repeated the fit using model 4E, thus allowing for an eccentric orbit. We obtained bound posteriors for all parameters, and compare the resulting scale factors in the right panel of Fig. \ref{fig: single_versus_four_scale_factors}. We also show full corner plots in the Fig. \ref{fig:4E_vs_1E}, and report best fit parameters and their errors in Table \ref{table:retrieval results}. We observe that the scale factors are uncorrelated with all other parameters at our level of precision. As a result, the posteriors of the phase-resolved scale factors are very similar to those of model 4C, and the posteriors of orbital parameters are close to those of model 1E. We conclude that, at our level of precision, the Doppler-shift and line shape of the exoplanet atmosphere spectra can be modelled separately without any loss of information.

Finally, we run MCMC chains using Lambert sphere models (L, Loff, Lbase). We achieve convergence on all parameters in these models, and we report the full posteriors in Appendix (Fig. \ref{fig:Slam_vs_Soff}; Fig. \ref{fig:Slam_vs_Sbase}), and best fit parameters with their errors in Table \ref{table:retrieval results}. The orbital properties ($K_\mathrm{p}$, $v_\mathrm{sys}$, $e$, $\omega$) are in excellent agreement across the three models, confirming that Doppler and line intensity information are uncorrelated, and confirming the preference for an eccentric orbit.

\begin{figure}
  \centering
  \includegraphics[width=\hsize]{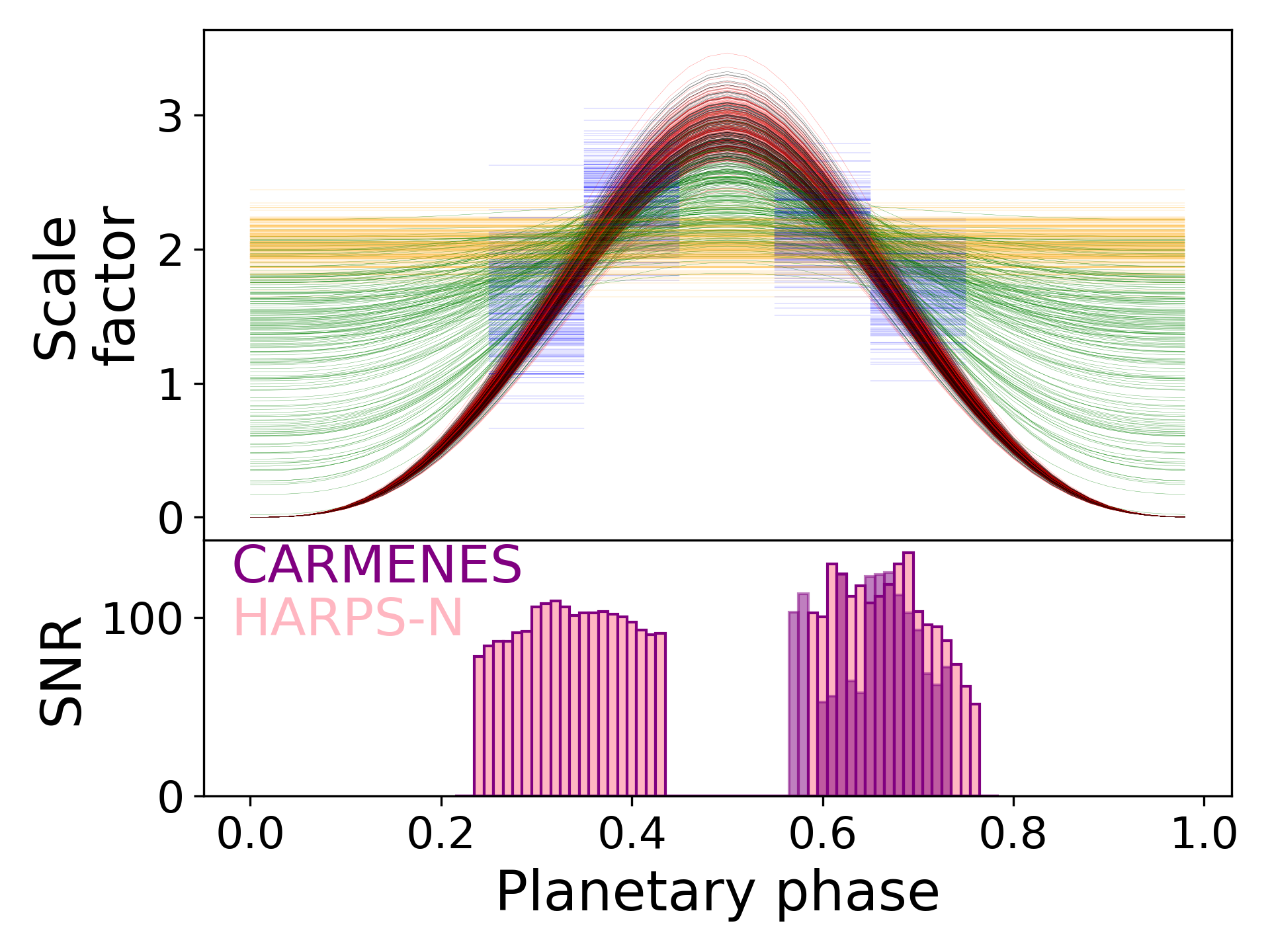}
      \caption{The upper panel shows 100 draws from the posterior for scale factors as a function of phase obtained by fitting models 1E (orange), 4E (blue), L (black), Loff (red) and Lbase (green) to the data. The phase coverage of the data is shown in the lower panel, in terms of the S/N of stellar spectra in each phase bin (HARPS-N: lighter pink; CARMENES: darker purple). Where multiple nights cover the same orbital phase bin their contributions are summed in quadrature.}
         \label{fig:collier_cameron_models}
  \end{figure}

We compare results from these models and results from model 1E and 4E in Fig. \ref{fig:collier_cameron_models}. A first clear result is that the peak of the emission is tightly constrained at the substellar point. This is supported by the posterior of the parameter $\varphi_0$ in model Loff, whose value is $\varphi = 0.00\pm0.01$,
which translates to an angular displacement from the substellar point of $0\pm5^\circ$. Secondly, model Lbase presents a degeneracy between $S_\mathrm{Lambert}$ and $S$. In other words, it is possible to explain the data both with an intensity profile which decreases moving towards the nightside (akin to model L, where contribution from the anti-stellar point is forced to 0), but also with a constant intensity profile independent of phase. Solutions with no $S_\mathrm{Lambert}$ contribution are also allowed, while the model supports at $>2\sigma$ significance the presence of a phase-independent contribution to emission. When comparing the intensity profiles (see Fig. \ref{fig:collier_cameron_models}), it is clear that both models L and Lbase agree in the range actually probed by observations (phases $0.25 < \varphi < 0.45$ and $0.55 < \varphi < 0.75$). In addition, all Lambert sphere models are in good agreement with the scale factors measured with model 4E in the probed planet phase range. We conclude that our data are not sufficient to support a strong evidence of a variation of the intensity of iron lines with longitude. This is further confirmed by the strong preference of model 1E over model L and Loff (e.g., $\mathrm{AIC_L - AIC_{1E} = 8.5}$, and same difference in BIC), and the marginal to strong preference over Lbase according to AIC and BIC ($\mathrm{AIC_{L_{base}} - AIC_{1E} = 2.4}$; $\mathrm{BIC_Lbase - BIC_{1E} = 18}$). We argue that Lbase is less disfavoured because it allows solutions with smaller to no dependence on orbital phase of the scale factor, further supporting our results. 

On the other hand, we can confidently conclude that there is no detectable asymmetry (e.g. due to a hot-spot offset) in the intensity of lines in pre- and post-eclipse spectra. Indeed, model Loff is strongly disfavoured in terms of BIC and AIC compared to model L (by 19 and 3, respectively), and its parameter $\varphi_0$ presents a tight posterior around 0. 

\section{Discussion}

\label{sec:discussion}

\subsection{HRS phase curves as a new tool for exoplanet atmosphere characterization}
\label{sec:discussion_phase_curves}
Our results show that HRS emission spectroscopy observations of hot gas giants have the potential to reveal much more information when treated as a phase-curve rather than a single phase-collapsed emission spectrum, as was done in most literature works (e.g., \citealt{Pino2020, Yan2020,Nugroho2020, Borsa2022}). Conceptually, an HRS phase curve is similar to a classic low-resolution (LRS) phase curve observed from space (e.g., \citealt{Stevenson2016}). The key difference between LRS and HRS phase curves is the observable they record. For LRS phase curves, this is simply the flux of the planet-plus-star system  as a function of planetary phase. On the other hand, HRS phase curves record two separate observable quantities.
%

The first observable quantity is the contrast of planet lines relative to the continuum generated by the star and planet as a function of phase (\citealt{Pino2020}; also noted and exploited by \citealt{Herman2022} and \citealt{vanSluijs2022}). Unlike in LRS phase curves, the spectral features of the planet continuum are lost because it is impossible to reconstruct the chromatic fibre losses, which are due to differential refraction from Earth atmosphere. Such losses introduce an additional, unknown wavelength and time dependence which need to be removed from the spectra in order to correct the stellar and telluric lines, and extract the planetary signal (e.g., Appendix A of \citealt{Pino2020}). This is typically performed with a normalization, which unfortunately also removes the exoplanet atmosphere continuum features. This may seem a strong limitation compared to LRS phase curves. However, when interpreted in an appropriate statistical framework \citep{Brogi2019}, HRS emission spectra still contain enough information to constrain the thermal profiles and the volume mixing ratios \citep{Pino2020, Kasper2021}, including their absolute values \citep{Line2021}. This extends to HRS phase curves. Indeed, both in this work and in \cite{Herman2022} and \cite{vanSluijs2022}, HRS phase curves are used to constrain the value of a scaling factor as a function of phase, which is a proxy for the steepness of the thermal gradient and volume mixing ratio of the species present in the cross-correlation template. In addition, compared to LRS phase curves, HRS phase curves probe higher up in the atmosphere, and, thanks to the simultaneous detection of lines of different strengths, a broader span of pressures.

The second observable quantity is the phase resolved Doppler-shift of planet lines. This is obtained by individually resolving the spectral lines (see section \ref{sec: results, eccentric_orbit}), and is only possible with HRS spectrographs. \cite{Herman2022} do not include this aspect in their analysis, and \cite{vanSluijs2022} indicate that at a resolution of 15000 this information is washed out. At the price of lower efficiency compared to lower resolution slit spectrographs, fibre-fed, $R\sim100000$ spectrographs open a new window into planetary climate and dynamics, which can not be obtained with any other technique (\citealt{Kawahara2012}; see sections \ref{sec: discussion_eccentricity}, \ref{sec: discussion_climate}).

Line contrasts and wind induced Doppler shifts are determined by the complex interplay between energy transport, chemistry and dynamics. They are thus intrinsically linked. An appropriate framework to interpret HRS phase curves needs to be able to reproduce both observable quantities simultaneously. This has already been shown by computing HRS spectra starting from GCMs of irradiated planets (e.g., \citealt{Zhang2017, Beltz2021, Beltz2022}). Unfortunately, such simulations are time-consuming, and it is currently not realistic to use them for data comparison over a large parameter space. However, our analysis demonstrates that, to first-order, the line contrasts (parameterized by the scale factor) and the positions (parameterized by $K_\mathrm{p}$, $v_\mathrm{sys}$, $h$, and $k$) are uncorrelated for KELT-9b. As a result, if Doppler shifts can be attributed to planetary climate, it is possible to build a first-order picture of winds (through line positions) and thermal/abundance structure (through line contrasts) even in the lack of a GCM analysis (see section \ref{sec: discussion_climate}).

\subsection{Sensitivity of HRS phase curves to slight eccentricities in UHJs}
\label{sec: discussion_eccentricity}

In section \ref{sec: results, eccentric_orbit} we have demonstrated that, with our new method based on the relative velocities of the planet probed by its atmospheric emission lines throughout the orbit, HRS phase curves have the potential to detect radial velocity deviations from a circular orbit of the order of a few kilometres-per-second. In section \ref{sec: results, eccentric_orbit}, we further demonstrated that, if due to a slightly eccentric orbit, deviations of this order would correspond to a solution of $e=0.016\pm0.003$ that we measured with high significance. In the specific case of KELT-9b, TESS photometry shows that this eccentricity is unlikely real (\citealt{Wong2020}, Sec. \ref{sec: results, eccentric_orbit}). However, it is interesting to note that, if KELT-9b had an eccentricity of this magnitude, we would have had the sensitivity to measure it with our new technique.

The challenge of measuring slight eccentricities of UHJs with traditional methods such as stellar radial velocities and transit photometry is manyfold. Indeed, many planets of this class orbit A-type or early F-type stars, which have fast rotational velocities (up to more than 100 $\mathrm{km s^{-1}}$). As a result, the stellar lines are significantly broadened, resulting in reduced stellar radial velocity precision. In addition, at the level of precision required, detailed modelling of second order effects, such as the radial velocity contamination by tidal bulges induced by the planet on the host star, need to be accounted for. Transit photometry offers an alternative avenue to measure slight eccentricities, but the interpretation of these measurements is not straightforward. For instance, an inhomogeneous day-side surface brightness of the planet could affect the retrieved mid-occultation timing \citep{Shporer2014}, whose accurate knowledge is necessary to measure a slight eccentricity with this method.

Our method is completely independent, and does not suffer from the same systematics as those based on stellar radial velocities and transit photometry. However, as we already discussed in Sec. \ref{sec: results, eccentric_orbit} and will discuss in more detail in Sec. \ref{sec: discussion_climate}, winds of moderate strength (a few kilometres-per-second) in UHJs can create signals of the same order of magnitude as those induced by a slightly eccentric orbit. Thus, while planetary radial velocities of UHJs could be used to measure or set tight upper limits to the potential slight eccentricities of UHJs, the interpretation of radial velocity anomalies in terms of orbital dynamics or climate could remain challenging.

It is out of the scope to explore in detail how to break such degeneracy. However, our work already shows that combination with independent measurements of eccentricities has the potential to discriminate between the two possibilities. In our case, \cite{Wong2020} provides a tight upper limit by combining information on the timing and duration of the primary and secondary eclipse of KELT-9b (measured by TESS). Indeed, these combined observables depend on the parameters $k$ and $h$, respectively \citep{Ragozzine2009}. Unfortunately, in many practical cases, only the timing difference can be measured to a useful precision, thus making it impossible to completely determine $e$ and $\omega$ \citep{Charbonneau2005}. Still, in the most favourable cases, this method has provided some of the best precision to date on eccentricities, pushing to measuring eccentricities smaller than 0.01 at $>5\sigma$ significance (e.g., \citealt{Nymeyer2011}). Extensive stellar radial velocities campaigns also demonstrated the capability to significantly measure eccentricities smaller than 0.01 in the best cases. For example, \citealt{Bonomo2017}, one of the largest homogeneous studies of eccentricities, significantly measured eccentricities smaller than 0.01 in 10 out of 231 planets. However, those cases all correspond to planets orbiting relatively slowly rotating host stars ($v\sin(i) < 20~\mathrm{km\ s^{-1}}$), and are the subject of multi-year radial velocity campaigns, and thus they still represent a minority. 

In addition, different planetary wind patterns produce a different shape for the radial velocity anomaly (see Sec. \ref{sec: discussion_climate}), and so does an eccentric orbit. Through detailed modelling, and using a wide planetary phase coverage, it could thus be possible to distinguish between the radial velocity anomaly morphology induced by an eccentric orbit or winds. For instance, \cite{Kawahara2012} shows that the additional modulation induced by a slightly eccentric orbit over a circular one has half the period of the corresponding circular orbit, while planetary rotation, and, by extension, simple wind patterns, induce a deviation with a full orbital period (see also Sec. \ref{sec: discussion_climate}). Planetary radial velocities obtained towards quadrature and beyond (pushing into the night-side) are likely best suited to disentangle these solutions, which we will study in future work.

Finally, combining information from multiple species could prove the key to ultimately break this degeneracy. Eccentricity would produce the same radial velocity anomaly independently of the species or specific lines analysed. On the other hand, winds are expected to be altitude dependent. Lines of different oscillator strength probe different altitudes in the atmosphere, spanning several orders of magnitude in pressure with HRS (e.g., \citealt{Pino2020}). As a result, the radial velocity measured through different subsets of lines or atmospheric species that probe different altitudes would change depending on the wind pattern experienced by them.

Thanks to the advent of new generation spectrographs mounted on 8-metre class telescopes, with better throughput and broader spectral coverage (e.g., ESPRESSO@VLT, MAROON-X@Gemini-N), the same precision that we achieve on KELT-9b using \ion{Fe}{i} lines only will be reached on systems at least 2.5 magnitudes dimmer than KELT-9b (pushing up to $\mathrm{V}=9$). Despite the challenges outlined above, the prospect of a survey of eccentricities for this class of planets is tantalizing, as it would open a new window onto UHJ evolution history, and potentially of their internal structure \citep{Fortney2021}.

\subsection{The climate of KELT-9b}
\label{sec: discussion_climate}

As discussed in section \ref{sec: results, eccentric_orbit}, we consider unlikely that KELT-9b's orbit is actually eccentric. To reconcile our results with \cite{Wong2020}, we thus explore an alternative and more likely scenario in which the neutral iron signal comes from a localized region of the planet day-side atmosphere. Indeed, in this scenario, this region will show changes in net Doppler shift due to rotation and winds, and the neutral iron lines Doppler shift would track them.

We build a toy model to capture this effect. We define two sets of spherical polar coordinate systems to represent a spherical planet: $(\beta, \alpha)$ represent the latitude and longitude in a reference frame centred in the planet, with the observer located along the x-axis ($\alpha=0$); $(\theta, \phi)$ represent the latitude and longitude in a reference frame centred in the planet, co-rotating with the planet and with the sub-stellar point located along the x-axis ($\phi=0$). Since KELT-9b is transiting, we assume that the orbital inclination is $90^{\circ}$, which leads to the relations $\theta=\beta$ and $\alpha=\phi+\varphi$, where $\varphi$ is the planet phase. We assume a solid-body rotation for the planet, and an obliquity of $0^\circ$. We additionally allow for a zonal wind flow: we model both an eastward wind and a day-to-night wind, which are the wind circulation patterns largely predicted to be dominant by General Circulation Models of KELT-9b \citep{Komacek2017}. Then, the line-of-sight (LOS) component of the velocity of a mass of iron gas located at coordinates $(\beta, \alpha)$ is:
\begin{equation}
\label{Eq: vlos}
v_\mathrm{LOS}(\beta, \alpha) = -\left(R_\mathrm{p}\Omega+v_{\mathrm{wind}}\right)~\sin\alpha~\sin\beta\ ,
\end{equation}
where $R_\mathrm{p}$ is the radius of the planet and $\Omega$ is the angular velocity derived from the orbital period, and $v_{\mathrm{wind}}$ is always positive in the eastward jet case, but is negative (contrary to the sense of planet rotation) for $\phi < \pi$ in the day-to-night flow case. In addition, we need to take into account that not every surface element of the planet contributes to the net Doppler shift in the same way. Following \cite{Zhang2017}, we weight every contribution by the flux emitted in the direction of the observer. The projection of the normal to the local surface along the direction of observation results in a factor of $\mu = \cos \alpha \sin\beta$, which is thus the visibility weight. Finally, we weight every surface element with 1 if it is located in the day-side of the planet, and 0 if it is located in the night-side of the planet to account for the uneven brightness distribution.

\begin{figure*}
  \sidecaption
      \includegraphics[trim={0 1cm 0 0}, width=12cm]{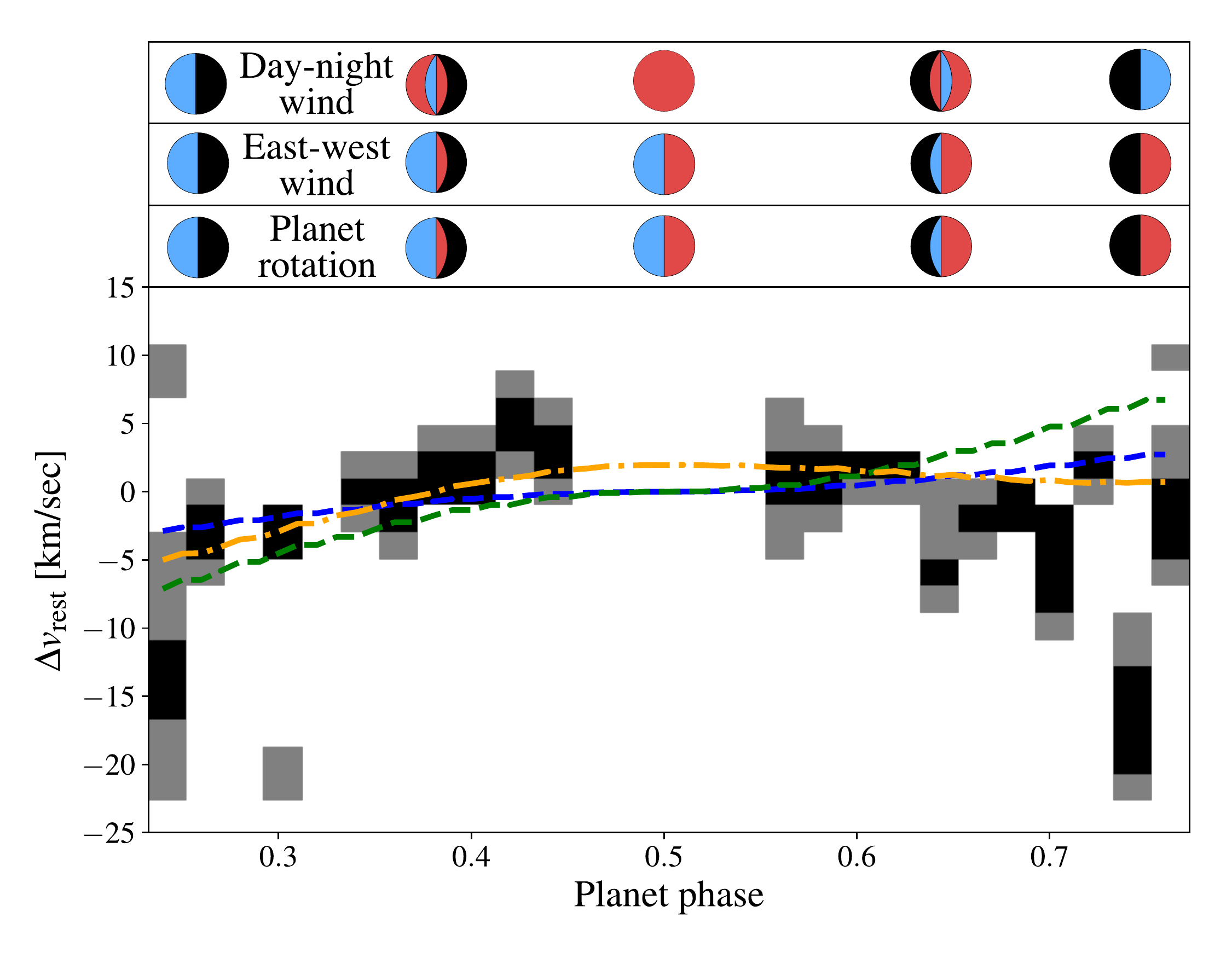}
      \caption{Planetary trail obtained in the best-fit circular orbit reference frame (same as left panel of Fig. \ref{fig:trail}, but zoomed on smaller radial velocity displacements from the best fit circular orbit rest frame), and comparison with predictions of net Doppler-shift in different cases: pure rotation (blue dashed curve), east-west wind of $10~\mathrm{km~s^{-1}}$ (green dashed curve), and day-to-night wind of $5~\mathrm{km~s^{-1}}$ (orange dot-dashed curve). The upper panel shows the contribution to Doppler shift from the portion of the planet disk seen at different phases due to the two wind geometries considered and planet rotation. Black portions of the disk correspond to the night-side, while red(blue) portions correspond to regions producing red(blue)shifted lines. Less weight should be given to phase bins close to quadrature ($\varphi < 0.35$; $\varphi>0.65$) because of the detrimental effect of our analysis on the signal-to-noise ratio in this phase range (see Fig. \ref{fig:processing}). \label{fig: trail_and_winds}}
  \end{figure*}

We compare the predictions of this toy model with the planet trail obtained in the best fit circular orbit rest frame of the planet in Fig. \ref{fig: trail_and_winds}. In the case of pure rotation, we can intuitively understand the curve. Right after transit ($\varphi=0$), the day-side starts to be visible as it rotates towards the observer. As a result, the net Doppler-shift is negative and close in value to the rotational velocity. While more of the blue-shifted day-side is visible with increasing planet phase, the Doppler-shift becomes dominated by regions of the planet that have a smaller projected velocity component towards the observer due to geometry, and thus its magnitude decreases. After phase 0.25 part of the visible day-side is rotating away from the observer. At $\varphi=0.5$ the two parts cancel out exactly as the day-side is completely visible. The effect is symmetric for $\varphi>0.5$. The discontinuity at $\varphi=0$ is due to our assumption that the night-side is not contributing to the emission \citep{Beltz2022}. This reasoning is easily extended to the case of eastward wind, which simply adds to the planet rotation velocity, while the day-to-night flow case is slightly more complicated, but can be understood with similar reasoning. 

The pure rotation curve (blue curve) is consistent with the trail at the $2\sigma$ level in the majority of the individual phase bins, both in pre-eclipse and post-eclipse. However, it does not seem to correctly capture the trends in neither phase range. In the pre-eclipse ($\varphi<0.5$) phase range, the trail appears to be steeper compared to the prediction of the pure rotation model. In other words, the rotation alone seems to be too slow to explain the pre-eclipse trail. The addition of zonal flows in the direction of rotation provides a possible solution. In this phase range, both an eastward (green curve) and a day-to-night (orange curve) flows are aligned with the planet rotation. We find that with the addition of an eastward wind model with $v_\mathrm{wind}=10~\mathrm{km~s^{-1}}$ the prediction matches the observed slope of the trail better compared to the pure rotation case. However, this model fails to reproduce the evidence for redshift observed right before eclipse, and it deviates from the trail in the post-eclipse range where the trail appears flatter. Both these trends are instead captured by a rotation plus day-to-night flow model with $v_\mathrm{wind}=5~\mathrm{km~s^{-1}}$. Indeed, in this configuration, the winds tend to redshift the signal close to secondary eclipse, and to counter-act the effect of rotation in post-eclipse phases, producing a nearly constant-with-phase trail.



Our results are broadly consistent with expectations from GCMs and \textit{Spitzer} phase curve observations of KELT-9b. Indeed, \cite{Mansfield2020} indicate a preference for strong drag GCMs to explain the day-side phase curve of this planet. In strong drag models, the drag timescale is short compared to the rotational timescale. As a result, the balance in the horizontal angular momentum is primarily between the drag force (e.g., dissipation due to the interaction between the planetary magnetic field and the ions in the atmosphere) and the day-night thermal forcing. This results in a day-to-night flow which is roughly symmetric around the substellar point \citep{Showman2011, Tan2019}. For KELT-9b, GCMs predict wind speeds of about $5~\mathrm{km~s^{-1}}$, a roughly symmetric thermal structure around the substellar point, and a day-to-night flow at the photospheric pressures probed by \textit{Spitzer} \citep{Tan2019, Mansfield2020}. This is in very good agreement with the results from our HRS phase curve, which shows symmetry in line intensities around the substellar point (Sec. \ref{sec: results_symmetry}), and shows a phase-dependent Doppler shift which is well reproduced by day-to-night wind flows with the speeds predicted by GCMs (this section). In addition, \cite{Wong2020} report a small eastward hot-spot offset of $5.2\pm0.9^{\circ}$, which is consistent with our upper limit to the scale factor asymmetry in the HRS phase curve. Indeed, in the presence of a hot-spot offset driven by an eastward super-rotating jet, we could expect the steepness of the thermal gradient and the iron ionization fraction to vary across the day-side in an asymmetric way, which would be picked by our retrievals as an asymmetry in the scale factor as a function of longitude \citep{vanSluijs2022}. However, \cite{Mansfield2020} report a larger hot-spot offset of $19\pm2^{\circ}$. If the vertical gradient in the thermal structure followed the same longitudinal pattern as the photospheric emission from \textit{Spitzer}, our retrieval should have the precision to detect it.  It is anyway important to note that a coherent explanation of the \textit{TESS} and \textit{Spitzer} phase curves for this planet based on GCM models is still missing, and that neutral iron lines in this planet probe atmospheric pressures as low as $10^{-5}~\mathrm{bar}$ \citep{Pino2020}, which is significantly lower than the \textit{TESS} and \textit{Spitzer} photosphere ($10^{-1}$--$10^{-2}~\mathrm{bar}$), and well within the region where the double-gray assumption common to many GCMs does not apply \citep{Rauscher2012}. As a consequence, dedicated modelling effort to simultaneously fit LRS and HRS phase curves is necessary, and we leave it for future work.

\subsection{Implications for wind dynamics measured in transmission spectroscopy}

Day-to-night wind flow in KELT-9b was also reported by \cite{PaiAsnodkar2022} and \cite{BelloArufe2022} based on the Doppler shift of \ion{Fe}{ii} lines seen in transmission compared to the stellar systemic velocity. While \ion{Fe}{i} traces gas motions in the gravitationally bound part of the atmosphere of a planet, \ion{Fe}{ii} is likely tracing the dynamics of the evaporating exosphere \citep{Zhang2022}. Therefore, a full analysis combining high spectral resolution phase curves and transmission spectroscopy of UHJs has the potential to illuminate the physical processes leading to mass loss in this class of exoplanets. The challenge is that this requires the comparison between different geometries, and accurate modelling over a broad range of pressures, possibly including hydrodynamics and non-local thermodynamic equilibrium effects \citep{Huang2017, Hoeijmakers2019}. This is out of the scope of this paper, and we limit ourselves to a qualitative comparison, and to highlight potential pathways and caveats towards combining the two techniques.

Both \cite{PaiAsnodkar2022} and \cite{BelloArufe2022} report winds directed from day-side to night-side, with speeds between $5$ and $10~\mathrm{km~s^{-1}}$. At a qualitative level, both the wind speed and pattern that they report is consistent with our physical interpretation of the \ion{Fe}{i} Doppler shifts emerging from the HRS phase curve. This may suggest that the transition in dynamical regimes is not sharp across the pressures probed by \ion{Fe}{i} and \ion{Fe}{ii} in KELT-9b. At face value, this is surprising, since the two sets of observations likely bridge from the gravitationally bound part of the atmosphere to the exosphere. However, both our analysis and the analyses by \cite{PaiAsnodkar2022} and \cite{BelloArufe2022} only account for Doppler shift of lines, and yet, line shapes are also distorted differently by different wind patterns \citep{Seidel2020}. A more detailed analysis that accounts for this effect is thus necessary to properly assess whether a transition in the dynamical regime experienced by \ion{Fe}{i} and \ion{Fe}{ii} takes place. 

Additionally, \cite{PaiAsnodkar2022} report potential variability in the measured wind speed at several $\mathrm{km~s^{-1}}$, although this does not reflect in strong brightness temperature variations for the planet \citep{Jones2022}. We leave the study of the effect of variability to future work, and only note that our observations likely probe deeper layers in the atmosphere, and that our results assume that no variability across our observed epochs is present. 

It is also important to note that the measurement of the systemic velocity from the stellar lines of KELT-9 appears to be challenging. 
This highlights a potential weakness of transmission spectra as probes of atmospheric dynamics \citep{BelloArufe2022}, since they must rely on the accuracy of the systemic velocity. On the other hand, our measurement only depends on the relative shift of planetary lines during the orbit, and it is therefore independent of the value of the systemic velocity. Our result thus constitutes an independent measurement of the likely presence of day-to-night winds at the several kilometre-per-second level, blowing at the altitudes probed by neutral atomic species in KELT-9b, which could indeed extend to the upper atmosphere probed by \ion{Fe}{ii} lines in transmission.

Looking forward, our technique could also be used to revisit results from transmission spectroscopy without using the systemic velocity measured through stellar lines, which is potentially inaccurate at the level of a few $\mathrm{km~s^{-1}}$ in this planet \citep{Pino2020}. For instance, our measured value of $v_\mathrm{sys}$ has a precision of about $0.5~\mathrm{km~s^{-1}}$, which could still be sufficient to compare with Doppler displacements measured in the transmission spectrum of the planet to reveal day-to-night winds of several $\mathrm{km~s^{-1}}$. However, our measurement is not guaranteed to be a completely accurate estimate of the systemic velocity itself, because its value could be affected by winds blowing neutral iron gas (as it happens in transmission). Still, $v_\mathrm{sys}$ values measured assuming a circular or an eccentric orbit (capturing winds in our interpretation) are compatible at $1\sigma$ (Table D.1). This provides some confidence that our systemic velocity is accurate at the level of $1~\mathrm{km~s^{-1}}$. It is also consistent with values measured on stellar lines by \cite{PaiAsnodkar2022}, \cite{Hoeijmakers2019}, and \cite{BelloArufe2022}, but seems more difficult to reconcile with the value measured by \cite{Borsa2019} and \cite{Gaudi2017}, from whose measurements it deviates by $3~\mathrm{km~s^{-1}}$ and $4~\mathrm{km~s^{-1}}$, respectively. Ultimately, while a more detailed analysis of the potential systematics and astrophysical effects impacting the measurement of $v_\mathrm{sys}$ with each technique is required, our analysis provides additional confidence to the literature estimates of wind speeds measured from transmission spectra in KELT-9b.


\subsection{KELT-9b in the context of ultra-hot Jupiters}

Our current understanding of the physical processes regulating the climate of hot and ultra-hot gas giants has mostly come from space-borne phase curve studies. One of the most striking results from these surveys is the relation between day-night temperature contrast, phase offset and level of stellar irradiation. Phase curve observations have found a systematic increase of day-night temperature contrast with increasing temperature up to about $2500~\mathrm{K}$, accompanied by a decrease in hot-spot offset, which is also a prediction of General Circulation Models (GCMs). More recently, it was found that hotter planets showcase a surprisingly low day-to-night contrast, which is interpreted as likely due to cooling of the day-side due to $\mathrm{H}_2$ dissociation, and warming of the night-side due to $\mathrm{H}_2$ recombination \citep{Bell2018, Tan2019}. It is less clear whether there is a trend in hot-spot offset: \cite{Zhang2018} suggest an increase of the offset for planets hotter than $3400~\mathrm{K}$, but this is based on a few data points and was challenged by the lack of a large offset in individual ultra-hot Jupiters (e.g., WASP-103b, \citealt{Kreidberg2018}).

KELT-9b constitutes the high end of the population of the most irradiated planets, and thus an ideal test-case for climate theories of ultra hot Jupiters. Our results (1) indicate that there is no strong asymmetry between east and west day-side thermal structure or volume mixing ratio of iron from KELT-9b, and (2) indicate a possible preference for day-to-night flow over eastward flow. Both results are consistent with predictions of GCMs, and partially agree with results from photometric phase curves of this planet (section \ref{sec: discussion_climate}). However, as noted in section \ref{sec: discussion_climate}, it is not trivial to directly compare results from LRS and HRS phase curves because they probe very different pressure regions in the atmosphere. 

It is however interesting to compare our results on KELT-9b with recent work studying the HRS phase curve of WASP-33b through neutral iron \citep{Herman2022} and carbon-monoxide \citep{vanSluijs2022} lines. Both studies report a phase-dependence of the scale factor, with larger scale factors found at phases $\varphi>0.5$ (post-eclipse). Indeed, using a model similar to our Loff and Lbase models, \cite{Herman2022} found that the scale factor peaks at $22\pm12^\circ$, which is westwards of the sub-stellar point. They refrain from interpreting this result in terms of hot-spot offset, but \cite{vanSluijs2022} points out that this is consistent with GCM predictions: the thermal profile tends to get more isothermal inside the hot-spot, leading to a smaller scale factor. In other words, the westward offset in the peak scale factor of iron lines found by \cite{Herman2022}, and confirmed with CO by \cite{vanSluijs2022}, can be interpreted as indirect evidence for an eastward hot-spot offset. Furthermore, based on HARPS-N emission spectroscopy, \cite{Borsa2022} report an evidence of asymmetry in thermal structure between pre- and post-eclipse spectra of UHJ KELT-20b, although it is not statistically significant. They also find that \ion{Fe}{ii} and \ion{Cr}{i} are only observable in post-eclipse, and not in pre-eclipse, despite achieving a similar total S/N across both phase ranges. \cite{Johnson2022} were not able to confirm the detection of \ion{Fe}{ii} and \ion{Cr}{i} with LBT-PEPSI, however, they did not investigate in detail the reason for the discrepancy which could be due to differences in analysis techniques. Therefore, even if additional work is required to understand results on KELT-20b, observations of this UHJ further reinforce the expectation that climate can induce phase-dependent signals in HRS phase curves.

With its symmetric neutral iron scale factor within $10^\circ$ of the substellar point at $2\sigma$, KELT-9b shows qualitatively different behaviour. Taken together, these first results with HRS phase curves mark the opening of a new observational window to study the relation between climate and irradiation, which eventually could help to shed light on questions that have been opened by LRS phase curve population studies. Excitingly, we can expect a population study of HRS phase curves in the coming years thanks to new generation high-resolution spectrographs. 

\section{Conclusions}
\label{sec:conclusions}

In this paper we have demonstrated for the first time that the optical HRS phase curve of UHJ KELT-9b, observed between quadratures ($0.25 <\varphi <0.75$) with 4-metre-class telescopes, has the necessary precision to measure two separate observable quantities: the phase-dependent \ion{Fe}{i} emission line strengths, and the phase-dependent \ion{Fe}{i} emission line Doppler shifts. This result sheds new light on the atmospheric circulation of KELT-9b. Our work has implications for the study of orbital dynamics, and of climate of UHJs, and allows us to draw a pathway towards optimal exploitation of HRS phase curves also in the context of complementary observations and in the JWST era. 

In terms of studies of orbital dynamics of UHJs, we draw the following conclusions:
\begin{enumerate}
    \item Thanks to the favourable planet-to-star flux ratio, and by exploiting state-of-the-art likelihood frameworks, 4-metre class telescopes are able to measure the orbital velocities of UHJs with sub-kilometre-per-second precision.
	\item We detect a radial velocity anomaly of a few $\mathrm{km~s^{-1}}$ in KELT-9b. If attributed to a slightly eccentric orbit, this corresponds to $e = 0.016\pm0.003$ ($5\sigma$). 
	\item This demonstrates that HRS phase curves are potentially sensitive to eccentricities smaller than 0.02 in UHJs, although there are potential degeneracies with atmospheric winds.
	\item To reconcile this result with the upper limit on eccentricity from \textit{TESS} transit timing and duration of KELT-9b by \cite{Wong2020}, we interpret this eccentricity as apparent, and due to atmospheric winds.  
\end{enumerate}

In terms of climate of KELT-9b and UHJs, we draw the following conclusions:
\begin{enumerate}
    \item KELT-9b has symmetric \ion{Fe}{i} line intensity around the sub-stellar point within $10^{\circ}$ ($2\sigma)$. The simplest explanation is the lack of a strong hot spot offset at the probed altitudes.
    \item We detect a radial velocity anomaly in the \ion{Fe}{i} emission lines that we attribute to atmospheric dynamics (winds). By accounting for planet rotation and a day-to-night wind flow blowing at a few kilometre-per-second, we can reproduce the observed radial velocity anomaly.
    \item Taken altogether, these observations favour predictions of GCMs with strong atmospheric drag for KELT-9b, confirming that the climate of UHJs is likely different from that of their cooler counterparts.
\end{enumerate}

This study demonstrates the potential role of HRS phase curves in the upcoming JWST era, and complementarity with other techniques:
\begin{enumerate}
    \item HRS phase curves provide unique information that complements spectrophotometric and photometric phase curves. Even in the JWST era, this technique will remain relevant. In addition, HRS phase curves extend HRS phase-resolved transmission spectroscopy to a broader range of phases, and to the photosphere.
    \item Current HRS optical spectrographs hosted on 8-metre class telescopes (e.g., ESPRESSO@VLT, MAROON-X@Gemini-N) could measure HRS phase curves of UHJs hosted around stars brighter than $V=9$ (at least 10) with a precision comparable to ours. In the short term future, we can expect a statistical survey of HRS optical phase curves, targeting neutral iron but also additional species that are spectrally active in this wavelength range (e.g., \ion{Fe}{ii}, TiO, \ion{Ti}{i}, \ion{Ti}{ii}, \ion{Ca}{ii}). 
\end{enumerate}

Finally, we highlight lessons learned, and areas where additional work is necessary:
\begin{enumerate}
    \item We recommend that future works on HRS phase curves consider both the phase-dependence of line intensities and shapes, and of Doppler-shifts, as they provide complementary information.
    \item Dedicated work is necessary to fully explore the optimal way to combine HRS phase curves of exoplanets with stellar radial velocities and transit photometry to determine the orbital properties of targeted systems. However, our results already show that the combination of these methods opens unprecedented opportunities to simultaneously constrain the orbital and atmospheric properties of exoplanets.
    \item The advantage of combining HRS phase curves, phase-resolved transmission spectroscopy at high spectral resolution and photometric and spectrophotometric phase curves of a planet is that they probe different altitudes and longitudes in UHJ atmospheres. This is also a challenge, as it requires a fully 3D GCM model with all key ingredients that are peculiar to UHJ atmospheres. Such a model is still not available.
    \item Much information in HRS phase curves is found close to quadrature, where Doppler shifts, thermal structure and chemistry deviate the most from pure HRS emission spectroscopy. These are the phases where PCA-based and other classic HRS data reduction techniques most strongly alter and reduce the planet signal. We encourage future studies to design alternative methods to correct or mask telluric and stellar lines that better preserve the planetary signal.
\end{enumerate}







\begin{acknowledgements}
We thank the anonymous referee, whose feedback improved the quality and clarity of the manuscript. Based on observations or data obtained in the framework of GAPS. We made use of the Astropy Project tools and resources \citep{astropy:2013,astropy:2018, astropy:2022}. We acknowledge financial contribution from PRIN INAF 2019. Based on observations made with the Italian Telescopio Nazionale Galileo (TNG) operated by the Fundaci\'on Galileo Galilei (FGG) of the Istituto Nazionale di Astrofisica (INAF) at the Observatorio del Roque de los Muchachos (La Palma, Canary Islands, Spain). J.M.D acknowledges support from the European Research Council (ERC) European Union’s Horizon 2020 research and innovation programme (grant agreement no. 679633; Exo-Atmos) and the research programme VIDI New Frontiers in Exoplanetary Climatology with project number 614.001.601, which is (partly) financed by the Dutch Research Council (NWO).

\end{acknowledgements}




\bibliographystyle{aa}
\bibliography{My_biblio}

\begin{thebibliography}{107}
\expandafter\ifx\csname natexlab\endcsname\relax\def\natexlab#1{#1}\fi

\bibitem[{{Addison} {et~al.}(2021){Addison}, {Knudstrup}, {Wong},
  {H{\'e}brard}, {Dorval}, {Snellen}, {Albrecht}, {Bello-Arufe}, {Almenara},
  {Boisse}, {Bonfils}, {Dalal}, {Demangeon}, {Hoyer}, {Kiefer}, {Santos},
  {Nowak}, {Luque}, {Stangret}, {Palle}, {Tronsgaard}, {Antoci}, {Buchhave},
  {G{\"u}nther}, {Daylan}, {Murgas}, {Parviainen}, {Esparza-Borges}, {Crouzet},
  {Narita}, {Fukui}, {Kawauchi}, {Watanabe}, {Rabus}, {Johnson}, {Otten}, {Jan
  Talens}, {Cabot}, {Fischer}, {Grundahl}, {Fredslund Andersen},
  {Jessen-Hansen}, {Pall{\'e}}, {Shporer}, {Ciardi}, {Clark}, {Wittenmyer},
  {Wright}, {Horner}, {Collins}, {Jensen}, {Kielkopf}, {Schwarz}, {Srdoc},
  {Yilmaz}, {Senavci}, {Diamond}, {Harbeck}, {Komacek}, {Smith}, {Wang},
  {Eastman}, {Stassun}, {Latham}, {Vanderspek}, {Seager}, {Winn}, {Jenkins},
  {Louie}, {Bouma}, {Twicken}, {Levine}, \& {McLean}}]{Addison2021}
{Addison}, B.~C., {Knudstrup}, E., {Wong}, I., {et~al.} 2021, \aj, 162, 292

\bibitem[{{Allart} {et~al.}(2020){Allart}, {Pino}, {Lovis}, {Sousa},
  {Casasayas-Barris}, {Zapatero Osorio}, {Cretignier}, {Palle}, {Pepe},
  {Cristiani}, {Rebolo}, {Santos}, {Borsa}, {Bourrier}, {Demangeon},
  {Ehrenreich}, {Lavie}, {Lendl}, {Lillo-Box}, {Micela}, {Oshagh}, {Sozzetti},
  {Tabernero}, {Adibekyan}, {Allende Prieto}, {Alibert}, {Amate}, {Benz},
  {Bouchy}, {Cabral}, {Dekker}, {D'Odorico}, {Di Marcantonio}, {Dumusque},
  {Figueira}, {Genova Santos}, {Gonz{\'a}lez Hern{\'a}ndez}, {Lo Curto},
  {Manescau}, {Martins}, {M{\'e}gevand}, {Mehner}, {Molaro}, {Nunes},
  {Poretti}, {Riva}, {Su{\'a}rez Mascare{\~n}o}, {Udry}, \&
  {Zerbi}}]{Allart2020}
{Allart}, R., {Pino}, L., {Lovis}, C., {et~al.} 2020, \aap, 644, A155

\bibitem[{{Astropy Collaboration} {et~al.}(2022){Astropy Collaboration},
  {Price-Whelan}, {Lim}, {Earl}, {Starkman}, {Bradley}, {Shupe}, {Patil},
  {Corrales}, {Brasseur}, {N{"o}the}, {Donath}, {Tollerud}, {Morris},
  {Ginsburg}, {Vaher}, {Weaver}, {Tocknell}, {Jamieson}, {van Kerkwijk},
  {Robitaille}, {Merry}, {Bachetti}, {G{"u}nther}, {Aldcroft},
  {Alvarado-Montes}, {Archibald}, {B{'o}di}, {Bapat}, {Barentsen}, {Baz{'a}n},
  {Biswas}, {Boquien}, {Burke}, {Cara}, {Cara}, {Conroy}, {Conseil}, {Craig},
  {Cross}, {Cruz}, {D'Eugenio}, {Dencheva}, {Devillepoix}, {Dietrich},
  {Eigenbrot}, {Erben}, {Ferreira}, {Foreman-Mackey}, {Fox}, {Freij}, {Garg},
  {Geda}, {Glattly}, {Gondhalekar}, {Gordon}, {Grant}, {Greenfield}, {Groener},
  {Guest}, {Gurovich}, {Handberg}, {Hart}, {Hatfield-Dodds}, {Homeier},
  {Hosseinzadeh}, {Jenness}, {Jones}, {Joseph}, {Kalmbach}, {Karamehmetoglu},
  {Ka{l}uszy{'n}ski}, {Kelley}, {Kern}, {Kerzendorf}, {Koch}, {Kulumani},
  {Lee}, {Ly}, {Ma}, {MacBride}, {Maljaars}, {Muna}, {Murphy}, {Norman},
  {O'Steen}, {Oman}, {Pacifici}, {Pascual}, {Pascual-Granado}, {Patil},
  {Perren}, {Pickering}, {Rastogi}, {Roulston}, {Ryan}, {Rykoff}, {Sabater},
  {Sakurikar}, {Salgado}, {Sanghi}, {Saunders}, {Savchenko}, {Schwardt},
  {Seifert-Eckert}, {Shih}, {Jain}, {Shukla}, {Sick}, {Simpson},
  {Singanamalla}, {Singer}, {Singhal}, {Sinha}, {Sip{H{o}}cz}, {Spitler},
  {Stansby}, {Streicher}, {{{S}}umak}, {Swinbank}, {Taranu}, {Tewary},
  {Tremblay}, {Val-Borro}, {Van Kooten}, {Vasovi{'c}}, {Verma}, {de Miranda
  Cardoso}, {Williams}, {Wilson}, {Winkel}, {Wood-Vasey}, {Xue}, {Yoachim},
  {Zhang}, {Zonca}, \& {Astropy Project Contributors}}]{astropy:2022}
{Astropy Collaboration}, {Price-Whelan}, A.~M., {Lim}, P.~L., {et~al.} 2022,
  apj, 935, 167

\bibitem[{{Astropy Collaboration} {et~al.}(2018){Astropy Collaboration},
  {Price-Whelan}, {Sip{\H{o}}cz}, {G{\"u}nther}, {Lim}, {Crawford}, {Conseil},
  {Shupe}, {Craig}, {Dencheva}, {Ginsburg}, {Vand erPlas}, {Bradley},
  {P{\'e}rez-Su{\'a}rez}, {de Val-Borro}, {Aldcroft}, {Cruz}, {Robitaille},
  {Tollerud}, {Ardelean}, {Babej}, {Bach}, {Bachetti}, {Bakanov}, {Bamford},
  {Barentsen}, {Barmby}, {Baumbach}, {Berry}, {Biscani}, {Boquien}, {Bostroem},
  {Bouma}, {Brammer}, {Bray}, {Breytenbach}, {Buddelmeijer}, {Burke},
  {Calderone}, {Cano Rodr{\'\i}guez}, {Cara}, {Cardoso}, {Cheedella}, {Copin},
  {Corrales}, {Crichton}, {D'Avella}, {Deil}, {Depagne}, {Dietrich}, {Donath},
  {Droettboom}, {Earl}, {Erben}, {Fabbro}, {Ferreira}, {Finethy}, {Fox},
  {Garrison}, {Gibbons}, {Goldstein}, {Gommers}, {Greco}, {Greenfield},
  {Groener}, {Grollier}, {Hagen}, {Hirst}, {Homeier}, {Horton}, {Hosseinzadeh},
  {Hu}, {Hunkeler}, {Ivezi{\'c}}, {Jain}, {Jenness}, {Kanarek}, {Kendrew},
  {Kern}, {Kerzendorf}, {Khvalko}, {King}, {Kirkby}, {Kulkarni}, {Kumar},
  {Lee}, {Lenz}, {Littlefair}, {Ma}, {Macleod}, {Mastropietro}, {McCully},
  {Montagnac}, {Morris}, {Mueller}, {Mumford}, {Muna}, {Murphy}, {Nelson},
  {Nguyen}, {Ninan}, {N{\"o}the}, {Ogaz}, {Oh}, {Parejko}, {Parley}, {Pascual},
  {Patil}, {Patil}, {Plunkett}, {Prochaska}, {Rastogi}, {Reddy Janga},
  {Sabater}, {Sakurikar}, {Seifert}, {Sherbert}, {Sherwood-Taylor}, {Shih},
  {Sick}, {Silbiger}, {Singanamalla}, {Singer}, {Sladen}, {Sooley},
  {Sornarajah}, {Streicher}, {Teuben}, {Thomas}, {Tremblay}, {Turner},
  {Terr{\'o}n}, {van Kerkwijk}, {de la Vega}, {Watkins}, {Weaver}, {Whitmore},
  {Woillez}, {Zabalza}, \& {Astropy Contributors}}]{astropy:2018}
{Astropy Collaboration}, {Price-Whelan}, A.~M., {Sip{\H{o}}cz}, B.~M., {et~al.}
  2018, \aj, 156, 123

\bibitem[{{Astropy Collaboration} {et~al.}(2013){Astropy Collaboration},
  {Robitaille}, {Tollerud}, {Greenfield}, {Droettboom}, {Bray}, {Aldcroft},
  {Davis}, {Ginsburg}, {Price-Whelan}, {Kerzendorf}, {Conley}, {Crighton},
  {Barbary}, {Muna}, {Ferguson}, {Grollier}, {Parikh}, {Nair}, {Unther},
  {Deil}, {Woillez}, {Conseil}, {Kramer}, {Turner}, {Singer}, {Fox}, {Weaver},
  {Zabalza}, {Edwards}, {Azalee Bostroem}, {Burke}, {Casey}, {Crawford},
  {Dencheva}, {Ely}, {Jenness}, {Labrie}, {Lim}, {Pierfederici}, {Pontzen},
  {Ptak}, {Refsdal}, {Servillat}, \& {Streicher}}]{astropy:2013}
{Astropy Collaboration}, {Robitaille}, T.~P., {Tollerud}, E.~J., {et~al.} 2013,
  \aap, 558, A33

\bibitem[{{Bard} {et~al.}(1991){Bard}, {Kock}, \& {Kock}}]{BKK}
{Bard}, A., {Kock}, A., \& {Kock}, M. 1991, Astron. and Astrophys., 248, 315,
  (BKK)

\bibitem[{{Bard} \& {Kock}(1994)}]{BK}
{Bard}, A. \& {Kock}, M. 1994, Astron. and Astrophys., 282, 1014, (BK)

\bibitem[{{Barklem} \& {Collet}(2016)}]{Barklem2016}
{Barklem}, P.~S. \& {Collet}, R. 2016, \aap, 588, A96

\bibitem[{{Barklem} {et~al.}(2000){Barklem}, {Piskunov}, \& {O'Mara}}]{BPM}
{Barklem}, P.~S., {Piskunov}, N., \& {O'Mara}, B.~J. 2000, Astron. and
  Astrophys. Suppl. Ser., 142, 467, (BPM)

\bibitem[{{Baxter} {et~al.}(2020){Baxter}, {D{\'e}sert}, {Parmentier}, {Line},
  {Fortney}, {Arcangeli}, {Bean}, {Todorov}, \& {Mansfield}}]{Baxter2020}
{Baxter}, C., {D{\'e}sert}, J.-M., {Parmentier}, V., {et~al.} 2020, \aap, 639,
  A36

\bibitem[{{Bell} \& {Cowan}(2018)}]{Bell2018}
{Bell}, T.~J. \& {Cowan}, N.~B. 2018, \apjl, 857, L20

\bibitem[{{Bello-Arufe} {et~al.}(2022){Bello-Arufe}, {Buchhave},
  {Mendon{\c{c}}a}, {Tronsgaard}, {Heng}, {Hoeijmakers}, \&
  {Mayo}}]{BelloArufe2022}
{Bello-Arufe}, A., {Buchhave}, L.~A., {Mendon{\c{c}}a}, J.~M., {et~al.} 2022,
  arXiv e-prints, arXiv:2203.04969

\bibitem[{{Beltz} {et~al.}(2021){Beltz}, {Rauscher}, {Brogi}, \&
  {Kempton}}]{Beltz2021}
{Beltz}, H., {Rauscher}, E., {Brogi}, M., \& {Kempton}, E. M.~R. 2021, \aj,
  161, 1

\bibitem[{{Beltz} {et~al.}(2022{\natexlab{a}}){Beltz}, {Rauscher}, {M. -R
  Kempton}, {Malsky}, {Ochs}, {Arora}, \& {Savel}}]{Beltz2022}
{Beltz}, H., {Rauscher}, E., {M. -R Kempton}, E., {et~al.} 2022{\natexlab{a}},
  arXiv e-prints, arXiv:2204.12996

\bibitem[{{Beltz} {et~al.}(2022{\natexlab{b}}){Beltz}, {Rauscher}, {Roman}, \&
  {Guilliat}}]{Beltz2022a}
{Beltz}, H., {Rauscher}, E., {Roman}, M.~T., \& {Guilliat}, A.
  2022{\natexlab{b}}, \aj, 163, 35

\bibitem[{{Bonomo} {et~al.}(2017){Bonomo}, {Desidera}, {Benatti}, {Borsa},
  {Crespi}, {Damasso}, {Lanza}, {Sozzetti}, {Lodato}, {Marzari}, {Boccato},
  {Claudi}, {Cosentino}, {Covino}, {Gratton}, {Maggio}, {Micela}, {Molinari},
  {Pagano}, {Piotto}, {Poretti}, {Smareglia}, {Affer}, {Biazzo}, {Bignamini},
  {Esposito}, {Giacobbe}, {H{\'e}brard}, {Malavolta}, {Maldonado}, {Mancini},
  {Martinez Fiorenzano}, {Masiero}, {Nascimbeni}, {Pedani}, {Rainer}, \&
  {Scandariato}}]{Bonomo2017}
{Bonomo}, A.~S., {Desidera}, S., {Benatti}, S., {et~al.} 2017, \aap, 602, A107

\bibitem[{{Borsa} {et~al.}(2022){Borsa}, {Giacobbe}, {Bonomo}, {Brogi}, {Pino},
  {Fossati}, {Lanza}, {Nascimbeni}, {Sozzetti}, {Amadori}, {Benatti}, {Biazzo},
  {Bignamini}, {Boschin}, {Claudi}, {Cosentino}, {Covino}, {Desidera},
  {Fiorenzano}, {Guilluy}, {Harutyunyan}, {Maggio}, {Maldonado}, {Mancini},
  {Micela}, {Molinari}, {Molinaro}, {Pagano}, {Pedani}, {Piotto}, {Poretti},
  {Rainer}, {Scandariato}, \& {Stoev}}]{Borsa2022}
{Borsa}, F., {Giacobbe}, P., {Bonomo}, A.~S., {et~al.} 2022, arXiv e-prints,
  arXiv:2204.04948

\bibitem[{{Borsa} {et~al.}(2019){Borsa}, {Rainer}, {Bonomo}, {Barbato},
  {Fossati}, {Malavolta}, {Nascimbeni}, {Lanza}, {Esposito}, {Affer},
  {Andreuzzi}, {Benatti}, {Biazzo}, {Bignamini}, {Brogi}, {Carleo}, {Claudi},
  {Cosentino}, {Covino}, {Damasso}, {Desidera}, {Garrido Rubio}, {Giacobbe},
  {Gonz{\'a}lez-{\'A}lvarez}, {Harutyunyan}, {Knapic}, {Leto}, {Ligi},
  {Maggio}, {Maldonado}, {Mancini}, {Fiorenzano}, {Masiero}, {Micela},
  {Molinari}, {Pagano}, {Pedani}, {Piotto}, {Pino}, {Poretti}, {Scandariato},
  {Smareglia}, \& {Sozzetti}}]{Borsa2019}
{Borsa}, F., {Rainer}, M., {Bonomo}, A.~S., {et~al.} 2019, arXiv e-prints,
  arXiv:1907.10078

\bibitem[{{Brogi} \& {Line}(2019)}]{Brogi2019}
{Brogi}, M. \& {Line}, M.~R. 2019, \aj, 157, 114

\bibitem[{{Brogi} {et~al.}(2013){Brogi}, {Snellen}, {de Kok}, {Albrecht},
  {Birkby}, \& {de Mooij}}]{Brogi2013}
{Brogi}, M., {Snellen}, I.~A.~G., {de Kok}, R.~J., {et~al.} 2013, \apj, 767, 27

\bibitem[{{Charbonneau} {et~al.}(2005){Charbonneau}, {Allen}, {Megeath},
  {Torres}, {Alonso}, {Brown}, {Gilliland}, {Latham}, {Mandushev}, {O'Donovan},
  \& {Sozzetti}}]{Charbonneau2005}
{Charbonneau}, D., {Allen}, L.~E., {Megeath}, S.~T., {et~al.} 2005, \apj, 626,
  523

\bibitem[{{Claudi} {et~al.}(2017){Claudi}, {Benatti}, {Carleo}, {Ghedina},
  {Guerra}, {Micela}, {Molinari}, {Oliva}, {Rainer}, {Tozzi}, {Baffa},
  {Baruffolo}, {Buchschacher}, {Cecconi}, {Cosentino}, {Fantinel}, {Fini},
  {Ghinassi}, {Giani}, {Gonzalez}, {Gonzalez}, {Gratton}, {Harutyunyan},
  {Hernandez}, {Lodi}, {Malavolta}, {Maldonado}, {Origlia}, {Sanna}, {Sanjuan},
  {Scuderi}, {Seemann}, {Sozzetti}, {Perez Ventura}, {Hernandez Diaz}, {Galli},
  {Gonzalez}, {Riverol}, \& {Riverol}}]{Claudi2017}
{Claudi}, R., {Benatti}, S., {Carleo}, I., {et~al.} 2017, European Physical
  Journal Plus, 132, 364

\bibitem[{{Collier Cameron} {et~al.}(1999){Collier Cameron}, {Horne}, {Penny},
  \& {James}}]{CollierCameron1999}
{Collier Cameron}, A., {Horne}, K., {Penny}, A., \& {James}, D. 1999, \nat,
  402, 751

\bibitem[{{Collier Cameron} {et~al.}(2002){Collier Cameron}, {Horne}, {Penny},
  \& {Leigh}}]{CollierCameron2002}
{Collier Cameron}, A., {Horne}, K., {Penny}, A., \& {Leigh}, C. 2002, \mnras,
  330, 187

\bibitem[{{Cont} {et~al.}(2021){Cont}, {Yan}, {Reiners}, {Casasayas-Barris},
  {Molli{\`e}re}, {Pall{\'e}}, {Henning}, {Nortmann}, {Stangret}, {Czesla},
  {L{\'o}pez-Puertas}, {S{\'a}nchez-L{\'o}pez}, {Rodler}, {Ribas},
  {Quirrenbach}, {Caballero}, {Amado}, {Carone}, {Khaimova}, {Kreidberg},
  {Molaverdikhani}, {Montes}, {Morello}, {Nagel}, {Oshagh}, \&
  {Zechmeister}}]{Cont2021}
{Cont}, D., {Yan}, F., {Reiners}, A., {et~al.} 2021, \aap, 651, A33

\bibitem[{{Cosentino} {et~al.}(2012){Cosentino}, {Lovis}, {Pepe}, {Collier
  Cameron}, {Latham}, {Molinari}, {Udry}, {Bezawada}, {Black}, {Born},
  {Buchschacher}, {Charbonneau}, {Figueira}, {Fleury}, {Galli}, {Gallie},
  {Gao}, {Ghedina}, {Gonzalez}, {Gonzalez}, {Guerra}, {Henry}, {Horne},
  {Hughes}, {Kelly}, {Lodi}, {Lunney}, {Maire}, {Mayor}, {Micela}, {Ordway},
  {Peacock}, {Phillips}, {Piotto}, {Pollacco}, {Queloz}, {Rice}, {Riverol},
  {Riverol}, {San Juan}, {Sasselov}, {Segransan}, {Sozzetti}, {Sosnowska},
  {Stobie}, {Szentgyorgyi}, {Vick}, \& {Weber}}]{Cosentino2012}
{Cosentino}, R., {Lovis}, C., {Pepe}, F., {et~al.} 2012, in Society of
  Photo-Optical Instrumentation Engineers (SPIE) Conference Series, Vol. 8446,
  Ground-based and Airborne Instrumentation for Astronomy IV, ed. I.~S.
  {McLean}, S.~K. {Ramsay}, \& H.~{Takami}, 84461V

\bibitem[{{Deeg} \& {Belmonte}(2018)}]{Deeg2018}
{Deeg}, H.~J. \& {Belmonte}, J.~A. 2018, {Handbook of Exoplanets}

\bibitem[{{Dobbs-Dixon} \& {Blecic}(2022)}]{DobbsDixon2022}
{Dobbs-Dixon}, I. \& {Blecic}, J. 2022, \apj, 929, 46

\bibitem[{{Eastman} {et~al.}(2013){Eastman}, {Gaudi}, \& {Agol}}]{Eastman2013}
{Eastman}, J., {Gaudi}, B.~S., \& {Agol}, E. 2013, \pasp, 125, 83

\bibitem[{{Eastman} {et~al.}(2010){Eastman}, {Siverd}, \&
  {Gaudi}}]{Eastman2010}
{Eastman}, J., {Siverd}, R., \& {Gaudi}, B.~S. 2010, \pasp, 122, 935

\bibitem[{{Ehrenreich} {et~al.}(2020){Ehrenreich}, {Lovis}, {Allart}, {Zapatero
  Osorio}, {Pepe}, {Cristiani}, {Rebolo}, {Santos}, {Borsa}, {Demangeon},
  {Dumusque}, {Gonz{\'a}lez Hern{\'a}ndez}, {Casasayas-Barris},
  {S{\'e}gransan}, {Sousa}, {Abreu}, {Adibekyan}, {Affolter}, {Allende Prieto},
  {Alibert}, {Aliverti}, {Alves}, {Amate}, {Avila}, {Baldini}, {Bandy}, {Benz},
  {Bianco}, {Bolmont}, {Bouchy}, {Bourrier}, {Broeg}, {Cabral}, {Calderone},
  {Pall{\'e}}, {Cegla}, {Cirami}, {Coelho}, {Conconi}, {Coretti}, {Cumani},
  {Cupani}, {Dekker}, {Delabre}, {Deiries}, {D'Odorico}, {Di Marcantonio},
  {Figueira}, {Fragoso}, {Genolet}, {Genoni}, {G{\'e}nova Santos}, {Hara},
  {Hughes}, {Iwert}, {Kerber}, {Knudstrup}, {Landoni}, {Lavie}, {Lizon},
  {Lendl}, {Lo Curto}, {Maire}, {Manescau}, {Martins}, {M{\'e}gevand},
  {Mehner}, {Micela}, {Modigliani}, {Molaro}, {Monteiro}, {Monteiro},
  {Moschetti}, {M{\"u}ller}, {Nunes}, {Oggioni}, {Oliveira}, {Pariani},
  {Pasquini}, {Poretti}, {Rasilla}, {Redaelli}, {Riva}, {Santana Tschudi},
  {Santin}, {Santos}, {Segovia Milla}, {Seidel}, {Sosnowska}, {Sozzetti},
  {Span{\`o}}, {Su{\'a}rez Mascare{\~n}o}, {Tabernero}, {Tenegi}, {Udry},
  {Zanutta}, \& {Zerbi}}]{Ehrenreich2020}
{Ehrenreich}, D., {Lovis}, C., {Allart}, R., {et~al.} 2020, \nat, 580, 597

\bibitem[{{Feng} {et~al.}(2016){Feng}, {Line}, {Fortney}, {Stevenson}, {Bean},
  {Kreidberg}, \& {Parmentier}}]{Feng2016}
{Feng}, Y.~K., {Line}, M.~R., {Fortney}, J.~J., {et~al.} 2016, \apj, 829, 52

\bibitem[{{Fortney} {et~al.}(2021){Fortney}, {Dawson}, \&
  {Komacek}}]{Fortney2021}
{Fortney}, J.~J., {Dawson}, R.~I., \& {Komacek}, T.~D. 2021, Journal of
  Geophysical Research (Planets), 126, e06629

\bibitem[{{Fossati} {et~al.}(2020){Fossati}, {Shulyak}, {Sreejith}, {Koskinen},
  {Young}, {Cubillos}, {Lara}, {France}, {Rengel}, {Cauley}, {Turner},
  {Wyttenbach}, \& {Yan}}]{Fossati2020}
{Fossati}, L., {Shulyak}, D., {Sreejith}, A.~G., {et~al.} 2020, \aap, 643, A131

\bibitem[{{Fossati} {et~al.}(2021){Fossati}, {Young}, {Shulyak}, {Koskinen},
  {Huang}, {Cubillos}, {France}, \& {Sreejith}}]{Fossati2021}
{Fossati}, L., {Young}, M.~E., {Shulyak}, D., {et~al.} 2021, \aap, 653, A52

\bibitem[{{Fuhr} {et~al.}(1988){Fuhr}, {Martin}, \& {Wiese}}]{FMW}
{Fuhr}, J.~R., {Martin}, G.~A., \& {Wiese}, W.~L. 1988, Journal of Physical and
  Chemical Reference Data, Volume 17, Suppl.~4.~New York: American Institute of
  Physics (AIP) and American Chemical Society, 1988, 17, (FMW)

\bibitem[{{Gaudi} {et~al.}(2017){Gaudi}, {Stassun}, {Collins}, {Beatty},
  {Zhou}, {Latham}, {Bieryla}, {Eastman}, {Siverd}, {Crepp}, {Gonzales},
  {Stevens}, {Buchhave}, {Pepper}, {Johnson}, {Colon}, {Jensen}, {Rodriguez},
  {Bozza}, {Novati}, {D'Ago}, {Dumont}, {Ellis}, {Gaillard}, {Jang-Condell},
  {Kasper}, {Fukui}, {Gregorio}, {Ito}, {Kielkopf}, {Manner}, {Matt}, {Narita},
  {Oberst}, {Reed}, {Scarpetta}, {Stephens}, {Yeigh}, {Zambelli}, {Fulton},
  {Howard}, {James}, {Penny}, {Bayliss}, {Curtis}, {Depoy}, {Esquerdo},
  {Gould}, {Joner}, {Kuhn}, {Labadie-Bartz}, {Lund}, {Marshall}, {McLeod},
  {Pogge}, {Relles}, {Stockdale}, {Tan}, {Trueblood}, \&
  {Trueblood}}]{Gaudi2017}
{Gaudi}, B.~S., {Stassun}, K.~G., {Collins}, K.~A., {et~al.} 2017, \nat, 546,
  514

\bibitem[{{Giacobbe} {et~al.}(2021){Giacobbe}, {Brogi}, {Gandhi}, {Cubillos},
  {Bonomo}, {Sozzetti}, {Fossati}, {Guilluy}, {Carleo}, {Rainer},
  {Harutyunyan}, {Borsa}, {Pino}, {Nascimbeni}, {Benatti}, {Biazzo},
  {Bignamini}, {Chubb}, {Claudi}, {Cosentino}, {Covino}, {Damasso}, {Desidera},
  {Fiorenzano}, {Ghedina}, {Lanza}, {Leto}, {Maggio}, {Malavolta}, {Maldonado},
  {Micela}, {Molinari}, {Pagano}, {Pedani}, {Piotto}, {Poretti}, {Scandariato},
  {Yurchenko}, {Fantinel}, {Galli}, {Lodi}, {Sanna}, \& {Tozzi}}]{Giacobbe2021}
{Giacobbe}, P., {Brogi}, M., {Gandhi}, S., {et~al.} 2021, \nat, 592, 205

\bibitem[{{Gibson} {et~al.}(2020){Gibson}, {Merritt}, {Nugroho}, {Cubillos},
  {de Mooij}, {Mikal-Evans}, {Fossati}, {Lothringer}, {Nikolov}, {Sing},
  {Spake}, {Watson}, \& {Wilson}}]{Gibson2020}
{Gibson}, N.~P., {Merritt}, S., {Nugroho}, S.~K., {et~al.} 2020, \mnras, 220

\bibitem[{{Gibson} {et~al.}(2022){Gibson}, {Nugroho}, {Lothringer}, {Maguire},
  \& {Sing}}]{Gibson2022}
{Gibson}, N.~P., {Nugroho}, S.~K., {Lothringer}, J., {Maguire}, C., \& {Sing},
  D.~K. 2022, \mnras, 512, 4618

\bibitem[{{Grimm} \& {Heng}(2015)}]{Grimm2015}
{Grimm}, S.~L. \& {Heng}, K. 2015, \apj, 808, 182

\bibitem[{{Guilluy} {et~al.}(2020){Guilluy}, {Andretta}, {Borsa}, {Giacobbe},
  {Sozzetti}, {Covino}, {Bourrier}, {Fossati}, {Bonomo}, {Esposito},
  {Giampapa}, {Harutyunyan}, {Rainer}, {Brogi}, {Bruno}, {Claudi}, {Frustagli},
  {Lanza}, {Mancini}, {Pino}, {Poretti}, {Scandariato}, {Affer}, {Baffa},
  {Baruffolo}, {Benatti}, {Biazzo}, {Bignamini}, {Boschin}, {Carleo},
  {Cecconi}, {Cosentino}, {Damasso}, {Desidera}, {Falcini}, {Martinez
  Fiorenzano}, {Ghedina}, {Gonz{\'a}lez-{\'A}lvarez}, {Guerra}, {Hernandez},
  {Leto}, {Maggio}, {Malavolta}, {Maldonado}, {Micela}, {Molinari},
  {Nascimbeni}, {Pagano}, {Pedani}, {Piotto}, \& {Reiners}}]{Guilluy2020}
{Guilluy}, G., {Andretta}, V., {Borsa}, F., {et~al.} 2020, \aap, 639, A49

\bibitem[{{Guilluy} {et~al.}(2022){Guilluy}, {Giacobbe}, {Carleo}, {Cubillos},
  {Sozzetti}, {Bonomo}, {Brogi}, {Gandhi}, {Fossati}, {Nascimbeni}, {Turrini},
  {Schisano}, {Borsa}, {Lanza}, {Mancini}, {Maggio}, {Malavolta}, {Micela},
  {Pino}, {Rainer}, {Bignamini}, {Claudi}, {Cosentino}, {Covino}, {Desidera},
  {Fiorenzano}, {Harutyunyan}, {Lorenzi}, {Knapic}, {Molinari}, {Pacetti},
  {Pagano}, {Pedani}, {Piotto}, \& {Poretti}}]{Guilluy2022}
{Guilluy}, G., {Giacobbe}, P., {Carleo}, I., {et~al.} 2022, arXiv e-prints,
  arXiv:2207.09760

\bibitem[{{Helling} {et~al.}(2019){Helling}, {Gourbin}, {Woitke}, \&
  {Parmentier}}]{Helling2019_w18}
{Helling}, C., {Gourbin}, P., {Woitke}, P., \& {Parmentier}, V. 2019, \aap,
  626, A133

\bibitem[{{Herman} {et~al.}(2022){Herman}, {de Mooij}, {Nugroho}, {Gibson}, \&
  {Jayawardhana}}]{Herman2022}
{Herman}, M.~K., {de Mooij}, E. J.~W., {Nugroho}, S.~K., {Gibson}, N.~P., \&
  {Jayawardhana}, R. 2022, \aj, 163, 248

\bibitem[{{Hoeijmakers} {et~al.}(2018){Hoeijmakers}, {Ehrenreich}, {Heng},
  {Kitzmann}, {Grimm}, {Allart}, {Deitrick}, {Wyttenbach}, {Oreshenko}, {Pino},
  {Rimmer}, {Molinari}, \& {Di Fabrizio}}]{Hoeijmakers2018_k9}
{Hoeijmakers}, H.~J., {Ehrenreich}, D., {Heng}, K., {et~al.} 2018, \nat, 560,
  453

\bibitem[{{Hoeijmakers} {et~al.}(2019){Hoeijmakers}, {Ehrenreich}, {Kitzmann},
  {Allart}, {Grimm}, {Seidel}, {Wyttenbach}, {Pino}, {Nielsen}, {Fisher},
  {Rimmer}, {Bourrier}, {Cegla}, {Lavie}, {Lovis}, {Patzer}, {Stock}, {Pepe},
  \& {Heng}}]{Hoeijmakers2019}
{Hoeijmakers}, H.~J., {Ehrenreich}, D., {Kitzmann}, D., {et~al.} 2019, \aap,
  627, A165

\bibitem[{{Huang} {et~al.}(2017){Huang}, {Arras}, {Christie}, \&
  {Li}}]{Huang2017}
{Huang}, C., {Arras}, P., {Christie}, D., \& {Li}, Z.-Y. 2017, \apj, 851, 150

\bibitem[{Hyndman(1996)}]{Hyndman1996}
Hyndman, R.~J. 1996, The American Statistician, 50, 120

\bibitem[{{Ivshina} \& {Winn}(2022)}]{Ivshina2022}
{Ivshina}, E.~S. \& {Winn}, J.~N. 2022, \apjs, 259, 62

\bibitem[{{John}(1988)}]{John1988}
{John}, T.~L. 1988, \aap, 193, 189

\bibitem[{{Johnson} {et~al.}(2022){Johnson}, {Wang}, {Pai Asnodkar}, {Bonomo},
  {Gaudi}, {Henning}, {Ilyin}, {Keles}, {Malavolta}, {Mallonn},
  {Molaverdikhani}, {Nascimbeni}, {Patience}, {Poppenhaeger}, {Scandariato},
  {Schlawin}, {Shkolnik}, {Sicilia}, {Sozzetti}, {Strassmeier}, {Veillet}, \&
  {Yan}}]{Johnson2022}
{Johnson}, M.~C., {Wang}, J., {Pai Asnodkar}, A., {et~al.} 2022, arXiv
  e-prints, arXiv:2205.12162

\bibitem[{{Jones} {et~al.}(2022){Jones}, {Morris}, {Demory}, {Heng}, {Hooton},
  {Billot}, {Ehrenreich}, {Hoyer}, {Simon}, {Lendl}, {Demangeon}, {Sousa},
  {Bonfanti}, {Wilson}, {Salmon}, {Csizmadia}, {Parviainen}, {Bruno},
  {Alibert}, {Alonso}, {Anglada}, {B{\'a}rczy}, {Barrado y Navascues},
  {Barros}, {Baumjohann}, {Beck}, {Beck}, {Benz}, {Bonfils}, {Brandeker},
  {Broeg}, {Cabrera}, {Charnoz}, {Collier Cameron}, {Davies}, {Deleuil},
  {Deline}, {Delrez}, {Erikson}, {Fortier}, {Fossati}, {Fridlund}, {Gandolfi},
  {Gillon}, {G{\"u}del}, {Isaak}, {Kiss}, {Laskar}, {Lecavelier des Etangs},
  {Lovis}, {Magrin}, {Maxted}, {Nascimbeni}, {Olofsson}, {Ottensamer},
  {Pagano}, {Pall{\'e}}, {Peter}, {Piotto}, {Pollacco}, {Queloz}, {Ragazzoni},
  {Rando}, {Ratti}, {Rauer}, {Reimers}, {Ribas}, {Santos}, {Scandariato},
  {S{\'e}gransan}, {Smith}, {Steller}, {Szab{\'o}}, {Thomas}, {Udry}, {Van
  Grootel}, {Walter}, {Walton}, \& {Jungo}}]{Jones2022}
{Jones}, K.~D., {Morris}, B.~M., {Demory}, B.~O., {et~al.} 2022, arXiv
  e-prints, arXiv:2208.04818

\bibitem[{{Kasper} {et~al.}(2021){Kasper}, {Bean}, {Line}, {Seifahrt},
  {St{\"u}rmer}, {Pino}, {D{\'e}sert}, \& {Brogi}}]{Kasper2021}
{Kasper}, D., {Bean}, J.~L., {Line}, M.~R., {et~al.} 2021, \apjl, 921, L18

\bibitem[{{Kawahara}(2012)}]{Kawahara2012}
{Kawahara}, H. 2012, \apjl, 760, L13

\bibitem[{{Kitzmann} {et~al.}(2018){Kitzmann}, {Heng}, {Rimmer}, {Hoeijmakers},
  {Tsai}, {Malik}, {Lendl}, {Deitrick}, \& {Demory}}]{Kitzmann2018}
{Kitzmann}, D., {Heng}, K., {Rimmer}, P.~B., {et~al.} 2018, \apj, 863, 183

\bibitem[{{Komacek} {et~al.}(2017){Komacek}, {Showman}, \& {Tan}}]{Komacek2017}
{Komacek}, T.~D., {Showman}, A.~P., \& {Tan}, X. 2017, \apj, 835, 198

\bibitem[{{Kreidberg} {et~al.}(2018){Kreidberg}, {Line}, {Parmentier},
  {Stevenson}, {Louden}, {Bonnefoy}, {Faherty}, {Henry}, {Williamson},
  {Stassun}, {Beatty}, {Bean}, {Fortney}, {Showman}, {D{\'e}sert}, \&
  {Arcangeli}}]{Kreidberg2018}
{Kreidberg}, L., {Line}, M.~R., {Parmentier}, V., {et~al.} 2018, \aj, 156, 17

\bibitem[{{Kupka} {et~al.}(1999){Kupka}, {Piskunov}, {Ryabchikova}, {Stempels},
  \& {Weiss}}]{Kupka1999}
{Kupka}, F., {Piskunov}, N., {Ryabchikova}, T.~A., {Stempels}, H.~C., \&
  {Weiss}, W.~W. 1999, \aaps, 138, 119

\bibitem[{{Kupka} {et~al.}(2000){Kupka}, {Ryabchikova}, {Piskunov}, {Stempels},
  \& {Weiss}}]{Kupka2000}
{Kupka}, F.~G., {Ryabchikova}, T.~A., {Piskunov}, N.~E., {Stempels}, H.~C., \&
  {Weiss}, W.~W. 2000, Baltic Astronomy, 9, 590

\bibitem[{{Kurucz}(2014)}]{K14}
{Kurucz}, R.~L. 2014, Robert L. Kurucz on-line database of observed and
  predicted atomic transitions

\bibitem[{{Line} {et~al.}(2021){Line}, {Brogi}, {Bean}, {Gandhi}, {Zalesky},
  {Parmentier}, {Smith}, {Mace}, {Mansfield}, {Kempton}, {Fortney}, {Shkolnik},
  {Patience}, {Rauscher}, {D{\'e}sert}, \& {Wardenier}}]{Line2021}
{Line}, M.~R., {Brogi}, M., {Bean}, J.~L., {et~al.} 2021, \nat, 598, 580

\bibitem[{{Lothringer} {et~al.}(2018){Lothringer}, {Barman}, \&
  {Koskinen}}]{Lothringer2018}
{Lothringer}, J.~D., {Barman}, T., \& {Koskinen}, T. 2018, \apj, 866, 27

\bibitem[{{Lothringer} {et~al.}(2021){Lothringer}, {Rustamkulov}, {Sing},
  {Gibson}, {Wilson}, \& {Schlaufman}}]{Lothringer2021}
{Lothringer}, J.~D., {Rustamkulov}, Z., {Sing}, D.~K., {et~al.} 2021, \apj,
  914, 12

\bibitem[{{Mansfield} {et~al.}(2020){Mansfield}, {Bean}, {Stevenson},
  {Komacek}, {Bell}, {Tan}, {Malik}, {Beatty}, {Wong}, {Cowan}, {Dang},
  {D{\'e}sert}, {Fortney}, {Gaudi}, {Keating}, {Kempton}, {Kreidberg}, {Line},
  {Parmentier}, {Stassun}, {Swain}, \& {Zellem}}]{Mansfield2020}
{Mansfield}, M., {Bean}, J.~L., {Stevenson}, K.~B., {et~al.} 2020, \apjl, 888,
  L15

\bibitem[{{Mansfield} {et~al.}(2021){Mansfield}, {Line}, {Bean}, {Fortney},
  {Parmentier}, {Wiser}, {Kempton}, {Gharib-Nezhad}, {Sing},
  {L{\'o}pez-Morales}, {Baxter}, {D{\'e}sert}, {Swain}, \&
  {Roudier}}]{Mansfield2021}
{Mansfield}, M., {Line}, M.~R., {Bean}, J.~L., {et~al.} 2021, Nature Astronomy,
  5, 1224

\bibitem[{{Mikal-Evans} {et~al.}(2022){Mikal-Evans}, {Sing}, {Barstow},
  {Kataria}, {Goyal}, {Lewis}, {Taylor}, {Mayne}, {Daylan}, {Wakeford},
  {Marley}, \& {Spake}}]{MikalEvans2022}
{Mikal-Evans}, T., {Sing}, D.~K., {Barstow}, J.~K., {et~al.} 2022, Nature
  Astronomy, 6, 471

\bibitem[{{Nugroho} {et~al.}(2020){Nugroho}, {Gibson}, {de Mooij}, {Herman},
  {Watson}, {Kawahara}, \& {Merritt}}]{Nugroho2020}
{Nugroho}, S.~K., {Gibson}, N.~P., {de Mooij}, E. J.~W., {et~al.} 2020, \apjl,
  898, L31

\bibitem[{{Nymeyer} {et~al.}(2011){Nymeyer}, {Harrington}, {Hardy},
  {Stevenson}, {Campo}, {Madhusudhan}, {Collier-Cameron}, {Loredo}, {Blecic},
  {Bowman}, {Britt}, {Cubillos}, {Hellier}, {Gillon}, {Maxted}, {Hebb},
  {Wheatley}, {Pollacco}, \& {Anderson}}]{Nymeyer2011}
{Nymeyer}, S., {Harrington}, J., {Hardy}, R.~A., {et~al.} 2011, \apj, 742, 35

\bibitem[{{O'Brian} {et~al.}(1991){O'Brian}, {Wickliffe}, {Lawler}, {Whaling},
  \& {Brault}}]{BWL}
{O'Brian}, T.~R., {Wickliffe}, M.~E., {Lawler}, J.~E., {Whaling}, W., \&
  {Brault}, J.~W. 1991, Journal of the Optical Society of America B Optical
  Physics, 8, 1185, (BWL)

\bibitem[{{Pai Asnodkar} {et~al.}(2022){Pai Asnodkar}, {Wang}, {Eastman},
  {Cauley}, {Gaudi}, {Ilyin}, \& {Strassmeier}}]{PaiAsnodkar2022}
{Pai Asnodkar}, A., {Wang}, J., {Eastman}, J.~D., {et~al.} 2022, \aj, 163, 155

\bibitem[{{Parmentier} \& {Crossfield}(2018)}]{ParmentierCrossfield2018}
{Parmentier}, V. \& {Crossfield}, I. J.~M. 2018, in Handbook of Exoplanets, ed.
  H.~J. {Deeg} \& J.~A. {Belmonte}, 116

\bibitem[{{Parmentier} {et~al.}(2018){Parmentier}, {Line}, {Bean}, {Mansfield},
  {Kreidberg}, {Lupu}, {Visscher}, {D{\'e}sert}, {Fortney}, {Deleuil},
  {Arcangeli}, {Showman}, \& {Marley}}]{Parmentier2018}
{Parmentier}, V., {Line}, M.~R., {Bean}, J.~L., {et~al.} 2018, \aap, 617, A110

\bibitem[{{Pelletier} {et~al.}(2021){Pelletier}, {Benneke}, {Darveau-Bernier},
  {Boucher}, {Cook}, {Piaulet}, {Coulombe}, {Artigau}, {Lafreni{\`e}re},
  {Delisle}, {Allart}, {Doyon}, {Donati}, {Fouqu{\'e}}, {Moutou}, {Cadieux},
  {Delfosse}, {H{\'e}brard}, {Martins}, {Martioli}, \&
  {Vandal}}]{Pelletier2021}
{Pelletier}, S., {Benneke}, B., {Darveau-Bernier}, A., {et~al.} 2021, \aj, 162,
  73

\bibitem[{{Perna} {et~al.}(2010){Perna}, {Menou}, \& {Rauscher}}]{Perna2010}
{Perna}, R., {Menou}, K., \& {Rauscher}, E. 2010, \apj, 719, 1421

\bibitem[{{Pino} {et~al.}(2020){Pino}, {D{\'e}sert}, {Brogi}, {Malavolta},
  {Wyttenbach}, {Line}, {Hoeijmakers}, {Fossati}, {Bonomo}, {Nascimbeni},
  {Panwar}, {Affer}, {Benatti}, {Biazzo}, {Bignamini}, {Borsa}, {Carleo},
  {Claudi}, {Cosentino}, {Covino}, {Damasso}, {Desidera}, {Giacobbe},
  {Harutyunyan}, {Lanza}, {Leto}, {Maggio}, {Maldonado}, {Mancini}, {Micela},
  {Molinari}, {Pagano}, {Piotto}, {Poretti}, {Rainer}, {Scandariato},
  {Sozzetti}, {Allart}, {Borsato}, {Bruno}, {Di Fabrizio}, {Ehrenreich},
  {Fiorenzano}, {Frustagli}, {Lavie}, {Lovis}, {Magazz{\`u}}, {Nardiello},
  {Pedani}, \& {Smareglia}}]{Pino2020}
{Pino}, L., {D{\'e}sert}, J.-M., {Brogi}, M., {et~al.} 2020, \apjl, 894, L27

\bibitem[{{Piskunov} {et~al.}(1995){Piskunov}, {Kupka}, {Ryabchikova}, {Weiss},
  \& {Jeffery}}]{Piskunov1995}
{Piskunov}, N.~E., {Kupka}, F., {Ryabchikova}, T.~A., {Weiss}, W.~W., \&
  {Jeffery}, C.~S. 1995, \aaps, 112, 525

\bibitem[{{Poretti} {et~al.}(2016){Poretti}, {Boccato}, {Claudi}, {Cosentino},
  {Covino}, {Desidera}, {Gratton}, {Lanza}, {Maggio}, {Micela}, {Molinari},
  {Pagano}, {Piotto}, {Smareglia}, {Sozzetti}, \& {GAPS
  Collaboration}}]{Poretti2016}
{Poretti}, E., {Boccato}, C., {Claudi}, R., {et~al.} 2016, \memsai, 87, 141

\bibitem[{{Quirrenbach} {et~al.}(2016){Quirrenbach}, {Amado}, {Caballero},
  {Mundt}, {Reiners}, {Ribas}, {Seifert}, {Abril}, {Aceituno},
  {Alonso-Floriano}, {Anwand-Heerwart}, {Azzaro}, {Bauer}, {Barrado},
  {Becerril}, {Bejar}, {Benitez}, {Berdinas}, {Brinkm{\"o}ller}, {Cardenas},
  {Casal}, {Claret}, {Colom{\'e}}, {Cortes-Contreras}, {Czesla}, {Doellinger},
  {Dreizler}, {Feiz}, {Fernandez}, {Ferro}, {Fuhrmeister}, {Galadi},
  {Gallardo}, {G{\'a}lvez-Ortiz}, {Garcia-Piquer}, {Garrido}, {Gesa},
  {G{\'o}mez Galera}, {Gonz{\'a}lez Hern{\'a}ndez}, {Gonzalez Peinado},
  {Gr{\"o}zinger}, {Gu{\`a}rdia}, {Guenther}, {de Guindos}, {Hagen}, {Hatzes},
  {Hauschildt}, {Helmling}, {Henning}, {Hermann}, {Hern{\'a}ndez Arabi},
  {Hern{\'a}ndez Casta{\~n}o}, {Hern{\'a}ndez Hernando}, {Herrero}, {Huber},
  {Huber}, {Huke}, {Jeffers}, {de Juan}, {Kaminski}, {Kehr}, {Kim}, {Klein},
  {Kl{\"u}ter}, {K{\"u}rster}, {Lafarga}, {Lara}, {Lamert}, {Laun},
  {Launhardt}, {Lemke}, {Lenzen}, {Llamas}, {Lopez del Fresno},
  {L{\'o}pez-Puertas}, {L{\'o}pez-Santiago}, {Lopez Salas}, {Magan
  Madinabeitia}, {Mall}, {Mandel}, {Mancini}, {Marin Molina}, {Maroto
  Fern{\'a}ndez}, {Mart{\'\i}n}, {Mart{\'\i}n-Ruiz}, {Marvin}, {Mathar},
  {Mirabet}, {Montes}, {Morales}, {Morales Mu{\~n}oz}, {Nagel}, {Naranjo},
  {Nowak}, {Palle}, {Panduro}, {Passegger}, {Pavlov}, {Pedraz}, {Perez},
  {P{\'e}rez-Medialdea}, {Perger}, {Pluto}, {Ram{\'o}n}, {Rebolo}, {Redondo},
  {Reffert}, {Reinhart}, {Rhode}, {Rix}, {Rodler}, {Rodr{\'\i}guez},
  {Rodr{\'\i}guez L{\'o}pez}, {Rohloff}, {Rosich}, {Sanchez Carrasco},
  {Sanz-Forcada}, {Sarkis}, {Sarmiento}, {Sch{\"a}fer}, {Schiller}, {Schmidt},
  {Schmitt}, {Sch{\"o}fer}, {Schweitzer}, {Shulyak}, {Solano}, {Stahl},
  {Storz}, {Tabernero}, {Tala}, {Tal-Or}, {Ulbrich}, {Veredas}, {Vico Linares},
  {Vilardell}, {Wagner}, {Winkler}, {Zapatero Osorio}, {Zechmeister},
  {Ammler-von Eiff}, {Anglada-Escud{\'e}}, {del Burgo}, {Garcia-Vargas},
  {Klutsch}, {Lizon}, {Lopez-Morales}, {Ofir}, {P{\'e}rez-Calpena}, {Perryman},
  {S{\'a}nchez-Blanco}, {Strachan}, {St{\"u}rmer}, {Su{\'a}rez}, {Trifonov},
  {Tulloch}, \& {Xu}}]{Quirrenbach2016}
{Quirrenbach}, A., {Amado}, P.~J., {Caballero}, J.~A., {et~al.} 2016, in
  Society of Photo-Optical Instrumentation Engineers (SPIE) Conference Series,
  Vol. 9908, Ground-based and Airborne Instrumentation for Astronomy VI, ed.
  C.~J. {Evans}, L.~{Simard}, \& H.~{Takami}, 990812

\bibitem[{{Quirrenbach} {et~al.}(2018){Quirrenbach}, {Amado}, {Ribas},
  {Reiners}, {Caballero}, {Seifert}, {Aceituno}, {Azzaro}, {Baroch}, {Barrado},
  {Bauer}, {Becerril}, {B{\`e}jar}, {Ben{\'\i}tez}, {Brinkm{\"o}ller}, {Cardona
  Guill{\'e}n}, {Cifuentes}, {Colom{\'e}}, {Cort{\'e}s-Contreras}, {Czesla},
  {Dreizler}, {Fr{\"o}lich}, {Fuhrmeister}, {Galad{\'\i}-Enr{\'\i}quez},
  {Gonz{\'a}lez Hern{\'a}ndez}, {Gonz{\'a}lez Peinado}, {Guenther}, {de
  Guindos}, {Hagen}, {Hatzes}, {Hauschildt}, {Helmling}, {Henning}, {Herbort},
  {Hern{\'a}ndez Casta{\~n}o}, {Herrero}, {Hintz}, {Jeffers}, {Johnson}, {de
  Juan}, {Kaminski}, {Klahr}, {K{\"u}rster}, {Lafarga}, {Sairam}, {Lamp{\'o}n},
  {Lara}, {Launhardt}, {L{\'o}pez del Fresno}, {L{\'o}pez-Puertas}, {Luque},
  {Mandel}, {Marfil}, {Mart{\'\i}n}, {Mart{\'\i}n-Ruiz}, {Mathar}, {Montes},
  {Morales}, {Nagel}, {Nortmann}, {Nowak}, {Pall{\'e}}, {Passegger}, {Pavlov},
  {Pedraz}, {P{\'e}rez-Medialdea}, {Perger}, {Rebolo}, {Reffert},
  {Rodr{\'\i}guez}, {Rodr{\'\i}guez L{\'o}pez}, {Rosich}, {Sabotta}, {Sadegi},
  {Salz}, {S{\'a}nchez-L{\'o}pez}, {Sanz-Forcada}, {Sarkis}, {Sch{\"a}fer},
  {Schiller}, {Schmitt}, {Sch{\"o}fer}, {Schweitzer}, {Shulyak}, {Solano},
  {Stahl}, {Tala Pinto}, {Trifonov}, {Zapatero Osorio}, {Yan}, {Zechmeister},
  {Abell{\'a}n}, {Abril}, {Alonso-Floriano}, {Ammler-von Eiff},
  {Anglada-Escud{\'e}}, {Anwand-Heerwart}, {Arroyo-Torres}, {Berdi{\~n}as},
  {Bergondy}, {Bl{\"u}mcke}, {del Burgo}, {Cano}, {Carro}, {C{\'a}rdenas},
  {Casal}, {Claret}, {D{\'\i}ez-Alonso}, {Doellinger}, {Dorda}, {Feiz},
  {Fern{\'a}ndez}, {Ferro}, {Gaisn{\'e}}, {Gallardo}, {G{\'a}lvez-Ortiz},
  {Garc{\'\i}a-Piquer}, {Garc{\'\i}a-Vargas}, {Garrido}, {Gesa}, {G{\'o}mez
  Galera}, {Gonz{\'a}lez-{\'A}lvarez}, {Gonz{\'a}lez-Cuesta}, {Grohnert},
  {Gr{\"o}zinger}, {Gu{\`a}rdia}, {Guijarro}, {Hedrosa}, {Hermann}, {Hermelo},
  {Hern{\'a}ndez Arab{\'\i}}, {Hern{\'a}ndez Hernando}, {Hidalgo}, {Holgado},
  {Huber}, {Huber}, {Huke}, {Kehr}, {Kim}, {Klein}, {Kl{\"u}ter}, {Klutsch},
  {Labarga}, {Labiche}, {Lamert}, {Laun}, {L{\'a}zaro}, {Lemke}, {Lenzen},
  {Llamas}, {Lizon}, {Lodieu}, {L{\'o}pez Gonz{\'a}lez}, {L{\'o}pez-Morales},
  {L{\'o}pez Salas}, {L{\'o}pez-Santiago}, {Mag{\'a}n Madinabeitia}, {Mall},
  {Mancini}, {Mar{\'\i}n Molina}, {Mart{\'\i}nez-Rodr{\'\i}guez}, {Maroto
  Fern{\'a}ndez}, {Marvin}, {Mirabet}, {Moreno-Raya}, {Moya}, {Mundt},
  {Naranjo}, {Panduro}, {Pascual}, {P{\'e}rez-Calpena}, {Perryman}, {Pluto},
  {Ram{\'o}n}, {Redondo}, {Reinhart}, {Rhode}, {Rix}, {Rodler}, {Rohloff},
  {S{\'a}nchez-Blanco}, {S{\'a}nchez Carrasco}, {Sarmiento}, {Schmidt},
  {Storz}, {Strachan}, {St{\"u}rmer}, {Su{\'a}rez}, {Tabernero}, {Tal-Or},
  {Tulloch}, {Ulbrich}, {Veredas}, {Vico Linares}, {Vidal-Dasilva},
  {Vilardell}, {Wagner}, {Winkler}, {Wolthoff}, {Xu}, \&
  {Zhao}}]{Quirrenbach2018}
{Quirrenbach}, A., {Amado}, P.~J., {Ribas}, I., {et~al.} 2018, in Society of
  Photo-Optical Instrumentation Engineers (SPIE) Conference Series, Vol. 10702,
  Ground-based and Airborne Instrumentation for Astronomy VII, ed. C.~J.
  {Evans}, L.~{Simard}, \& H.~{Takami}, 107020W

\bibitem[{{Ragozzine} \& {Wolf}(2009)}]{Ragozzine2009}
{Ragozzine}, D. \& {Wolf}, A.~S. 2009, \apj, 698, 1778

\bibitem[{{Rainer} {et~al.}(2021){Rainer}, {Borsa}, {Pino}, {Frustagli},
  {Brogi}, {Biazzo}, {Bonomo}, {Carleo}, {Claudi}, {Gratton}, {Lanza},
  {Maggio}, {Maldonado}, {Mancini}, {Micela}, {Scandariato}, {Sozzetti},
  {Buchschacher}, {Cosentino}, {Covino}, {Ghedina}, {Gonzalez}, {Leto}, {Lodi},
  {Martinez Fiorenzano}, {Molinari}, {Molinaro}, {Nardiello}, {Oliva},
  {Pagano}, {Pedani}, {Piotto}, \& {Poretti}}]{Rainer2021}
{Rainer}, M., {Borsa}, F., {Pino}, L., {et~al.} 2021, \aap, 649, A29

\bibitem[{{Rauscher} \& {Menou}(2012)}]{Rauscher2012}
{Rauscher}, E. \& {Menou}, K. 2012, \apj, 750, 96

\bibitem[{{Ryabchikova} {et~al.}(2015){Ryabchikova}, {Piskunov}, {Kurucz},
  {Stempels}, {Heiter}, {Pakhomov}, \& {Barklem}}]{Ryabchikova2015}
{Ryabchikova}, T., {Piskunov}, N., {Kurucz}, R.~L., {et~al.} 2015, \physscr,
  90, 054005

\bibitem[{{Ryabchikova} {et~al.}(1997){Ryabchikova}, {Piskunov}, {Kupka}, \&
  {Weiss}}]{Ryabchikova1997}
{Ryabchikova}, T.~A., {Piskunov}, N.~E., {Kupka}, F., \& {Weiss}, W.~W. 1997,
  Baltic Astronomy, 6, 244

\bibitem[{{Savel} {et~al.}(2022){Savel}, {Kempton}, {Malik}, {Komacek}, {Bean},
  {May}, {Stevenson}, {Mansfield}, \& {Rauscher}}]{Savel2022}
{Savel}, A.~B., {Kempton}, E. M.~R., {Malik}, M., {et~al.} 2022, \apj, 926, 85

\bibitem[{{Scandariato} {et~al.}(2021){Scandariato}, {Borsa}, {Sicilia},
  {Malavolta}, {Biazzo}, {Bonomo}, {Bruno}, {Claudi}, {Covino}, {Di
  Marcantonio}, {Esposito}, {Frustagli}, {Lanza}, {Maldonado}, {Maggio},
  {Mancini}, {Micela}, {Nardiello}, {Rainer}, {Singh}, {Sozzetti}, {Affer},
  {Benatti}, {Bignamini}, {Biliotti}, {Capuzzo-Dolcetta}, {Carleo},
  {Cosentino}, {Damasso}, {Desidera}, {Garcia de Gurtubai}, {Ghedina},
  {Giacobbe}, {Giani}, {Harutyunyan}, {Hernandez}, {Hernandez Diaz}, {Knapic},
  {Leto}, {Mart{\'\i}nez Fiorenzano}, {Molinari}, {Nascimbeni}, {Pagano},
  {Pedani}, {Piotto}, {Poretti}, \& {Stoev}}]{Scandariato2021}
{Scandariato}, G., {Borsa}, F., {Sicilia}, D., {et~al.} 2021, \aap, 646, A159

\bibitem[{{Seidel} {et~al.}(2020){Seidel}, {Ehrenreich}, {Pino}, {Bourrier},
  {Lavie}, {Allart}, {Wyttenbach}, \& {Lovis}}]{Seidel2020}
{Seidel}, J.~V., {Ehrenreich}, D., {Pino}, L., {et~al.} 2020, \aap, 633, A86

\bibitem[{{Showman} \& {Polvani}(2011)}]{Showman2011}
{Showman}, A.~P. \& {Polvani}, L.~M. 2011, \apj, 738, 71

\bibitem[{{Shporer} {et~al.}(2014){Shporer}, {O'Rourke}, {Knutson},
  {Szab{\'o}}, {Zhao}, {Burrows}, {Fortney}, {Agol}, {Cowan}, {Desert},
  {Howard}, {Isaacson}, {Lewis}, {Showman}, \& {Todorov}}]{Shporer2014}
{Shporer}, A., {O'Rourke}, J.~G., {Knutson}, H.~A., {et~al.} 2014, \apj, 788,
  92

\bibitem[{{Snellen} {et~al.}(2014){Snellen}, {Brandl}, {de Kok}, {Brogi},
  {Birkby}, \& {Schwarz}}]{Snellen2014}
{Snellen}, I.~A.~G., {Brandl}, B.~R., {de Kok}, R.~J., {et~al.} 2014, \nat,
  509, 63

\bibitem[{{Southworth}(2008)}]{Southworth2008}
{Southworth}, J. 2008, \mnras, 386, 1644

\bibitem[{{Stephan} {et~al.}(2022){Stephan}, {Wang}, {Cauley}, {Gaudi},
  {Ilyin}, {Johnson}, \& {Strassmeier}}]{Stephan2022}
{Stephan}, A.~P., {Wang}, J., {Cauley}, P.~W., {et~al.} 2022, arXiv e-prints,
  arXiv:2203.02546

\bibitem[{{Stevenson}(2016)}]{Stevenson2016}
{Stevenson}, K.~B. 2016, \apjl, 817, L16

\bibitem[{{Stevenson} {et~al.}(2014){Stevenson}, {D{\'e}sert}, {Line}, {Bean},
  {Fortney}, {Showman}, {Kataria}, {Kreidberg}, {McCullough}, {Henry},
  {Charbonneau}, {Burrows}, {Seager}, {Madhusudhan}, {Williamson}, \&
  {Homeier}}]{Stevenson2014}
{Stevenson}, K.~B., {D{\'e}sert}, J.-M., {Line}, M.~R., {et~al.} 2014, Science,
  346, 838

\bibitem[{{Stock} {et~al.}(2018){Stock}, {Kitzmann}, {Patzer}, \&
  {Sedlmayr}}]{Stock2018}
{Stock}, J.~W., {Kitzmann}, D., {Patzer}, A. B.~C., \& {Sedlmayr}, E. 2018,
  \mnras, 479, 865

\bibitem[{{Tan} \& {Komacek}(2019)}]{Tan2019}
{Tan}, X. \& {Komacek}, T.~D. 2019, \apj, 886, 26

\bibitem[{{Toon} {et~al.}(1989){Toon}, {McKay}, {Ackerman}, \&
  {Santhanam}}]{Toon1989}
{Toon}, O.~B., {McKay}, C.~P., {Ackerman}, T.~P., \& {Santhanam}, K. 1989,
  \jgr, 94, 16287

\bibitem[{{van Sluijs} {et~al.}(2022){van Sluijs}, {Birkby}, {Lothringer},
  {Lee}, {Crossfield}, {Parmentier}, {Brogi}, {Kulesa}, {McCarthy}, {Powell},
  \& {Charbonneau}}]{vanSluijs2022}
{van Sluijs}, L., {Birkby}, J.~L., {Lothringer}, J., {et~al.} 2022, arXiv
  e-prints, arXiv:2203.13234

\bibitem[{{von Essen} {et~al.}(2020){von Essen}, {Mallonn}, {Borre}, {Antoci},
  {Stassun}, {Khalafinejad}, \& {Tautvai{\v{s}}ien{\.{e}}}}]{VonEssen2020}
{von Essen}, C., {Mallonn}, M., {Borre}, C.~C., {et~al.} 2020, \aap, 639, A34

\bibitem[{{Wardenier} {et~al.}(2022){Wardenier}, {Parmentier}, \&
  {Lee}}]{Wardenier2022}
{Wardenier}, J.~P., {Parmentier}, V., \& {Lee}, E. K.~H. 2022, \mnras, 510, 620

\bibitem[{{Wardenier} {et~al.}(2021){Wardenier}, {Parmentier}, {Lee}, {Line},
  \& {Gharib-Nezhad}}]{Wardenier2021}
{Wardenier}, J.~P., {Parmentier}, V., {Lee}, E. K.~H., {Line}, M.~R., \&
  {Gharib-Nezhad}, E. 2021, \mnras, 506, 1258

\bibitem[{{Wong} {et~al.}(2020){Wong}, {Shporer}, {Kitzmann}, {Morris}, {Heng},
  {Hoeijmakers}, {Demory}, {Ahlers}, {Mansfield}, {Bean}, {Daylan},
  {Fetherolf}, {Rodriguez}, {Benneke}, {Ricker}, {Latham}, {Vanderspek},
  {Seager}, {Winn}, {Jenkins}, {Burke}, {Christiansen}, {Essack}, {Rose},
  {Smith}, {Tenenbaum}, \& {Yahalomi}}]{Wong2020}
{Wong}, I., {Shporer}, A., {Kitzmann}, D., {et~al.} 2020, \aj, 160, 88

\bibitem[{{Yan} {et~al.}(2020){Yan}, {Pall{\'e}}, {Reiners}, {Molaverdikhani},
  {Casasayas-Barris}, {Nortmann}, {Chen}, {Molli{\`e}re}, \&
  {Stangret}}]{Yan2020}
{Yan}, F., {Pall{\'e}}, E., {Reiners}, A., {et~al.} 2020, \aap, 640, L5

\bibitem[{{Zhang} {et~al.}(2017){Zhang}, {Kempton}, \& {Rauscher}}]{Zhang2017}
{Zhang}, J., {Kempton}, E. M.~R., \& {Rauscher}, E. 2017, \apj, 851, 84

\bibitem[{{Zhang} {et~al.}(2018){Zhang}, {Knutson}, {Kataria}, {Schwartz},
  {Cowan}, {Showman}, {Burrows}, {Fortney}, {Todorov}, {Desert}, {Agol}, \&
  {Deming}}]{Zhang2018}
{Zhang}, M., {Knutson}, H.~A., {Kataria}, T., {et~al.} 2018, \aj, 155, 83

\bibitem[{{Zhang} {et~al.}(2022){Zhang}, {Snellen}, {Wyttenbach}, {Nielsen},
  {Lendl}, {Casasayas-Barris}, {Chaverot}, {Kesseli}, {Lovis}, {Pepe},
  {Psaridi}, {Seidel}, {Udry}, \& {Ulmer-Moll}}]{Zhang2022}
{Zhang}, Y., {Snellen}, I. A.~G., {Wyttenbach}, A., {et~al.} 2022, arXiv
  e-prints, arXiv:2208.11427

\end{thebibliography}

\begin{appendix} 
\section{The choice between weighted mask cross-correlation, and cross-correlation-likelihood mapping}
\label{sec_appendix: mask_vs_mapping}
We refer the reader to \cite{Pino2020} for a detailed explanation of the weighted mask method. \cite{Pino2020} also introduced a statistical framework to interpret this flavour of cross-correlation, and compared it to the cross-correlation-likelihood mapping. Here, we summarize the main differences between both techniques (see also \citealt{Allart2020}), and provide our rationale for using the cross-correlation-likelihood mapping in this work.

The weighted mask builds a `weighted average line profile' for the planet normalized to the stellar plus planetary continuum. It is more model-independent, since it does not rely on the shape of the lines, but only on their position and strength. Indeed, to compare models to data, the mask has to be separately applied to both, and the quality of the fit can be intuitively established also by visually comparing the average profiles found in data and models. For example, \cite{Pino2020} used this method to verify that the broadening of the average line profile in HARPS-N N1 is reproduced assuming the planetary rotational period derived from the tidally-locked assumption. Similarly, \cite{Ehrenreich2020} and \cite{Rainer2021} observed the variation of strength and broadening of the average iron line-profile during the transits of WASP-76b and KELT-20b, which could be physically interpreted without relying on a specific model.

The cross-correlation-likelihood mapping scheme by \cite{Brogi2019}, on the other hand, exploits the full information contained in the line profiles. To each model, it associates a likelihood, and the goodness of fit of different models can be compared by applying classic statistical tools. The precision reached on model parameters is, in general, superior compared to the weighted mask method, because more information is used to constrain them \citep{Pino2020}. However, it is not possible to visually judge the quality of a fit to data, since the signal is buried in the noise.

Establishing whether a model is a good fit to data (e.g., determining the significance of a detection) and determining confidence intervals on parameters of a given model are two conceptually separated problems. When calculating confidence intervals, the hidden assumption is that the parametric model employed provides an adequate description of the data. For this reason, a blind application of cross-correlation-likelihood mappings similar to \cite{Brogi2019} and \cite{Gibson2020} can be dangerous: while the determined confidence intervals could be precise, they may also be inaccurate. We argue that an additional assessment of the adequacy of the model employed within such scheme is crucial (see also \citealt{Pelletier2021}). In this work, \cite{Pino2020} provide the confidence in our model to satisfactorily reproduce the weighted mask profile. This foundational work justifies the use of the \cite{Brogi2019} cross-correlation-likelihood mapping scheme to maximize the precision reached on the retrieved parameters, which is key to the goals of this paper.


\section{Testing the sensitivity to an asymmetric phase curve}
\label{sec_appendix: injection}
In this section, we perform an injection-retrieval test of the Loff model (see Table \ref{table:models adopted} and Section \ref{sec:modelling}) with an eccentric orbital solution and a non-zero offset from the substellar point. This test is aimed at demonstrating that an asymmetric phase curve is indeed detectable despite the loss of sensitivity towards quadrature, and therefore our claimed lack of asymmetry is robust.

We inject a signal with $\varphi_0$=0.1 ($\sim$36$^\circ)$, $e$=0.02, $\omega$=135$^\circ$ (corresponding to $h$=0.1, $k$=$-0.1$), $S_\mathrm{lambert}$=2.85, $v_\mathrm{sys}$=$-16.5$ km s$^{-1}$, and $K_\mathrm{p}$=$-242$ km s$^{-1}$. The negative $K_\mathrm{p}$ allows us to test the same rate of planetary RV change as the actual signal (thus sampling noise similarly), albeit with the opposite slope in order not to interfere with the actual signal.  The injection is done on the pipeline-extracted spectra, which are subsequently passed through the data analysis described in Section \ref{Sec: data reduction} and thus processed in the same way as the real data. Sampling of the posterior of the Loff model is performed via MCMC to verify that the retrieved parameters are consistent with the injection.

We correctly recover the injected signal, albeit at a lower precision in velocity space than the actual observations ($v_\mathrm{sys}=-15.2^{+0.8}_{-0.7}$ km s$^{-1}$, $K_\mathrm{p}=-244^{+3}_{-2}$ km s$^{-1}$), and we also recover the exact amplitude scaling factor ($S_\mathrm{lambert}=3.0^{+0.2}_{-0.15}$). Furthermore, we successfully measure a phase shift of the maximum emission from the sub-stellar point ($\varphi_0=0.13\pm0.02$), compatible at 1.4$\sigma$ with the injected value and more importantly incompatible at $\ge4\sigma$ with a symmetric phase curve. We are therefore confident on our ability to detect an asymmetry, and on the lack of such asymmetry in the phase curve of KELT-9\,b.

\begin{figure*}
  \centering
  \resizebox{\hsize}{!}
  {\includegraphics[width=\hsize]{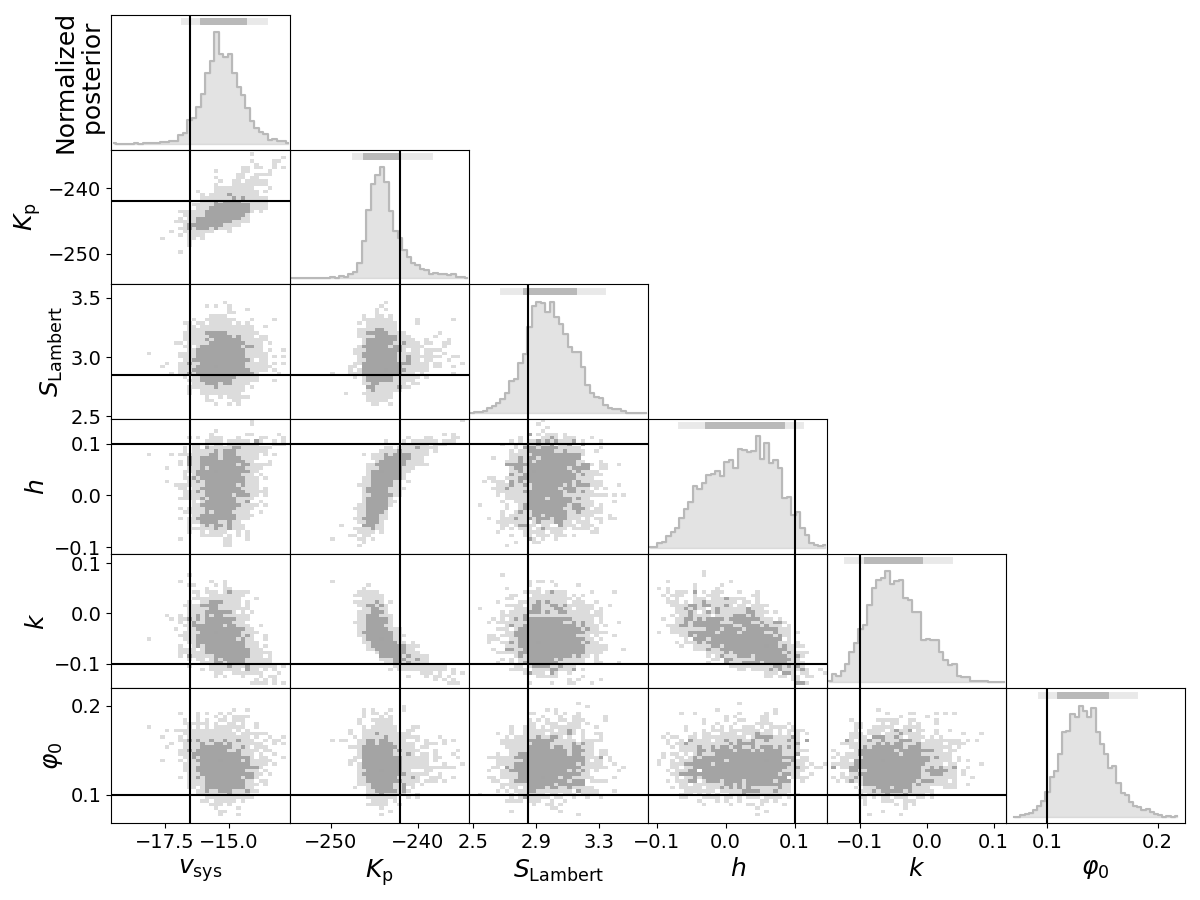}}
      \caption{Corner plot showing posteriors for our injection test of model Loff. Vertical and horizontal black lines represent the true values at which our model was injected into the data. Despite the correlation among the parameters, our analysis is capable of retrieving the correct answer within $2\sigma$ for every parameter.}
         \label{fig:corner_injection}
  \end{figure*}

Lastly, we do observe a moderate correlation between the eccentric parameters ($h,k$) and the two velocities ($v_\mathrm{sys},K_\mathrm{p}$; see Fig. \ref{fig:corner_injection}). Such correlation marginally biases the retrieved values ($h=0.03^{+0.05}_{-0.06}$, compatible at 1.4$\sigma$; $k=-0.06^{+0.05}_{-0.04}$, compatible at 1$\sigma$), resulting in a marginal preference for an eccentric solution ($e=0.006^{+0.011}_{-0.004}$, $\omega=154^{+81}_{-34}$ degrees). At the level of this test, it is hard to tell whether the lower precision compared to the real observations is due to the particular combination of $h,k$ chosen for the injection, an unfortunate noise pattern in the data, or the fact that the real signal lacks a strong phase variation, and it is therefore stronger than this injected signal. We leave this analysis for future work. However, we still successfully demonstrated the ability to measure a (shifted) phase curve at high spectral resolution.

\section{Comparison of intrinsic information content in HARPS-N and CARMENES}
\label{sec_appendix: intrisic_information_content}

In this section, we investigate the distribution of strong neutral iron lines across the wavelength range probed by HARPS-N and CARMENES.

We downloaded the VALD3 line list of neutral iron, and the partition function by \cite{Barklem2016}. We then computed the strength of each \ion{Fe}{i} line at $4000~\mathrm{K}$, a temperature representative of the day-side atmosphere of KELT-9b, using standard formulas (e.g., \citealt{Grimm2015}), and assuming LTE. We argue that the line strength is a good proxy for the signal carried by each line, which is approximately true in the lack of strong blending, assuming that all lines probe similar pressures and that they are not saturated. The spectrum of KELT-9b should be dominated by atomic lines, so we argue that the effect of blending should not be too dramatic. \cite{Pino2020} show the range of pressures probed by \ion{Fe}{i} lines observed with HARPS-N, which we expect to be similar to the range probed by CARMENES.

In addition, we take into account the planet-to-star flux ratio variation with wavelength, which is more favourable towards the red wavelengths probed by CARMENES. For this analysis, we weight the line strengths by the ratio of two blackbodies at temperatures of $4000~\mathrm{K}$, to represent the planet, and $10000~\mathrm{K}$, to represent the star.

\begin{figure}
  \centering
  \includegraphics[width=\hsize]{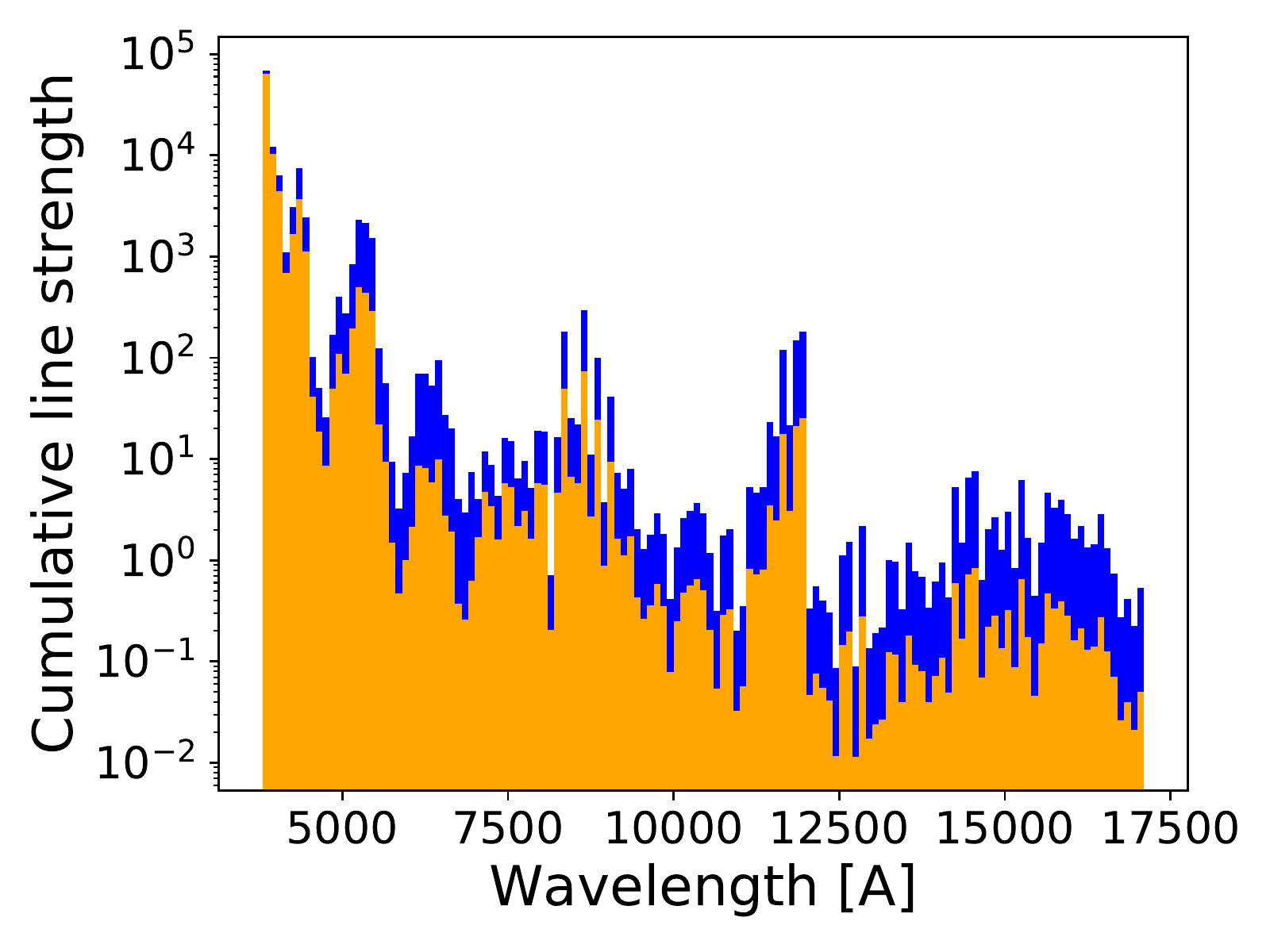}
      \caption{Cumulative line strength of neutral iron lines at $4000~\mathrm{K}$ across the wavelength range probed by HARPS-N and CARMENES. The orange histogram only accounts for the line strengths, while the blue histogram additionally accounts for the planet-to-star flux ratio.}
         \label{fig:line_strenghts}
  \end{figure}

Fig. \ref{fig:line_strenghts} shows a histogram of the line list weighted by the line strength at $4000~\mathrm{K}$, in arbitrary units. The blue histogram is additionally weighted by the planet-to-star flux ratio. While this additional factor increases the importance of the wavelength range probed by CARMENES compared to HARPS-N, blue wavelengths still seem to contain more information about \ion{Fe}{i} for KELT-9b. However, this conclusion can not be generalized to other planets (e.g., see \citealt{Herman2022}), since it depends on the location of the contribution functions of lines (hence, on the temperature-pressure profile and volume mixing ratio), and on blanketing and blending effects by other species which are likely weaker in KELT-9b. 

\section{Corner plots for individual night retrievals with model 1C}
\label{sec_appendix: individual nights}
In this section, we show corner plots displaying all the posteriors obtained for individual nights.

  \begin{figure}
  \centering
  \includegraphics[width=\hsize]{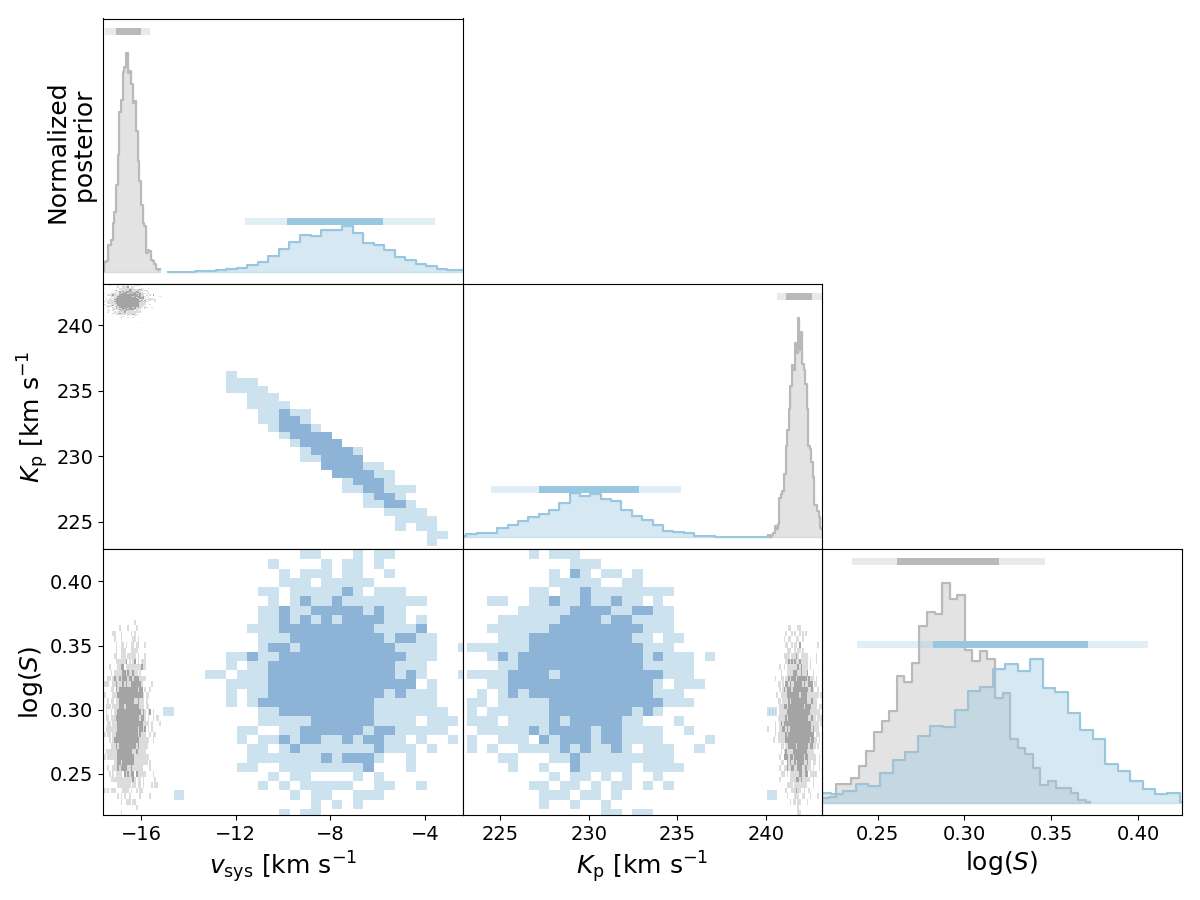}
      \caption{Posterior for the MCMC fit to HARPS-N N1 using model 1C (blue), compared to 5 nights combined (grey).}
         \label{fig:night 1}
  \end{figure}

  \begin{figure}
  \centering
  \includegraphics[width=\hsize]{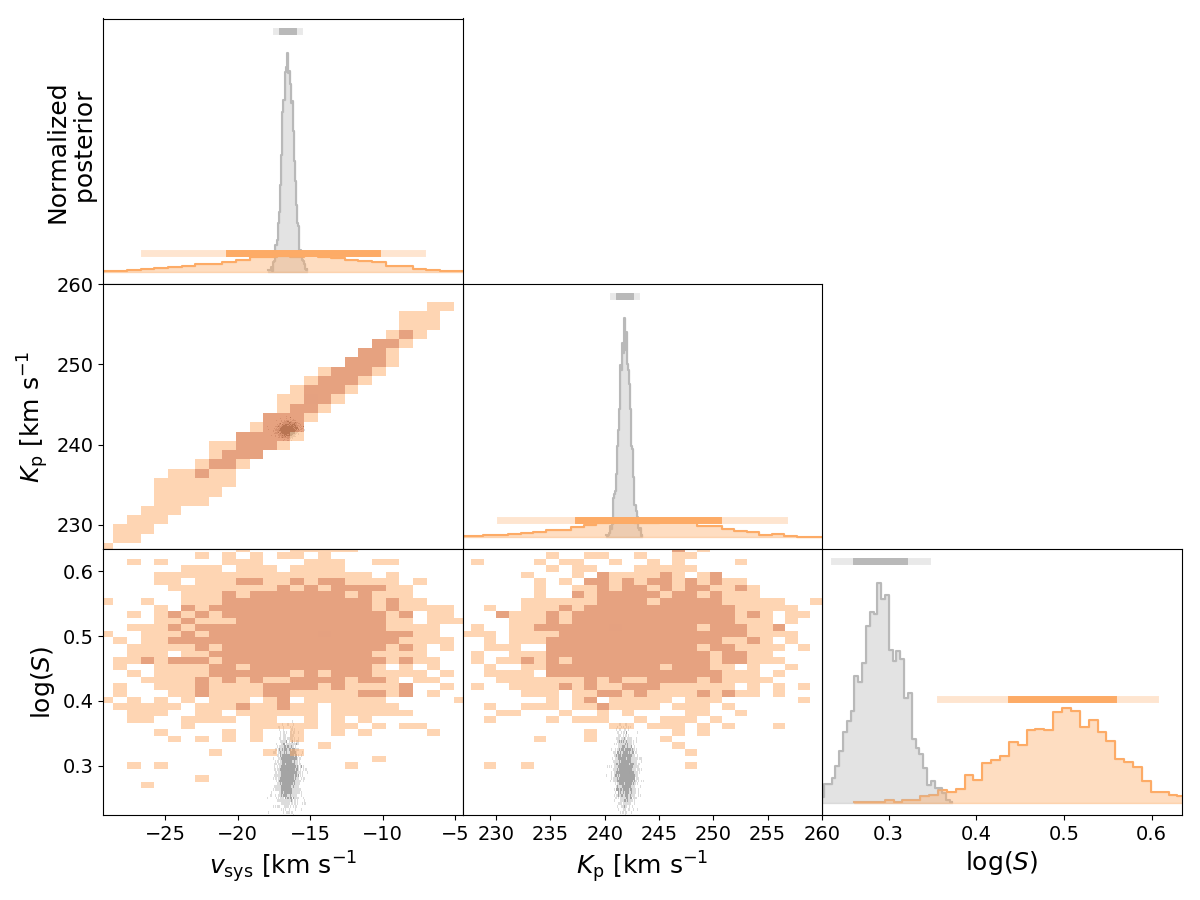}
      \caption{Posterior for the MCMC fit to HARPS-N N2 using model 1C (orange), compared to 5 nights combined (grey).}
         \label{fig:night 2}
  \end{figure}
  
    \begin{figure}
  \centering
  \includegraphics[width=\hsize]{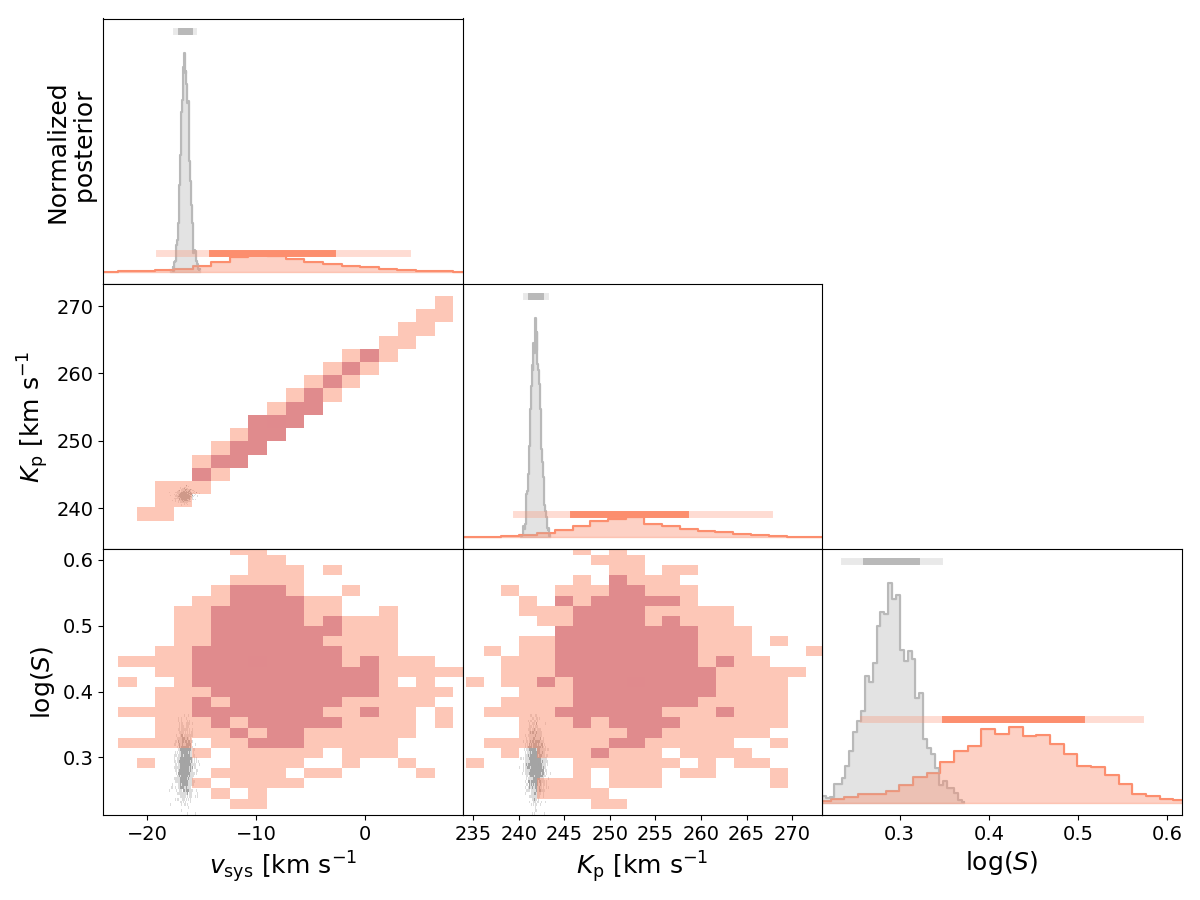}
      \caption{Posterior for the MCMC fit to CARMENES N1 using model 1C (red), compared to 5 nights combined (grey).}
         \label{fig:night 3}
  \end{figure}
  
    \begin{figure}
  \centering
  \includegraphics[width=\hsize]{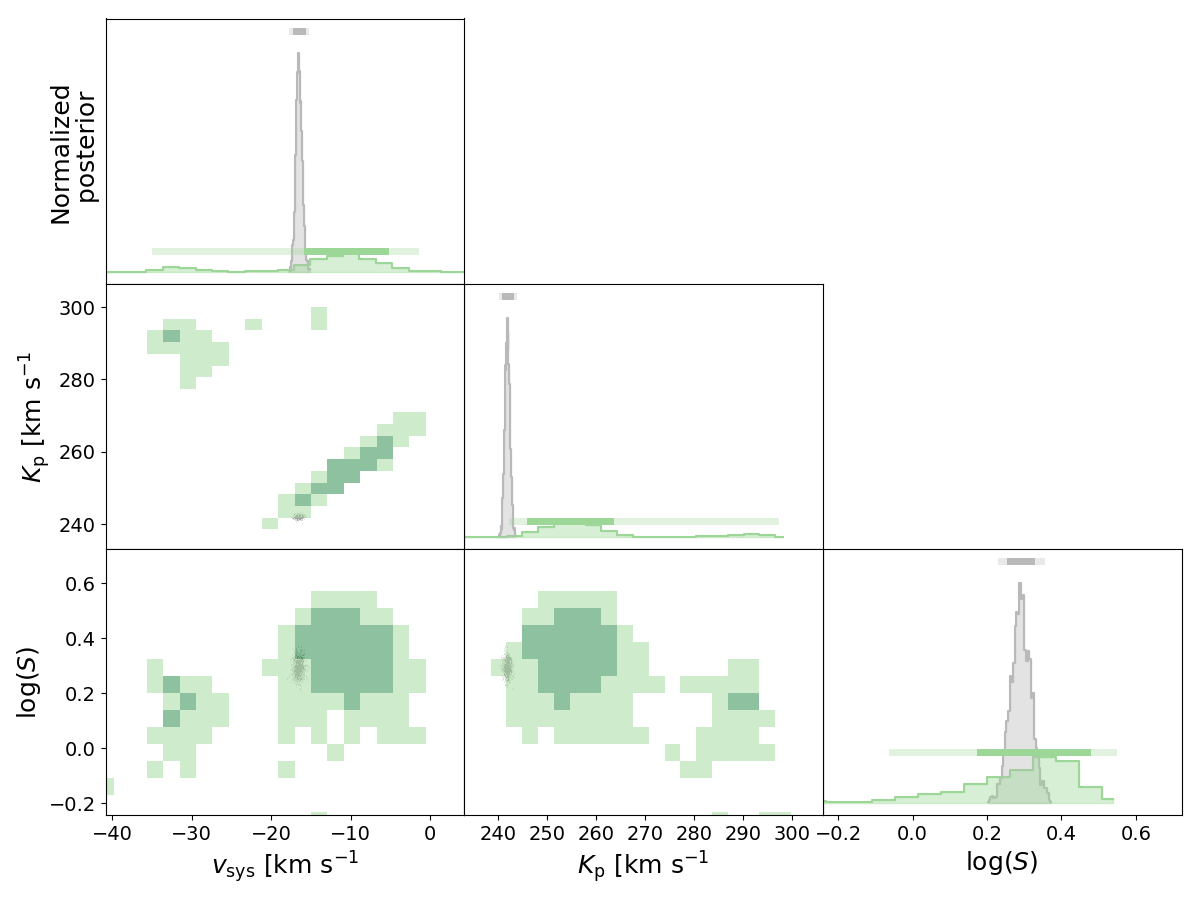}
      \caption{Posterior for the MCMC fit to CARMENES N2 using model 1C (green), compared to 5 nights combined (grey).}
         \label{fig:night 4}
  \end{figure}
  
    \begin{figure}
  \centering
  \includegraphics[width=\hsize]{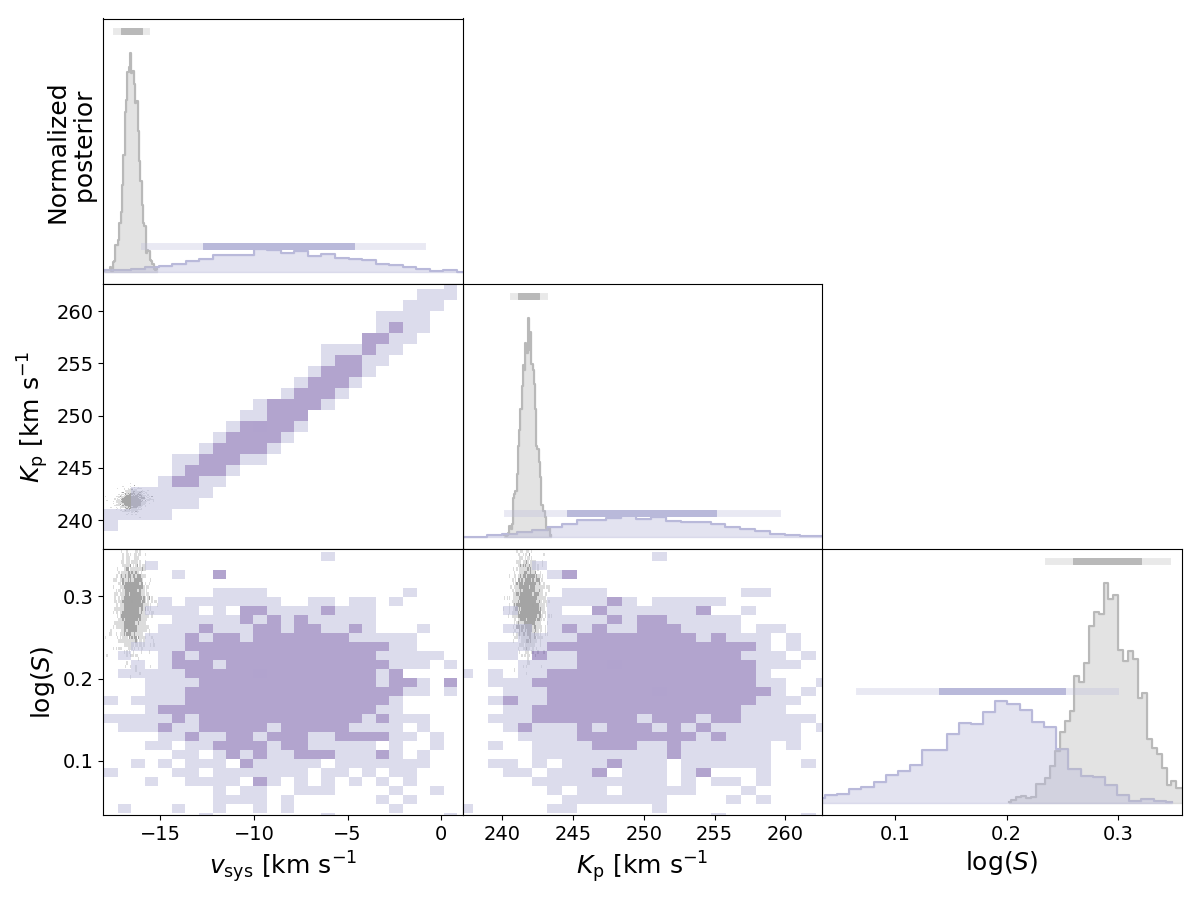}
      \caption{Posterior for the MCMC fit to HARPS-N N3 using model 1C (purple), compared to 5 nights combined (grey).}
         \label{fig:night 5}
  \end{figure}

\section{Corner plots for models 4C and 4E, and Slam, Soff and Sbase}
\label{sec_appendix:phase_dependent_intensity_corner_plots}
In this section, we show corner plots displaying all the posteriors obtained for 4 scale factor models and for the Lambert sphere model variations.

  \begin{figure*}
  \resizebox{\hsize}{!}
            {\includegraphics{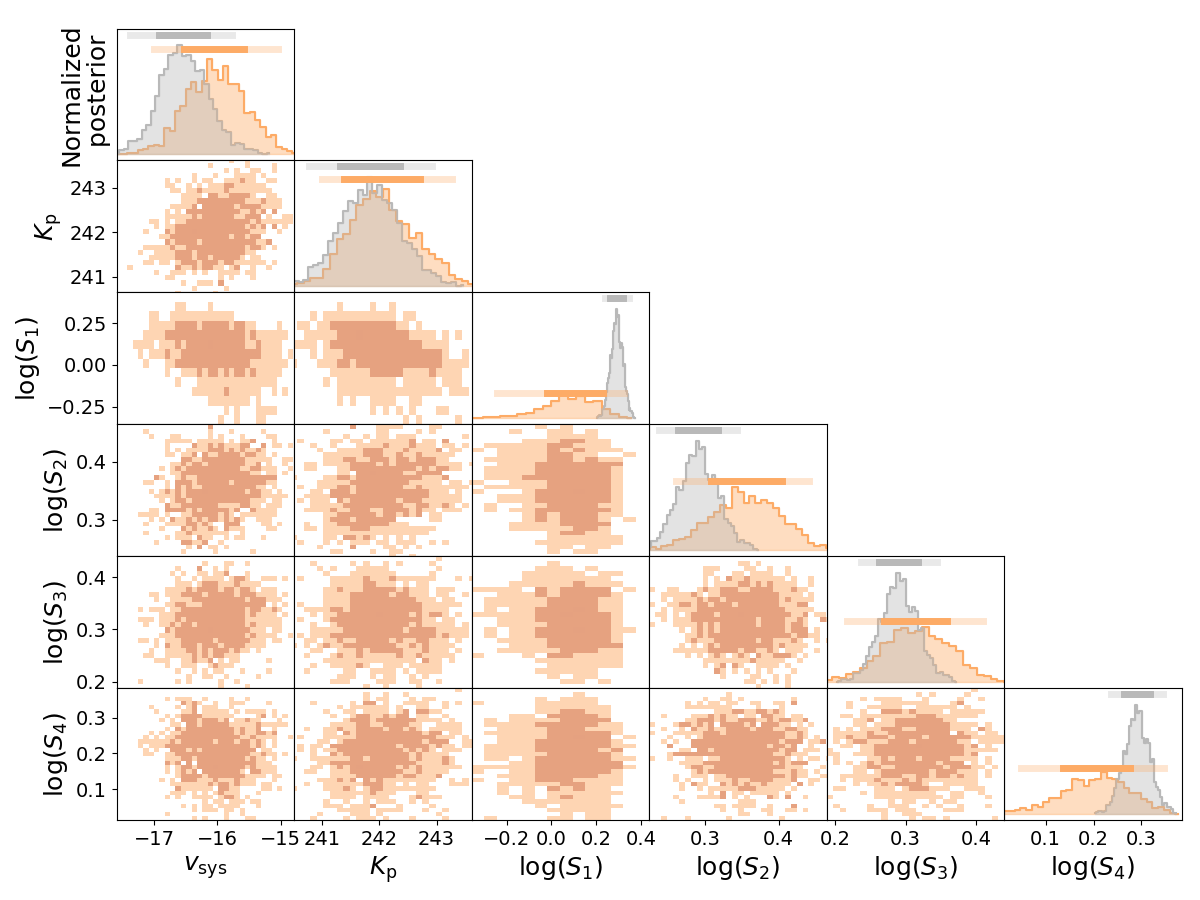}}
      \caption{Corner plot showing posteriors for model 4C (orange). We overlay 1D posteriors for model 1C in grey. The posteriors for $\log S_1 \dots \log S_4$ are all compared to the posterior for the individual scaling factor $\log S$ of model 1C. We zoom around the $1\sigma$ and $2\sigma$ confidence interval regions for each parameter.}\label{fig:4C_vs_1C}
  \end{figure*}

  \begin{figure*}
  \resizebox{\hsize}{!}
            {\includegraphics{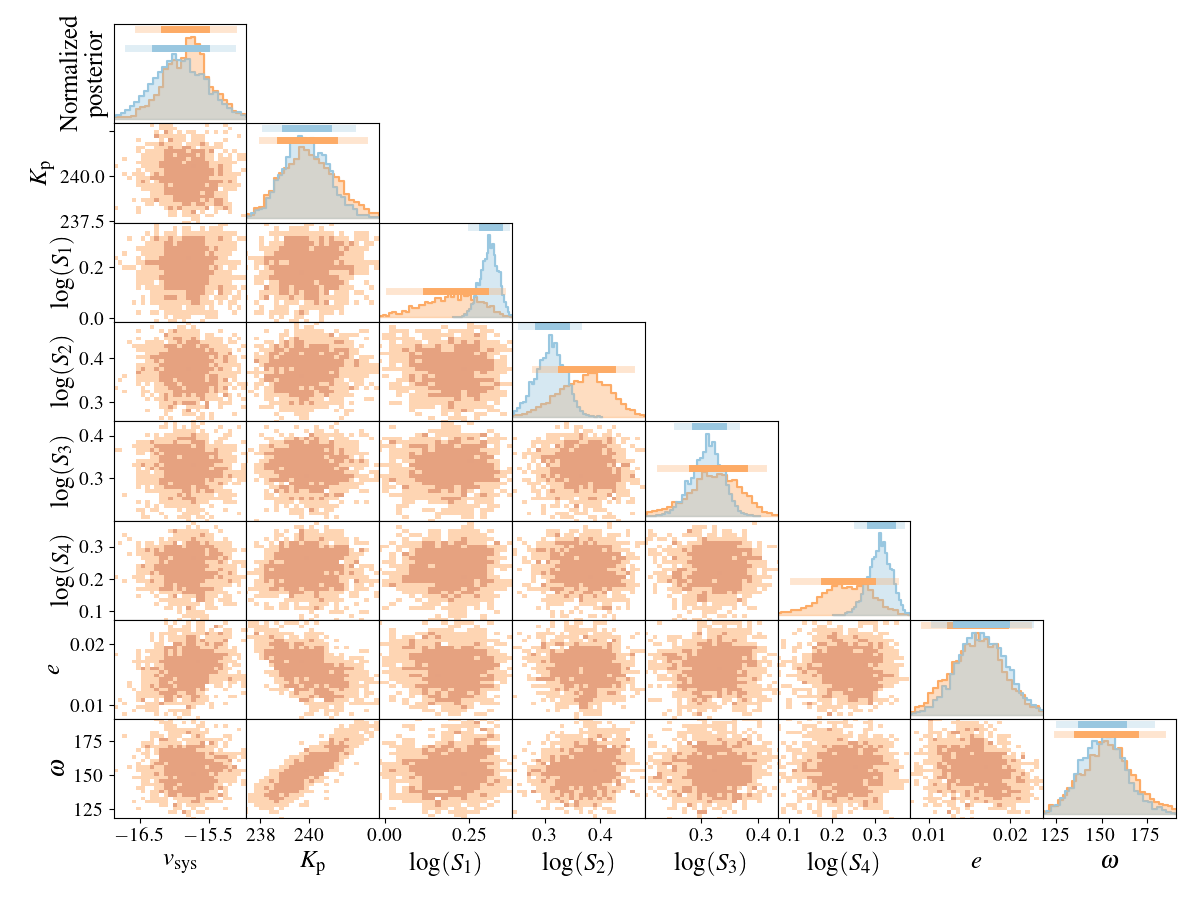}}
      \caption{Corner plot showing posteriors for model 4E (orange). We overlay 1D posteriors for model 1E in grey. The posteriors for $\log S_1 \dots \log S_4$ are all compared to the posterior for the individual scaling factor $\log S$ of model 1E. We zoom around the $1\sigma$ and $2\sigma$ confidence interval regions for each parameter.}\label{fig:4E_vs_1E}
  \end{figure*}

  \begin{figure*}
  \resizebox{\hsize}{!}
            {\includegraphics{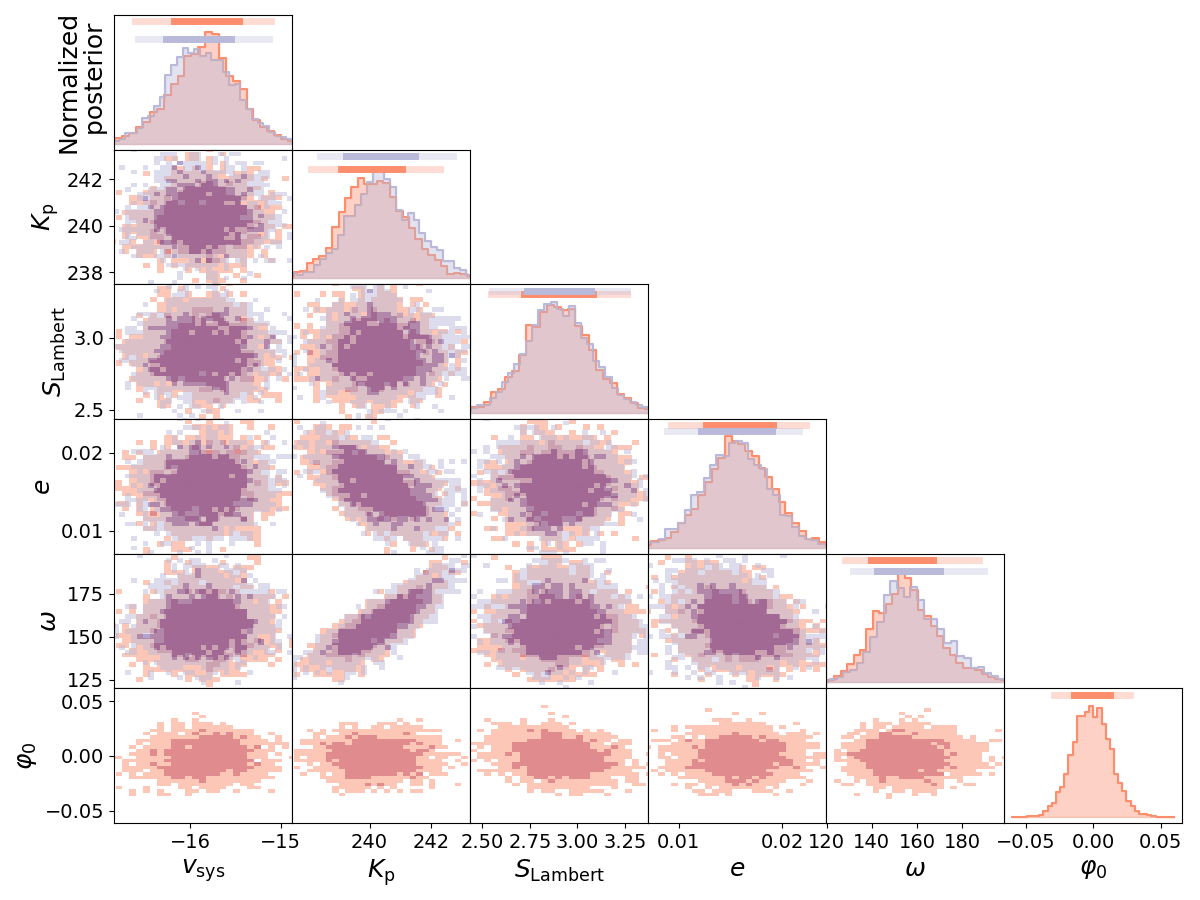}}
      \caption{Corner plot showing posteriors for model Loff (red). We overlay posteriors for model L in purple. We zoom around the $1\sigma$ and $2\sigma$ confidence interval regions for each parameter.}\label{fig:Slam_vs_Soff}
  \end{figure*}

  \begin{figure*}
  \resizebox{\hsize}{!}
            {\includegraphics{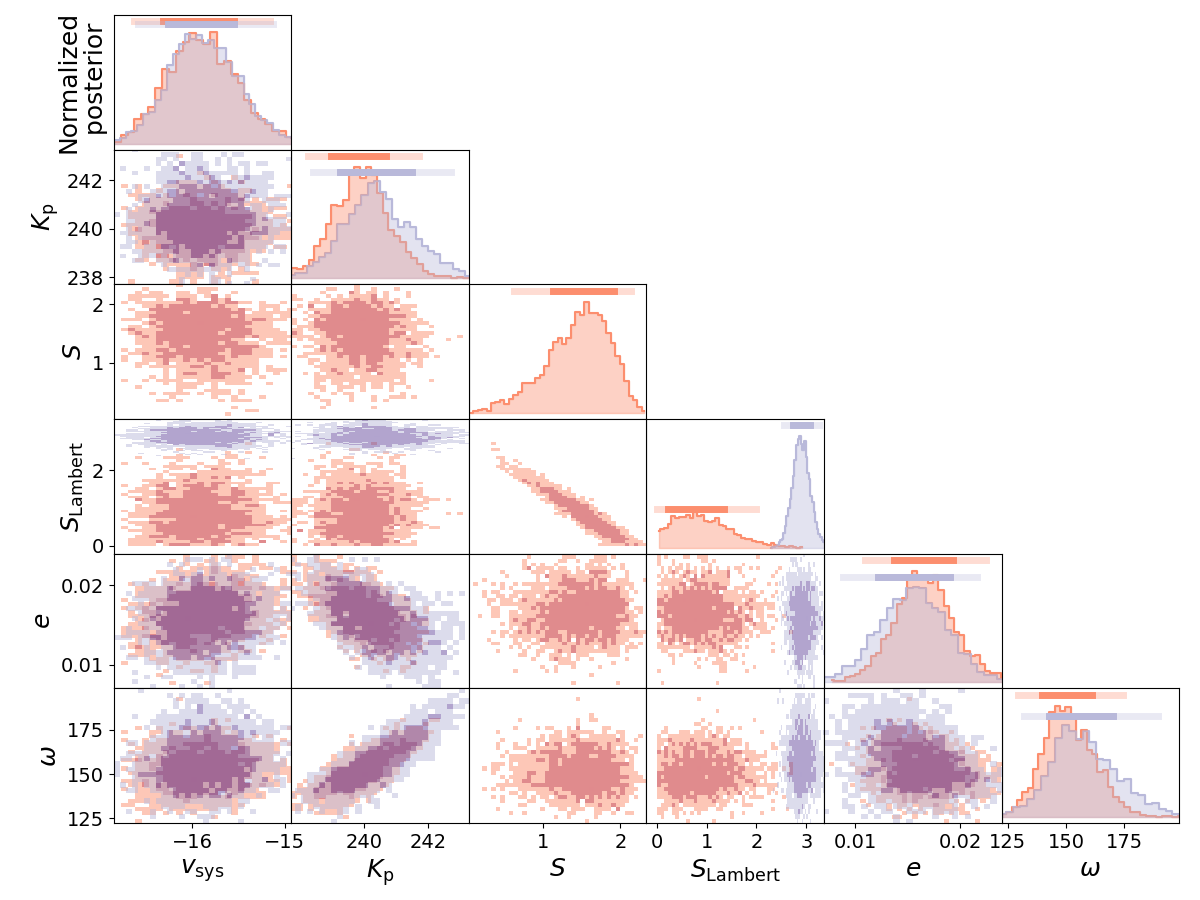}}
            \caption{Corner plot showing posteriors for model Lbase (red). We overlay posteriors for model Slam in purple. We zoom around the $1\sigma$ and $2\sigma$ confidence interval regions for each parameter. The limit of our prior ($S_\mathrm{Lambert}>0$) is visible for the L model. We discuss the clear correlation between $S$ and $S_\mathrm{Lambert}$ in section \ref{sec: results_symmetry}.\label{fig:Slam_vs_Sbase}}
  \end{figure*}

\section{Best-fit model parameter values and errors.}
In Table \ref{table:retrieval results} we report the full results of our retrievals on the 5 HARPS-N and CARMENES combined nights analysed. Each entry represents the fiftieth percentile of the Monte Carlo samples for each parameter, while $1\sigma$ intervals define the highest density region of the parameter range such that 95\% of the sample are found in that region.

\begin{sidewaystable*}
\caption{Model name (see Table \ref{table:models adopted}), best fit parameters and errors. A value of ``X'' indicates that the parameter is not part of the model. \label{table:retrieval results}}
\centering
\begin{tabular}{cccccccccccc}
\hline \hline
Model name & $K_\mathrm{p}~[\mathrm{km~s^{-1}}]$ & $v_\mathrm{sys}~[\mathrm{km~s^{-1}}]$ & $\log S$ (*$S$) & $e$ & $\omega~[^{\circ}]$ & $\log S_1$ & $\log S_2$ & $\log S_3$ & $\log S_4$ & $S_\mathrm{lambert}$ & $\varphi_0$\\ 
\hline
\textbf{1C} & $241.8^{+0.5}_{-0.5}$ & $-16.5^{+0.4}_{-0.4}$ & $0.29^{+0.03}_{-0.03}$ & X & X & X & X & X & X & X & X\\
%
\textbf{1E} & $239.9^{+0.8}_{-1}$ & $-15.9^{+0.3}_{-0.4}$ & $0.31^{+0.02}_{-0.03}$ & $0.016^{+0.003}_{-0.003}$ & $150^{+10}_{-10}$ & X & X & X & X & X & X\\
%
\textbf{4C} & $241.0^{+0.65}_{-0.7}$ & $-16.0^{+0.5}_{-0.45}$ & X & X & X & $0.1^{+0.1}_{-0.1}$ & $0.35^{+0.05}_{-0.05}$ & $0.31^{+0.05}_{-0.04}$ & $0.21^{+0.07}_{-0.07}$ & X & X\\ 
\textbf{4E} & $240^{+1}_{-1}$ & $-15.8^{+0.4}_{-0.25}$ & X & $0.016^{+0.003}_{-0.003}$ & $150^{+15}_{-10}$ & $0.20^{+0.08}_{-0.1}$ & $0.37^{+0.04}_{-0.05}$ & $0.32^{+0.04}_{-0.05}$ & $0.23^{+0.05}_{-0.06}$ & X & X\\
\textbf{L} & $240^{+1}_{-1}$ & $-15.8^{+0.3}_{-0.3}$ & X & $0.016^{+0.003}_{-0.003}$ & $150^{+10}_{-10}$ & X & X & X & X & $2.9^{+0.2}_{-0.2}$ & X\\
\textbf{Loff} & $240.1^{+1}_{-0.9}$ & $-15.8^{+0.35}_{-0.4}$ & X & $0.016^{+0.003}_{-0.003}$ & $150^{+15}_{-10}$ & X & X & X & X & $2.9^{+0.2}_{-0.2}$ & $0.00^{+0.01}_{-0.01}$\\
\textbf{Lbase} & $239.9^{+1}_{-0.8}$ & $-15.9^{+0.4}_{-0.35}$ & \tablefootmark{*}$1.5^{+0.4}_{-0.4}$ & $0.016^{+0.003}_{-0.003}$ & $150^{+10}_{-10}$ & X & X & X & X & $0.9^{+0.6}_{-0.5}$ & X\\ 
\hline
\end{tabular} 
\tablefoot{\tablefoottext{*}{For model Lbase, S is displayed in linear space and not in logarithmic space. This allows a direct comparison with the parameter $S_\mathrm{lambert}$, to compare the intensity of the constant and phase-dependent portions of the scale factor.}}

\end{sidewaystable*}

\end{appendix}

\end{document}